\documentclass[reqno,a4paper,11pt]{amsart}
\usepackage[english]{babel}
\renewcommand{\baselinestretch}{1.1}
\usepackage{graphicx,mathrsfs,tikz,latexsym,ifthen,amsmath,amsfonts,amssymb,amsthm,stmaryrd,fancyhdr}
\usepackage{enumerate,enumitem,empheq,mathtools,times,tikz}
\mathtoolsset{showonlyrefs,showmanualtags}
\usepackage[utf8x]{inputenc}
\usepackage[toc,page]{appendix}
\usepackage[pdftex]{hyperref}

\allowdisplaybreaks

\usepackage{todonotes}

\input xy
\xyoption{all}

\usepackage{color}
\definecolor{MyDarkBlue}{rgb}{0.15,0.25,0.45}
\usepackage{hyperref}
\hypersetup{
colorlinks=true,
citecolor=black,
linkcolor=black,
urlcolor=MyDarkBlue,
breaklinks=true
}

\newcommand{\hilb}[2]{\mathrm{Hilb}^{#1}(#2)}

\newcommand{\mbf}{\boldsymbol}

\newcommand{\Ecal}{{\mathcal E}}

\newcommand{\Ncal}{{\mathcal N}}
\newcommand{\Mcal}{\mathcal{M}}

\newcommand{\Zcal}{{\mathcal Z}}
\newcommand{\Ocal}{{\mathcal O}}

\newcommand{\Fcal}{{\mathcal F}}

\newcommand{\Rcal}{{\mathcal R}}
\newcommand{\Lcal}{{\mathcal L}}

\newcommand{\Pcal}{{\mathcal P}}

\newcommand{\hfrak}{\mathfrak{h}}

\newcommand{\Qfrak}{\mathfrak{Q}}
\newcommand{\Pfrak}{\mathfrak{P}}

\newcommand{\slfrak}{\mathfrak{sl}}
\newcommand{\glfrakhat}{\widehat{\mathfrak{gl}}}

\newcommand{\Ufrak}{\mathfrak{U}}

\newcommand{\R}{{\mathbb{R}}}

\newcommand{\C}{{\mathbb{C}}}
\newcommand{\Z}{{\mathbb{Z}}}
\newcommand{\Q}{{\mathbb{Q}}}
\newcommand{\T}{{\mathbb{T}}}
\newcommand{\PP}{{\mathbb{P}}}
\newcommand{\A}{{\mathbb{A}}}

\newcommand{\F}{{\mathbb{F}}}

\newcommand{\Tscr}{\mathscr{T}}

\newcommand{\Xscr}{{\mathscr{X}}}

\newcommand{\Dscr}{{\mathscr{D}}}
\newcommand{\Bscr}{{\mathscr{B}}}

\newcommand{\qsf}{\mathsf{q}}

\newcommand{\Csf}{\mathsf{C}}
\newcommand{\Dsf}{\mathsf{D}}

\newcommand{\quiv}{{\tt Q}}
\newcommand{\source}{{\tt s}}
\newcommand{\tail}{{\tt t}}

\newcommand{\Ebf}{\boldsymbol{E}}

\newcommand{\Vbf}{\boldsymbol{V}}

\newcommand{\mubf}{\boldsymbol{\mu}}
\newcommand{\qbf}{\boldsymbol{\mathsf{q}}}

\newcommand{\xibf}{\boldsymbol{\xi}}

\newcommand{\zsf}{\mathsf{z}}

\newcommand{\rk}{\operatorname{rk}}

\newcommand{\Hom}{\operatorname{Hom}}

\newcommand{\crm}{\operatorname{c}}
\newcommand{\End}{\operatorname{End}}

\newcommand{\Pic}{\operatorname{Pic}}
\newcommand{\Ext}{\operatorname{Ext}}

\newcommand{\eu}{\operatorname{eu}}
\newcommand{\ch}{\operatorname{ch}}
\def\e{{\,\rm e}\,}
\def\ii{{\,{\rm i}\,}}

\newcommand{\splus}{S^+}
\newcommand{\dplus}{d^+}
\newcommand{\sminus}{S^-}
\newcommand{\dminus}{d^-}

\newcommand{\triend}{\parbox{2mm}{\hfill} \hfill\mbox{\hspace{0.2mm}}\hfill$\triangle$}
\newcommand{\ocend}{\parbox{2mm}{\hfill} \hfill\mbox{\hspace{0.2mm}}\hfill$\oslash$}

\newtheorem*{corollary*}{Corollary}
\newtheorem*{theorem*}{Theorem}
\newtheorem*{proposition*}{Proposition}

\numberwithin{equation}{section}

\theoremstyle{remark}

\theoremstyle{remark}
\newtheorem{rem}[equation]{Remark}

\theoremstyle{definition}
\newtheorem{defin}[equation]{Definition}

\begin{document}

\begin{flushright}
EMPG--14--18
\end{flushright}

\vskip 1cm

\title[$\Ncal=2$ \ quiver gauge theories 
    on A-type ALE spaces]{\large{$\boldsymbol{\Ncal=2}$ \ quiver gauge theories on A-type ALE spaces}}

\vskip 1cm

\maketitle \thispagestyle{empty}

\begin{center}
{\large \sc Ugo Bruzzo$^{\S\ddag}$,
  Francesco Sala$^{\bullet}$} \ and \ {\large \sc Richard J. Szabo$^{\P\star\circ}$} \\[8pt] 
$^\S$ Scuola Internazionale Superiore di Studi Avanzati {\sc (SISSA),}\\ Via Bonomea 265, 34136
Trieste, Italia \\[3pt] $^\ddag$ Istituto Nazionale di Fisica
Nucleare, Sezione di Trieste \\[3pt]
$^\bullet$ Department of Mathematics, The University of Western Ontario,\\
Middlesex College, London  N6A 5B7, Ontario, Canada  \\[3pt]
$^\P$ Department of Mathematics, Heriot-Watt University,\\ Colin
Maclaurin Building, Riccarton, Edinburgh EH14 4AS, United Kingdom; \\[3pt]
$^\star$ Maxwell Institute for Mathematical Sciences, Edinburgh,
United Kingdom \\[3pt] $^\circ$ The Tait Institute, Edinburgh, United Kingdom
 \end{center}

\par\vfill

\noindent \begin{quote} 
{\sc Abstract.} \small
We survey and compare recent approaches to the computation of the partition
functions and correlators of chiral BPS observables in $\Ncal=2$ gauge
theories on ALE spaces based on quiver varieties and the minimal
resolution $X_k$ of the $A_{k-1}$ toric singularity $\C^2/\Z_k$, in light of
their recently conjectured duality with two-dimensional coset conformal field
theories. We review and elucidate the rigorous constructions of 
gauge theories for a particular family of ALE spaces,
using their relation to the cohomology of moduli spaces of framed torsion
free sheaves on a suitable orbifold compactification of $X_k$. We extend these computations to generic $\Ncal=2$
superconformal quiver gauge theories, obtaining in these instances new constraints on
fractional instanton charges, a rigorous proof of the Nekrasov master formula,
and new quantizations of Hitchin systems based on the underlying
Seiberg-Witten geometry.
\end{quote}

\par
\vfill
\parbox{.95\textwidth}{\hrulefill}\par
\noindent \begin{minipage}[c]{\textwidth}\parindent=0pt \renewcommand{\baselinestretch}{1.2}
\small
\emph{Date:}  October 2014  \par 
 \par
\emph{E-Mail:} \texttt{bruzzo@sissa.it, salafra83@gmail.com, R.J.Szabo@hw.ac.uk} \par
\end{minipage}

\newpage

\setlength{\parskip}{0.2ex}

\tableofcontents

\setlength{\parskip}{0.6ex plus 0.3ex minus 0.2ex}

\section{Introduction and summary}

Asymptotically locally Euclidean (ALE) spaces have received a great deal of attention over the years as
examples of self-dual solutions to the Einstein equations in four
dimensions with vanishing cosmological constant \cite{art:gibbonshawking1978}, and as spaces on which the construction of (anti-)selfdual
solutions to the Yang-Mills equations (instantons) can be explicitly
carried out in much the same way as on flat Euclidean space
$\R^4$ \cite{art:kronheimernakajima1990}. They are diffeomorphic to resolutions of the complex quotient
singularities $\C^2/\Gamma$ for $\Gamma$ a finite subgroup of ${\rm
  SU}(2)$ \cite{art:kronheimer1989}; in this paper we focus on the special case
$\Gamma=\Z_k$. They became an important testing ground for Montonen-Olive S-duality
conjectures with the realisation \cite{art:vafawitten1994} that the partition functions of
topologically twisted $\Ncal=4$ supersymmetric Yang-Mills theory on
ALE spaces, which are generating functions for the Euler
characteristics of instanton moduli spaces, reproduce (modular invariant) characters of
affine Lie algebras in accordance with Nakajima's geometric
construction of highest weight representations of Kac-Moody algebras based on quiver varieties \cite{art:nakajima1994-3}. These instanton moduli spaces find a natural realisation in Type~II string
theory where they arise as Higgs
branches of certain quiver gauge theories (with eight real supercharges) which appear as worldvolume
field theories on D$p$-branes in D$p$-D$(p+4)$ systems with the
D$(p+4)$-branes located at the fixed point of the orbifold
$\R^4/\Z_k$ \cite{art:douglasmoore1996}.

Gauge theories on ALE spaces with $\Ncal=2$ supersymmetry have received renewed impetus in recent
years as their partition functions extend the connection to Nakajima's construction beyond the level of affine characters, and hence provide
extensions to curved manifolds of the duality between 
$\Ncal=2$ supersymmetric Yang-Mills theory on $\R^4$ and
two-dimensional conformal field theories \cite{art:belavinfeigin2011,art:nishiokatachikawa2011,art:belavinbelavinbershtein2011,art:tan2013}. A crucial ingredient in all these correspondences is the role played
by \emph{fractional instantons}, which are those nonperturbative
configurations of the gauge theory which are stuck at the orbifold
singularity of $\C^2/\Z_k$; on the resolution of the singularity they
correspond to magnetic monopoles on the exceptional divisors of the
ALE space which carry fractional topological charge. For the topologically twisted $\Ncal=4$
gauge theory their contributions have been addressed from various
points of view \cite{art:fucitomoralespoghossian2006,art:griguoloseminaraszabotanzini2007,art:ciraficikashanipoorszabo2011}. For topologically twisted $\Ncal=2$ gauge theories
most of the literature has been concerned with their interpretation as
gauge theories on the orbifold $\R^4/\Z_k$
\cite{art:fucitomoralespoghossian2004,art:fujiiminabe2005}. Until very
recently \cite{art:bonellimaruyoshitanziniyagi2012,art:bruzzopedrinisalaszabo2013} there has been no first
principles analysis of them from the perspective of moduli spaces of
instantons directly on the resolutions.

It is by now customary to consider not a conventional gauge theory but rather a
noncommutative deformation of it, with the noncommutativity related to
the supergravity two-form $B$-field on the D-branes in the Type~II string theory
picture. Depending on the choice of stability parameters, the
instantons of this gauge theory can correspond to torsion free sheaves
\cite{art:nekrasovschwarz1998}; the resulting moduli spaces then provide a
smooth completion of the instanton moduli spaces. This point of view
will be adopted in the present paper.

The purpose of this paper is threefold. Firstly, we shall survey the various approaches to computing the partition
functions of $\Ncal=2$ gauge theories on the $A_{k-1}$-type ALE
spaces, comparing and contrasting the different results. Secondly, we
present a summary of the main results from
\cite{art:bruzzopedrinisalaszabo2013} which provided the first
rigorous computations of $\Ncal=2$ gauge theory partition functions on
ALE spaces from a moduli theory of framed sheaves on the minimal
resolution $X_k$ and presented the first step in providing an interpretation of Nakajima quiver varieties with real stability parameter in the chamber associated with $X_k$ in terms of sheaves on toric Deligne-Mumford stacks (in the complex analytic setting, a similar description by using sheaves on complex V-manifolds was provided in \cite{art:nakajima2007}). The main technical
ingredient in this approach is the construction of moduli spaces of framed torsion free
sheaves on a suitable orbifold compactification of $X_k$, which is the proper mathematical arena to incorporate
the contributions from fractional instantons; in the following we
shall present these constructions in a somewhat more informal manner that
we hope will be more palatable to physicists. The duality between
abelian $\Ncal=2$
quiver gauge theories in this setting and two-dimensional conformal
field theory was recently formulated and proved in \cite{art:pedrinisalaszabo2014}. Thirdly, we extend the
computations of \cite{art:bruzzopedrinisalaszabo2013} to calculate the
partition functions of generic $\Ncal=2$ superconformal quiver gauge
theories on $X_k$, obtaining many new results that are
summarised below.

The outline of this paper is as follows. In Section
\ref{sec:ALEinstantons} we review the relations between Nakajima
quiver varieties and moduli spaces of instantons on ALE spaces. In
Section \ref{sec:levelzero} we review the calculations of
\cite{art:fucitomoralespoghossian2004,art:fujiiminabe2005} which are
based on the interpretation of moduli spaces of ALE instantons as a
resolution of the moduli space of instantons on $\C^2/\Z_k$. In
Section \ref{sec:levelinfinity} we review the calculations of
\cite{art:bonellimaruyoshitanziniyagi2012} which are based on the
minimal resolution $X_k$ of the toric singularity $\C^2/\Z_k$ and use
the conjectural Nekrasov master formula \cite{art:nekrasov2006}. In
Section \ref{sec:rootstacks} we review the constructions from
\cite{art:bruzzopedrinisalaszabo2013} of moduli spaces parameterizing
framed sheaves on an orbifold compactification of $X_k$; these moduli
spaces are used to rigorously define the partition functions and
correlators of chiral BPS operators for $\Ncal=2$ gauge theories on
$X_k$, thus proving the factorization formulas involving partition
functions on the affine torus-invariant open subsets of $X_k$ as
conjectured by
\cite{art:bonellimaruyoshitanzini2011,art:bonellimaruyoshitanzini2012,art:bonellimaruyoshitanziniyagi2012}. In
Section \ref{sec:quivergaugetheory} we extend the calculations of
\cite{art:bruzzopedrinisalaszabo2013} to arbitrary $\Ncal=2$
superconformal quiver gauge theories on $X_k$, obtaining new
constraints on fractional instanton charges in these instances; we
also present a new computation of the perturbative partition functions
and hence provide a rigorous proof of the Nekrasov master formula for
$X_k$. We further explore the Seiberg-Witten geometry and elaborate on the interpretations of these gauge
theories as quantizations of certain Hitchin systems. Two appendices at the end of the paper summarise some technical details which are used in the main text: in Appendix \ref{app:eqcoh} we discuss some aspects of equivariant cohomology which are used to compute instanton partition functions, while in Appendix \ref{app:edgecontributions} we list some of the factors which enter the explicit expressions for the partition functions.

\subsection*{Acknowledgements} 
This work was supported in part by PRIN ``Geometria delle variet\`a algebriche,'' by GNSAGA-INdAM, by the Consolidated Grant ST/J000310/1 from the
UK Science and Technology Facilities Council and by AG Laboratory SU-HSE, RF Government Grant ag.11.G34.31.0023. Most of this paper was written while the second author was staying at Laboratory of Algebraic Geometry and its Applications, SU-HSE, and at the Department of Mathematics, Heriot-Watt University; he thanks these institutions for hospitality and support.

\bigskip
\section{Quiver varieties and instantons on ALE spaces\label{sec:ALEinstantons}}

ALE spaces are hyper-K\"ahler
four-manifolds with one end at infinity which resembles a quotient
$\R^4/\Gamma$, for $\Gamma\subset \mathrm{SU}(2)$ a finite group
acting isometrically on $\R^4\simeq\C^2$. A large class of ALE spaces was
discovered by Gibbons and Hawking
\cite{art:gibbonshawking1978}. Kronheimer \cite{art:kronheimer1989}
realized them as hyper-K\"ahler quotients. Kronheimer and
Nakajima~\cite{art:kronheimernakajima1990} constructed moduli spaces
of $\mathrm{U}(r)$ instantons on ALE spaces with fixed Chern character and flat connection at
infinity by means of an ADHM-type construction; they are also
hyper-K\"ahler manifolds. These moduli spaces were completed by
Nakajima \cite{art:nakajima1994-3} via a modification of the ADHM
equations, obtaining the Nakajima quiver varieties.
In this section we describe the relation between Nakajima quiver varieties of type
$\widehat{A}_{k-1}$ and moduli spaces of instantons on ALE spaces of type $A_{k-1}$ for an integer
$k\geq 2$.

\subsection{Quiver varieties of type $\widehat{A}_{k-1}$\label{sec:quivervar}}

Let $\quiv=(\quiv_0,\quiv_1,\source,\tail)$ be a quiver, i.e. an oriented graph
with a finite set of vertices $\quiv_0$, a finite set of edges
$\quiv_1$, and two projection maps
$\source,\tail\colon\quiv_1\rightrightarrows \quiv_0$ which assign to
each oriented edge its source and target vertex respectively. We
denote by $\quiv^{\mathrm{op}}$ the opposite quiver obtained from
$\quiv$ by reversing the orientation of the edges, and by
$\bar{\quiv}$ the double of $\quiv$, i.e. the quiver with the same
vertex set as $\quiv$ and whose edge set is the disjoint union of
the edge sets of $\quiv$ and of $\quiv^{\mathrm{op}}$. 

Let $k\geq 2$ be an integer. For the rest of this section $\quiv$ is the quiver whose
underlying graph is that of the affine Dynkin diagram of type
$\widehat{A}_{k-1}$, i.e. for $k=2$
\begin{equation}
  \begin{tikzpicture}[xscale=2,yscale=-1]
\node (A0_3) at (0,1) {$\circ$};
\node (A1_1) at (1,1) {$\circ$};
\node (0) at (0,0.7) {\footnotesize 0};
\node (1) at (1,0.7) {\footnotesize 1}; 
\node (starting0) at (0,0.9) {$ $};
\node (ending0) at (1,0.9) {$ $};
\node (starting1) at (0,1.1) {$ $};
\node (ending1) at (1,1.1) {$ $};
    \path (starting0) edge [->]node [auto] {$\scriptstyle{}$} (ending0);
    \path (ending1) edge [->]node [auto] {$\scriptstyle{}$} (starting1);
  \end{tikzpicture}
\end{equation}
and for $k\geq 3$
\begin{equation}
  \begin{tikzpicture}[xscale=2,yscale=-1]
\node (A0_3) at (3,0.3) {$\circ$};
\node (A1_1) at (1,1.8) {$\circ$}; 
\node (A1_2) at (2,1.8) {$\circ$}; 
\node (A1_3) at (3,1.8) {$\ldots$}; 
\node (A1_4) at (4,1.8) {$\circ$}; 
\node (A1_5) at (5,1.8) {$\circ$}; 
\node (0) at (3,0) {\footnotesize 0};
\node (1) at (1,1.5) {\footnotesize 1};
\node (2) at (2,1.5) {\footnotesize 2};
\node (k-2) at (4,1.5) {\footnotesize $k-2$};
\node (k-1) at (5,1.5) {\footnotesize $k-1$};
    \path (A1_1) edge [->]node [auto] {$\scriptstyle{}$} (A1_2);
    \path (A1_2) edge [->]node [auto] {$\scriptstyle{}$} (A1_3);
    \path (A1_3) edge [->]node [auto] {$\scriptstyle{}$} (A1_4);
    \path (A1_4) edge [->]node [auto] {$\scriptstyle{}$} (A1_5);
    \path (A0_3) edge [->]node [auto] {$\scriptstyle{}$} (A1_1);
    \path (A0_3) edge [->]node [auto] {$\scriptstyle{}$} (A1_5);
  \end{tikzpicture}
\end{equation}
Let $\vec{v},\vec{w}\in(\Z_{\geq 0})^{k}$ and let $V:=\bigoplus_{i\in\quiv_0}
\, V_i$,  $W:=\bigoplus_{i\in\quiv_0} \, W_i$ be $\quiv_0$-graded complex
vector spaces with $\dim V_i=v_i$, $\dim W_i=w_i$. Let
\begin{equation}
M(\vec{v},\vec{w}\, ):=\Big(\, \bigoplus_{e\in
  \bar{\quiv}_1}\,\Hom\big(V_{\source(e)}, V_{\tail(e)}\big)\, \Big) \
\oplus \ \Big(\, \bigoplus_{i\in\bar{\quiv}_0}\,\Hom\big(W_i,
V_i\big)\oplus\Hom\big(V_i,W_i\big)\, \Big)
\end{equation}
be a representation space for the double quiver $\bar\quiv$.
The group
$\mathrm{GL}_{\vec{v}}:=\prod_{i\in\quiv_0}\,\mathrm{GL}(V_i)$ acts on
the affine space $M(\vec{v},\vec{w}\, )$ by
\begin{equation} \label{eq:quivergroup}
(g_j)_{j\in\quiv_0} \triangleright (B_e, a_i, b_i):=\big(g_{\tail(e)}\,B_e\,
g_{\source(e)}^{-1}\,,\, g_i\, a_i\,,\, b_i\, g_i^{-1} \big)
\end{equation}
for $g_i\in \mathrm{GL}(V_i), B_e\in \Hom\big(V_{\source(e)},
V_{\tail(e)}\big), a_i\in \Hom\big(W_i, V_i\big)$ and $b_i\in
\Hom\big(V_i,W_i\big)$. One can define a symplectic form on
$M(\vec{v},\vec{w}\, )$ which is preserved by this
$\mathrm{GL}_{\vec{v}}\,$-action (see e.g. \cite[Section
1(i)]{art:nakajima2007}). Then the corresponding moment map
$\mu_\C\colon M(\vec{v},\vec{w}\, ) \to \mathfrak{gl}_{\vec{v}}:=
\bigoplus_{i\in\quiv_0}\, \End(V_i)$ which vanishes at the origin is given by
\begin{equation}
\mu_\C(B,a,b):=\Bigg(\, \sum_{\stackrel{\scriptstyle e\in\bar\quiv_1}{\scriptstyle \tail(e)=i}}\, \epsilon(e)\, B_e\,
B_{\bar{e}}+a_i\,b_i\, \Bigg)_{i\in\quiv_0} \ ,
\end{equation}
where $\bar{e}\in \quiv^{\mathrm{op}}_1$ is the reverse edge of $e$
with $\epsilon(e)=1$ and $\epsilon(\bar{e})=-1$ for $e\in\quiv_1$.

Let us fix $\xi=(\xi_\C,\xi_\R)\in \C^{k}\oplus \R^{k}$. We define an
element corresponding to $\xi_\C$ in the center of
$\mathfrak{gl}_{\vec{v}}$ by $\bigoplus_{i\in\quiv_0} \,
\xi_{\C}^{i}\, \mathrm{id}_{V_i}$, where we delete the summand
corresponding to node $i$ if $V_i=\{0\}$. One can introduce the notion of $\xi_\R$-(semi)stable points in
$M(\vec{v},\vec{w}\, )$ (see e.g. \cite[Definition
1.1]{art:nakajima2007}). One says that two $\xi_\R$-semistable points
$(B,a,b)$ and $(B',a',b'\, )$ are $S$-equivalent, and writes
$(B,a,b)\sim (B',a',b'\, )$, when the closures of their
$\mathrm{GL}_{\vec{v}}\,$-orbits intersect inside the
$\xi_\R$-semistable locus $\big(\mu_\C^{-1}(\xi_\C)\big)^{\rm ss}$ of
$\mu_\C^{-1}(\xi_\C)$. Then we define the \emph{Nakajima quiver
  variety} associated with $\vec{v}$, $\vec{w}$ and $\xi$ as the quotient
\begin{equation}
\Mcal_\xi(\vec{v},\vec{w}\,):=\big(\mu_\C^{-1}(\xi_\C)\big)^{\rm ss}\, \big/\, \sim\ .
\end{equation}
Below we also use the superscript `s' to denote $\xi_\R$-stable points.

The Cartan matrix of the affine Dynkin diagram of type $\widehat{A}_{k-1}$ is  
$\widehat{C}= 2\, \mathrm{id}_{\C^k} -A$, where $A=(a_{i, j})_{i, j\in\bar{\quiv}_0}$ is the adjacency matrix of the double quiver $\bar{\quiv}$, i.e. $a_{i, j}$ is the number of edges from $i$ to $j$ in $\bar{\quiv}_1$. Explicitly, 
for $k=2$ the matrix $\widehat{C}$ is given by
\begin{equation}
\widehat{C}=\begin{pmatrix}
2 & -2 \\
-2 & 2
\end{pmatrix}
\end{equation}
and for $k\geq 3$ 
\begin{equation}
\widehat{C}=\begin{pmatrix}
2 & -1 & 0 & \cdots & -1 \\
-1 & 2& -1&   \cdots & 0 \\
0 & -1& 2& \cdots & 0 \\
\vdots & \vdots & \vdots & \ddots & \vdots \\
-1 &0 &0 & \cdots & 2
\end{pmatrix}.
\end{equation}

Let $\widehat{\Qfrak}_+$ be the set of positive roots of the affine Dynkin diagram of type $\widehat{A}_{k-1}$, i.e. the set
\begin{equation}
\widehat{\Qfrak}_+:=\big\{\theta=(\theta_i)_{i\in\quiv_0} \in (\Z_{\geq 0})^{k}\
\big\vert\ \theta\cdot \widehat{C}\theta\leq 2 \big\}\,
\setminus\, \{0\}\ .
\end{equation}
Define the \emph{wall} $\Dsf_\theta$ associated with the
root $\theta \in \widehat{\Qfrak}_+$ by
\begin{equation}
\Dsf_\theta:=\Big\{x=\big(x^i \big)_{i\in\quiv_0} \in \R^{k}\ \Big\vert\
\mbox{$\sum\limits_{i\in\quiv_0}$} \, x^i\,\theta_i=0 \Big\} \ \subset
\ \R^{k}\ .
\end{equation}
For $\vec{v}\in (\Z_{\geq0})^{k}$ we set
\begin{equation}
\widehat{\Qfrak}_+(\vec{v}\, ):= \big\{\theta\in \widehat{\Qfrak}_+\
\big\vert\ \theta_i\leq v_i\quad \forall i\in\quiv_0 \big\}\ .
\end{equation}
Although $\widehat{\Qfrak}_+$ is an infinite set, the set $\widehat{\Qfrak}_+(\vec{v}\, )$ is always finite.

An element $\xi\in \C^{k}\oplus \R^{k}$ is \emph{generic} with respect
to $\vec{v}\in(\Z_{\geq0})^k$ if for any root $\theta\in
\widehat{\Qfrak}_+(\vec{v}\, )$ one has
\begin{equation}
\xi \ \notin \ \Dsf_\theta\otimes \R^3 \subset \R^{k}\otimes \R^3\simeq \C^{k}\oplus \R^{k}\ .
\end{equation}
By \cite[Theorem 2.8]{art:nakajima1994-3}, if $\xi$ is generic with respect to $\vec{v}\in(\Z_{\geq0})^k$ then $\Mcal_\xi(\vec{v}, \vec{w}\, )$ is a smooth connected variety of dimension
\begin{equation}
\dim \Mcal_\xi(\vec{v}, \vec{w}\, )=2\vec{w}\cdot \vec{v}-\vec{v}\cdot\widehat{C}\vec{v} \ .
\end{equation}

Fix a complex parameter $\xi_\C$. A connected component of the affine space
\begin{equation}\label{eq:realpararameterspace}
\R^{k}\ \big\backslash \ \bigcup_{\stackrel{\scriptstyle \theta\in
    \widehat{\Qfrak}_+(\vec{v}\, )}{\scriptstyle \xi_\C\cdot \theta=0}}\, \Dsf_\theta
\end{equation}
is called a \emph{chamber}. It is known that \cite[Section 1(i)]{art:nakajima2007}:
\begin{itemize}
\item If $\xi_\R, \xi_\R'$ lie in two distinct chambers such that $\xi=(\xi_\C, \xi_\R)$ and $\xi'=(\xi_\C, \xi_\R')$ are
  generic with respect to $\vec{v}\in (\Z_{\geq 0})^k$, then the quiver
  varieties $\Mcal_\xi(\vec{v}, \vec{w}\, )$ and
  $\Mcal_{\xi'}(\vec{v}, \vec{w}\, )$ are diffeomorphic.

\smallskip

\item If $\xi_\R, \xi_\R'$ lie in the same chamber such that $\xi=(\xi_\C, \xi_\R)$ and $\xi'=(\xi_\C, \xi_\R')$ are generic with respect to $\vec{v}\in (\Z_{\geq 0})^k$, then the quiver varieties $\Mcal_\xi(\vec{v}, \vec{w}\, )$ and  $\Mcal_{\xi'}(\vec{v}, \vec{w}\, )$ are isomorphic (as algebraic varieties).\end{itemize}

\subsection{ALE spaces of type $A_{k-1}$}

Fix an integer $k\geq 2$ and let
$\omega$ be a primitive $k$-th root of unity. We
define an action of the cyclic group $\Z_k$ of order $k$ on
$\R^4\simeq \C^2$ as
$\omega\triangleright (z,w):= (\omega\, z, \omega^{-1}\, w)$; it has a
unique fixed point at the origin. The
quotient $\C^2/\Z_k$ is a normal affine toric surface,
with a Kleinian singularity at the origin.
If
$\varphi_k\colon X_k\to\C^2/\Z_k$ is the minimal resolution of the
singularity of $\C^2/\Z_k$, then $X_k$ is a smooth toric surface.

Let $\delta=(1, \ldots, 1)\in (\Z_{\geq0})^k$. The wall
$\Dsf_\delta=\big\{x=(x^i)\in \R^k\,\big\vert\, \sum_{i\in\quiv_0} \,
x^i=0 \big\}$ is
called the \emph{level zero hyperplane}. Let $\xi^\circ=(\xi_\C^\circ,
\xi_\R^\circ)$ be an element in $\Dsf_\delta\otimes\R^3$ which is not
contained in any $\Dsf_\theta\otimes\R^3$ for $\theta\in
\widehat{\Qfrak}_+\setminus \{\delta\}$; we shall say that $\xi^\circ$
is generic. Then we define the \emph{ALE space $X_{\xi^\circ}$ of type
  $A_{k-1}$ with parameter $\xi^\circ$} by the geometric quotient
\begin{equation}
X_{\xi^\circ}:=\big(\mu_\C^{-1}(-\xi_\C)\big)^{\rm s}\, \big/\,\big(\mathrm{GL}_\delta/\C^\ast \big)\ ,
\end{equation}
where we consider $(-\xi_\R)$-stable points of $\mu_\C^{-1}(-\xi_\C)$.

It is proven in \cite{art:kronheimer1989} (see also \cite[Proposition 2.12]{art:nakajima1994-3}) that
$X_{\xi^\circ}$ is a four-dimensional hyper-K\"ahler manifold which is
diffeomorphic to $X_k$; it furthermore carries a hyper-K\"ahler ALE
metric $g$ of order four, i.e., there is a compact subset $K\subset
X_{\xi^\circ}$ and a diffeomorphism $X_{\xi^\circ}\setminus K\to
(\R^4\setminus B_r(0))/\Z_k$ under which the metric $g$ is
approximated by the standard Euclidean metric on $\R^4/\Z_k$, and in local
Euclidean coordinates $(x_\alpha)_{\alpha=1}^4$ one has 
\begin{equation}
g_{\alpha \beta}=\delta_{\alpha\beta}+b_{\alpha \beta}
\end{equation}
with $\partial^p b_{\alpha\beta}=O(r^{-4-p})$ for $p\gg 0$, where
$r^2=\sum_{\alpha=1}^4\, x_\alpha^2$ and $\partial$ denotes
differentiation with respect to the coordinates $x_\alpha$. Here
$B_r(0)$ is the ball of radius $r$ centered at the origin of $\R^4$. We call $X_{\xi^\circ}\setminus K$ the \emph{end} of $X_{\xi^\circ}$.

\subsection{Instanton moduli spaces}

A $\mathrm{U}(r)$ instanton on $X_{\xi^\circ}$ is a pair $(E, A)$
where $E$ is a Hermitean vector bundle on $X_{\xi^\circ}$ of rank $r$,
and $A$ is a unitary connection on $E$ which is anti-selfdual and
square-integrable. At the end of $X_{\xi^\circ}$, $A$ is approximated by a flat unitary connection $A_0$, i.e.
\begin{equation}
A-A_0= O\big(r^{-3} \big) \ ,\qquad \nabla A-\nabla A_0=O\big(r^{-4} \big) \ ,\qquad\ldots\ ,
\end{equation}
where $\nabla$ is the covariant derivative with respect to $A_0$. The
connection $A_0$ is determined by its holonomy, and hence by a unitary
representation of the fundamental group of the end of $X_{\xi^\circ}$,
which is isomorphic to $\Z_k$. 

Let $\rho_i$ for $i=0, 1, \ldots, k-1$ be the one-dimensional
representation $z\mapsto\omega^i \, z$ of
$\Z_k$ with weight $i$, and let $\vec{w}:=(w_0, w_1, \ldots,$
$w_{k-1})\in (\Z_{\geq 0})^k$ with $\sum_{i=0}^{k-1}\, w_i =r$. The main result of
\cite{art:kronheimernakajima1990} is that the moduli space of
$\mathrm{U}(r)$ instantons on $X_{\xi^\circ}$, with fixed Chern
character and flat connection at infinity characterised by the
representation $\bigoplus_{i=0}^{k-1}\,\rho_i^{\oplus w_i}$, is
isomorphic to the quotient space 
\begin{equation}
\Mcal_{\xi^\circ}^{\mathrm{reg}}(\vec{v},\vec{w}\,
):=\big(\mu_\C^{-1}(\xi_\C^\circ)\big)^{\rm s}\, \big/ \,
\mathrm{GL}_{\vec{v}} \ \subseteq \
\Mcal_{\xi^\circ}(\vec{v},\vec{w}\, )
\end{equation}
for a suitable dimension vector $\vec{v}\in(\Z_{\geq 0})^k$.
The quiver variety $\Mcal_{\xi^\circ}(\vec{v},\vec{w}\, )$ is a type of Uhlenbeck space of
$\Mcal_{\xi^\circ}^{\mathrm{reg}}(\vec{v},\vec{w}\, )$, in the sense that
one has (cf.\ \cite[Proposition 9.2]{art:kronheimernakajima1990} and \cite[Proposition 1.10]{art:nakajima2007})
\begin{equation}
\Mcal_{\xi^\circ}(\vec{v},\vec{w}\, )=\coprod_{p\geq 0}\,
\Mcal_{\xi^\circ}^{\mathrm{reg}}(\vec{v}-p\,\delta,\vec{w}\, ) \
\times \ \mathrm{Sym}^p\,X_{\xi^\circ}\ ,
\end{equation}
where $\mathrm{Sym}^p\,X_{\xi^\circ}$ is the $p$-th symmetric product
of $X_{\xi^\circ}$; moreover,
$\Mcal_{\xi^\circ}^{\mathrm{reg}}(\vec{v},\vec{w}\, )$ is the smooth
locus of $\Mcal_{\xi^\circ}(\vec{v},\vec{w}\, )$.

Let us fix $\xi_\C^\circ$ and $\vec{v}\in (\Z_{\geq 0})^k$. Choose a
parameter $\xi_\R$ in the chamber inside \eqref{eq:realpararameterspace} with $\xi_\R\cdot \delta <0$ which
contains $\xi_\R^\circ$ in its closure (this chamber is uniquely
determined by these conditions and by $\vec{v}$ ). Then
$\xi=(\xi_\C^\circ, \xi_\R)$ is generic with respect to $\vec{v}$,
hence $\Mcal_\xi(\vec{v}, \vec{w}\, )$ is smooth and there is a resolution of singularities (cf.\ \cite[Section 2(i)]{art:nakajima2007} and \cite[Section 4]{art:nakajima1994-3})
\begin{equation}\label{eq:projectivemorphism}
\Mcal_\xi(\vec{v}, \vec{w}\, ) \ \longrightarrow \
\Mcal_{\xi^\circ}(\vec{v},\vec{w}\, )
\end{equation}
which is an isomorphism over
$\Mcal_{\xi^\circ}^{\mathrm{reg}}(\vec{v},\vec{w}\, )$ for any choice of $\vec{w}\in(\Z_{\geq 0})^k$.

In the following we introduce two distinguished chambers and describe the partition functions of
supersymmetric gauge theories on ALE spaces with parameters in these chambers.

\bigskip
\section{$\Ncal=2$ gauge theory for the level zero chamber\label{sec:levelzero}}

Following \cite[Section 4.2]{art:nagao2009}, we define
\begin{equation}
\Csf_0:=\big\{\xi_\R=\big(\xi_\R^i \big)\in \R^k\ \big\vert\ \xi_\R^i<0\quad
\forall i\in \quiv_0 \big\}\ .
\end{equation}
Then any parameter $\xi_0=(0,\xi_\R)\in \C^k\oplus\R^k$ with
$\xi_\R\in \Csf_0$ is generic for any choice of $\vec{v}\in(\Z_{\geq
  0})^k$. Hence the Nakajima quiver variety $\Mcal_{\xi_0}(\vec{v},
\vec{w}\, )$ is smooth for any choice of dimension vectors
$\vec{v},\vec{w}\in(\Z_{\geq 0})^k$. We call $\Csf_0$ the \emph{level zero chamber}.

By \cite[Section
2.3]{art:varagnolovasserot1999}, the quiver variety
$\Mcal_{\xi_0}(\vec{v}, \vec{w}\, )$ is isomorphic to the moduli space
parameterizing isomorphism classes of framed sheaves $(E, \phi_E\colon
E\vert_{\ell_\infty}\xrightarrow{\sim} \Ocal_{\ell_\infty}\otimes W)$ on
the projective plane $\PP^2$, where $E$ is a $\Z_k$-equivariant
torsion free sheaf on $\PP^2$ with a $\Z_k$-invariant isomorphism $H^1(\PP^2; E\otimes
\Ocal_{\PP^2}(-\ell_\infty))\simeq V$ and
$\phi_E$ is a $\Z_k$-invariant isomorphism; here $\ell_\infty$ is a
line at infinity. In this sense $\Mcal_{\xi_0}(\vec{v}, \vec{w}\, )$
can be regarded as a resolution of the Uhlenbeck space of instantons on
$\C^2/\Z_k$; see \cite{art:ivanovalechtenfeldpopovszabo2013} for an
analysis of explicit ALE instanton solutions in this context. In this section we first introduce the instanton part of Nekrasov's partition function for pure $\Ncal=2$ gauge theory on $\C^2$, and then consider its $\Z_k$-invariant projection.

\subsection{Instanton counting on $\R^4$}

Nekrasov's instanton partition function \cite{art:nekrasov2003} is the generating function for integrals of equivariant cohomology classes over the moduli space $\Mcal_{r,n}$ of framed torsion free sheaves $(E,\phi_E:E|_{\ell_\infty}\xrightarrow{\sim} \Ocal_{\ell_\infty}^{\oplus r})$ over $\PP^2$ of rank $r$ and second Chern class ${\rm c}_2(E)= n$. The moduli space $\Mcal_{r,n}$ is a smooth quasi-projective variety of dimension $2\,r\,n$ which can be regarded as a resolution of the Uhlenbeck space of ${\rm U}(r)$ instantons on $\R^4$; in the rank one case $r=1$ it is isomorphic to the Hilbert scheme of $n$ points on $\C^2$: $\Mcal_{1,n}\simeq {\rm Hilb}^n(\C^2)$. There is an action of the torus $T:=(\C^\ast)^2\times (\C^\ast)^r$ on $\Mcal_{r,n}$, where the first factor of $T$ acts by pullback on $E$ from the natural $(\C^\ast)^2$-action on $\PP^2$ and the second factor acts by diagonal multiplication on $\phi_E$, for any point $(E,\phi_E)$ of the moduli space. Let $\Pcal_{r,n}$ be the set of $r$-tuples of Young tableaux $\mbf Y=(Y_1,\dots,Y_r)$ of total weight $|\mbf Y|:=\sum_{\alpha=1}^r\, |Y_\alpha|=n$, where $|Y_\alpha|$ is the number of boxes in the Young tableau $Y_\alpha$. Then the $T$-fixed points in $\Mcal_{r,n}$ are parameterized by $\Pcal_{r,n}$ \cite{art:nakajimayoshioka2005-I}.

Denote by $\varepsilon_1, \varepsilon_2, a_1,\dots,a_r$ the generators of $H^*_{T}({\rm pt})$ (see Appendix \ref{app:eqcoh}). Then the instanton partition function for pure $\Ncal=2$ supersymmetric ${\rm U}(r)$ gauge theory on $\R^4$ is given by
\begin{equation}
\Zcal_{\C^2}^{\rm inst}(\varepsilon_1, \varepsilon_2, \mbf a;\qsf) := \sum_{n=0}^\infty\, \qsf^n \ \int_{\Mcal_{r,n}}\, \big[\Mcal_{r,n}\big]_T \ ,
\end{equation}
where $\qsf$ is a formal variable which weighs the different topological sectors. The localization theorem in $T$-equivariant cohomology then gives the combinatorial expansion \cite{art:nekrasov2003,art:flumepoghossian2003,art:bruzzofucitomoralestanzini2003}
\begin{equation}
\Zcal_{\C^2}^{\rm inst}(\varepsilon_1, \varepsilon_2, \mbf a;\qsf) = \sum_{n=0}^\infty\,\qsf^n \ \sum_{\mbf Y\in\Pcal_{r,n}} \ \prod_{\alpha,\beta=1}^r \, \frac1{m_{Y_\alpha,Y_\beta}(\varepsilon_1,\varepsilon_2,a_{\beta\alpha})} \ ,
\end{equation}
where $a_{\beta\alpha}:=a_\beta-a_\alpha$ and
\begin{multline} \label{eq:mYfactor}
m_{Y_\alpha,Y_\beta}(\varepsilon_1,\varepsilon_2,a)
=\prod_{s\in Y_\alpha}\, \Big(a -L_{Y_\beta}(s)\, \varepsilon_1+ \big(A_{Y_\alpha}(s)+1\big)\, \varepsilon_2\Big) \\ 
\times \ \prod_{s'\in Y_\beta}\, \Big(a+\big(L_{Y_\alpha}(s'\, )+1\big)\, \varepsilon_1-A_{Y_\beta}(s'\, )\, \varepsilon_2 \Big)
\end{multline}
with $A_{Y_\alpha}(s)$ the number of boxes to the right of box $s$ in the tableau $Y_\alpha$ (the {\em arm length} of the box) and $L_{Y_\alpha}(s)$ the number of boxes on top of it (the {\em leg length} of the box).

\subsection{$\Z_k$-projection\label{sec:Zkproj}}

After fixing a lift of the $\Z_k$-action to
$\Ocal_{\ell_\infty}^{\oplus r}$ by $\vec w\in(\Z_{\geq0})^k$, with
$r=\sum_{i=0}^{k-1}\, w_i$, there is a natural $\Z_k$-action on
$\Mcal_{r,n}$ induced by the $\Z_k$-action on $\PP^2$; we write $W=\bigoplus_{\alpha=1}^r\, \rho_{p_\alpha}$ so that $w_i$ is the number of times that $i\in\{0,1,\dots,k-1\}$ appears in the vector $\mbf p=(p_1,\dots,p_r)$ with $p_\alpha\in\Z_k$ for $\alpha=1,\dots,r$. Then the fixed
point set $(\Mcal_{r,n})^{\Z_k}$ is a disjoint union of quiver
varieties $\Mcal_{\xi_0}(\vec v, \vec w\,)$ over all dimension vectors
$\vec v \in(\Z_{\geq0})^k$ with $\sum_{i=0}^{k-1}\, v_i=n$; in the
rank one case $\Mcal_{\xi_0}(\vec v, \vec w_0)$ is isomorphic to the
moduli space ${\rm Hilb}^n(\C^2)^{\Z_k,\vec v}$ parameterizing
$\Z_k$-invariant zero-dimensional subschemes $Z$ of $\C^2$ of length
$n$ such that $v_i$ is the multiplicity of the representation $\rho_i$
in $H^0(Z;\Ocal_Z)$, for $\vec
w_0:=(1,0,\dots,0)$ and $i=0,1,\dots,k-1$, where $\Ocal_Z$ is the
sheaf of regular functions on $Z$. To define
the instanton partition function, we introduce an action of the torus
$T$ on $\Mcal_{\xi_0}(\vec v, \vec w\, )$ in exactly the same way that
we defined the $T$-action on $\Mcal_{r,n}$. The resulting fixed point
locus consists of $r$-tuples of \emph{coloured} Young tableaux. A
colouring by $\gamma\in \{0,1, \ldots, k-1\}$ of a Young tableau
$Y\subset \Z_{>0}\times \Z_{>0}$
is an assignment to each box $s=(i,j)$ of $Y$ of an element
$\mathrm{res}(s):=\gamma-i+j\in \Z_k$, called the $k$-residue of
$s$. An $r$-tuple of Young tableaux $\mbf{Y}=(Y_1, \ldots, Y_r)$ is
said to be coloured by $\vec{w}$ if $Y_\alpha$ is coloured by
$i\in\{0,1, \ldots, k-1\}$ for $\sum_{j=0}^{i-1}\, w_j < \alpha \leq
\sum_{j=0}^i\, w_j$. From a geometric perspective, the colours and the
$k$-residues $\mathrm{res}(s)$, as $s$ varies over boxes in $Y_\alpha$ for $\alpha=1, \ldots, r$, specify the representations of $\Z_k$ acting on the fixed point corresponding to $\mbf{Y}$, so that the colouring encodes the holonomy at infinity of the instanton corresponding to the fixed point.

Denote by $\Pcal(\vec v, \vec w\, )$ the set of $r$-tuples of coloured Young tableaux $\mbf Y$ corresponding to the torus-fixed points of $\Mcal_{\xi_0}(\vec v, \vec w\,)$. Then the instanton partition function for pure $\Ncal=2$ supersymmetric ${\rm U}(r)$ gauge theory on the ALE space $X_{\xi_0^\circ}$ with stability parameter $\xi_0^\circ=(0, \xi_{\R}^\circ)$, where $\xi_{\R}^\circ$ lies in the closure of $\Csf_0$, is given by
\begin{equation}
\mathcal Z^{\rm inst}_{X_{\xi_0^\circ}}\big(
\varepsilon_1,\varepsilon_2;\mbf a; \zsf, \vec y \ \big)_{\vec{w}}
:= \sum_{\vec v\in (\Z_{\ge0})^k} \, 
\zsf^{v} \ \prod_{l=1}^{k-1}\, y_l^{u_l} \ \int_{\Mcal_{\xi_0}(\vec{v}, \vec{w}\, )} \, \big[\Mcal_{\xi_0}(\vec{v}, \vec{w}\, )\big]_T
\end{equation}
where $\zsf$ and $\vec\xi=(\xi_1, \dots,\xi_{k-1})$ are formal variables that weigh the topological invariants and the holonomies at infinity; here we have set
\begin{equation}
u_l = w_l + v_{l+1}+v_{l-1}-2v_l\qquad\mbox{and}\qquad
v =\sum_{l=0}^{k-1}\, \Big(v_l +\frac12\, (k-l)\, l\,u_l \Big) \ ,
\end{equation}
where indices are read modulo $k$. By the localization theorem one obtains
\begin{equation}\label{eq:wyllard}
\mathcal Z^{\rm inst}_{X_{\xi_0^\circ}}\big(
\varepsilon_1,\varepsilon_2;\mbf a; \zsf, \vec y \ \big)_{\vec{w}}= 
\sum_{\vec v\in (\Z_{\ge0})^k} \, \zsf^{v} \ \prod_{l=1}^{k-1}\, y_l^{u_l} \ \sum_{\mbf Y\in \Pcal (\vec{v}, \vec{w}\,)} \ \prod_{\alpha,\beta=1}^r \, \frac1{m_{Y_\alpha,Y_\beta}^{\Z_k}(\varepsilon_1,\varepsilon_2,a_{\beta\alpha})}\ ,
\end{equation}
where
\begin{multline} \label{eq:mYfactorZk}
m_{Y_\alpha,Y_\beta}^{\Z_k}(\varepsilon_1,\varepsilon_2, a)
=\prod_{s\in Y_\alpha}\, \Big(a-L_{Y_\beta}(s)\, \varepsilon_1+\big(A_{Y_\alpha}(s)+1 \big)\,\varepsilon_2 \Big) \,\delta^{(k)}_{h_{Y_\beta,Y_\alpha}(s),p_\alpha- p_\beta}\\ 
\times \ \prod_{s'\in Y_\beta}\, \Big(a+\big(L_{Y_\alpha}(s'\, )+1 \big)\, \varepsilon_1-A_{Y_\beta}(s'\, )\, \varepsilon_2\Big)\,\delta^{(k)}_{h_{Y_\alpha,Y_\beta}(s'\, ),p_\beta-p_\alpha}
\end{multline}
with $h_{Y_\beta,Y_\alpha}(s)= L_{Y_\alpha}(s)+ A_{Y_\beta}(s) +1$ the {\em hook length} of the box $s$ in the tableau $Y_\alpha$, and 
\begin{equation}
\delta^{(k)}_{i,j} := \left\{
\begin{array}{cl} 1 & \mbox{for}\quad i=j \!\!\mod k\ , \\
0 & \mbox{otherwise}\ .
\end{array}\right.
\end{equation}
The delta function factors in \eqref{eq:mYfactorZk} select weight zero
assignments modulo $k$ with respect to the induced action of the orbifold
group $\Z_k$ on the torus $T$.
This partition function was first computed by Fucito, Morales and Poghossian in \cite{art:fucitomoralespoghossian2004}, where
however only
the case $r=2$, $k=2$ was considered and they wrote down the contributions involving coloured Young tableaux with up to four boxes. A complete and more systematic analysis was developed in \cite{art:fujiiminabe2005} (see also \cite[Section 2]{art:wyllard2011}).  

In an analogous way one can define partition functions for
supersymmetric gauge theories with matter fields in fundamental or adjoint representations of the gauge group ${\rm U}(r)$.
In particular, by taking the zero mass limit of the gauge theory with a single adjoint hypermultiplet, one obtains the Vafa-Witten partition function $\Zcal^{\mathrm{VW}}_{X_{\xi_0^\circ}} \big(\zsf, \vec y \ 
\big)_{\vec w}$ \cite{art:vafawitten1994} for topologically twisted $\Ncal=4$ supersymmetric Yang-Mills theory on the ALE space $X_{\xi_0^\circ}$, which is the generating function for the Euler characteristics of instanton moduli spaces \cite[Section 4.2.2]{art:fujiiminabe2005}. In \cite{art:fujiiminabe2005} it is also
verified that the resulting partition function is the character of a highest weight representation of the affine Lie algebra $\glfrakhat(k)$. 

\bigskip
\section{$\Ncal=2$ gauge theory for the level infinity chamber\label{sec:levelinfinity}}

Define 
\begin{equation}
\Csf=\big\{\xi_\R=\big(\xi_\R^i \big)\in \Dsf_\delta\ \big\vert\ \xi_\R^i>0\quad \forall
i\in \quiv_0\setminus\{0\} \big\} \ \subset \ \Dsf_\delta\ .
\end{equation}
For any $\vec{v}\in (\Z_{\geq0})^k$, let $\Csf_\infty(\vec{v}\, )$ be the unique chamber inside \eqref{eq:realpararameterspace} which is contained in the set $\{\xi_\R=(\xi_\R^i)\in \R^k\,\vert\, \sum_{i\in\quiv_0}\, \xi_\R^i >0\}$ and has $\Csf$ as its face. Then any parameter $\xi_\infty(\vec{v}\,)=(0,\xi_\R)\in \C^k\oplus\R^k$ with $\xi_\R\in \Csf_\infty(\vec{v}\, )$ is generic with respect to $\vec{v}$. By \cite{art:kuznetsov2007} (see also  \cite[Section 4.4]{art:nagao2009}), for any $\xi_\R\in \Csf_\infty(\delta)$ there is an isomorphism of ALE spaces $\Mcal_{\xi_\infty(\delta)}(\delta, \vec{w}_0\, )\simeq X_k$ for $\vec{w}_0:=(1, 0, \ldots, 0)$. We call $\Csf_\infty(\vec v)$ the \emph{level infinity chamber}.

The first approach to the study of $\Ncal=2$ gauge theories in the level
infinity chamber is due to Bonelli, Maruyoshi, Tanzini and Yagi
\cite{art:bonellimaruyoshitanziniyagi2012}. It does not rely on the
formalism of the Nakajima quiver varieties $\Mcal_{\xi_\infty(\vec{v}\, )}(\vec{v}, \vec{w}\, )$, but is instead based on the conjectural
\emph{master formula} due to Nekrasov \cite{art:nekrasov2006}.

\subsection{Nekrasov master formula}

Let $X$ be a smooth toric surface with torus $T_t:=\C^\ast\times
\C^\ast$. One can associate with $X$ a smooth fan $\Sigma_X$ in
$N_\Q:=N\otimes_\Z \Q$, where $N$ is the lattice of one-parameter
subgroups of $T_t$ \cite[Chapter 3]{book:coxlittleschenck2011}. By the orbit-cone correspondence, the
two-dimensional cones of $\Sigma_X$ correspond to the $T_t$-fixed
points of $X$ and the one-dimensional cones correspond to the
$T_t$-invariant lines of $X$. If $X^{T_t}=\{p_1, \ldots, p_k\}$ with $k\geq 1$,  then
for each fixed point $p_i$ there exists a $T_t$-invariant affine open
neighbourhood $U_i\simeq \C^2$ of $p_i$ for $i=1, \ldots, k$. On the
other hand, the Picard group $\Pic(X)$ is a free abelian group of
finite rank (say $ m$) \cite[Proposition 4.25]{book:coxlittleschenck2011}; denote by $\Lcal_j$ a set of generators, with $j=1, \ldots, m$.

Let $\varepsilon_1, \varepsilon_2$ be the generators of
$H_{T_t}^\ast(\mathrm{pt})$ (see Appendix \ref{app:eqcoh}). For $i=1, \ldots, k$ the tangent space to
$X$ at a fixed point $p_i$ is a representation of the torus $T_t$,
hence it decomposes into two irreducible representations of
$T_t$. Denote by $\varepsilon^{(i)}_1(\varepsilon_1,\varepsilon_2)$
and $\varepsilon^{(i)}_2(\varepsilon_1,\varepsilon_2)$ the weights of
the characters corresponding to these representations. The
fibre of the line bundle $\Lcal_j\to X$ over the point $p_i$ is also a representation of $T_t$ for any $j=1, \ldots, m$, and we denote by $\phi_j^{(i)}(\varepsilon_1,\varepsilon_2)$ the weight of the associated character.

The $\Omega$-deformation of $\Ncal=2$ supersymmetric $\mathrm{U}(r)$ gauge
theory on $X$ is obtained as a
reduction of six-dimensional $\Ncal=1$ gauge theory on a flat
$X$-bundle $M$ over $\T^2$ in the limit where the torus $\T^2$
collapses to a point~\cite[Section~3.1]{art:nekrasov2006}. The bundle $M$ can be realised as the quotient of $\C\times X$ by the $\Z^2$-action
\begin{equation}
(n_1,n_2)\triangleright (w,x)=\big(w+(n_1+\sigma \, n_2)\,,\, g_1^{n_1} \,
g_2^{n_2} (x) \big)\ ,
\end{equation}
where $x\in X$, $w\in\C$, $(n_1,n_2)\in\Z^2$, $g_1,g_2$ are two commuting isometries of $X$ and $\sigma$ is the complex structure modulus of $\T^2$. In the collapsing limit, fields of the gauge theory which are charged under the
R-symmetry group are sections of the pullback to $M$ of a flat
$T_t$-bundle over $\T^2$. As pointed out in \cite[Section~2.2.2]{art:nekrasovokounkov2006},
the chiral observables of the $\Omega$-deformed $\Ncal=2$ gauge theory
become closed forms on the moduli spaces of framed
instantons which are equivariant with respect to the action of the
torus $T:=T_t\times T_{\boldsymbol a}$, where $T_{\boldsymbol a}$ is the maximal torus of
the group ${\rm GL}(r, \C)$ of constant gauge transformations which rotates the framing. Thus correlation functions of chiral
BPS operators become integrals of equivariant Chern classes of natural bundles over
the moduli spaces. 

Denote again by
$\varepsilon_1, \varepsilon_2, a_1, \ldots, a_r$ the generators of
$H_T^\ast(\mathrm{pt})$; in gauge theory,
$a_1,\ldots,a_r$ are the expectation values of the complex scalar field $\phi$
of the $\Ncal=2$ vector multiplet and $\varepsilon_1, \varepsilon_2$
parameterize the holonomy of a flat connection on the $T_t$-bundle over
$\T^2$ used to define the $\Omega$-deformation. Nekrasov's conjecture
for the {master formula} is stated in \cite[Section
4]{art:nekrasov2006} in the presence of 2-observables and by
considering instead of $T_t$ the torus associated with the \emph{Cox
  ring} of $X$, which is the polynomial ring with variables associated to
the rays of $\Sigma_X$. Here we state the conjecture
for gauge theories without 2-observables and in a form which will be
suitable for describing the results of
\cite{art:bonellimaruyoshitanziniyagi2012}: the full partition
function $\Zcal_{X}^{\mathrm{full}}\big(\varepsilon_1,
\varepsilon_2, \boldsymbol{a} ; \qsf\big)$ of the $\Omega$-deformed pure $\Ncal=2$ $\mathrm{U}(r)$
gauge theory on $X$ factorises into a product of partition functions on the open affine subsets $U_i$ of $X$ as
\begin{equation}\label{eq:nekrasov}
\Zcal_{X}^{\mathrm{full}}\big(\varepsilon_1, \varepsilon_2,
\boldsymbol{a} ; \qsf\big)=\sum_{(\boldsymbol{h}_1,\ldots,
    \mbf{h}_m)\in (\Z^r)^m} \ \prod_{i=1}^k\,
\Zcal_{\C^2}^{\mathrm{full}}\Big(\varepsilon_1^{(i)}\,,\,
\varepsilon_2^{(i)}\,,\, \mbf{a}+
\mbox{$\sum\limits_{j=1}^m$}\,\phi_j^{(i)}\,\mbf{h}_j\, ; \, \qsf\Big)\ .
\end{equation}

\subsection{Master formula for ALE spaces\label{sec:MasterALE}}

The minimal resolution $X_k$ is a smooth toric surface with $k$
torus-fixed points $p_1, \ldots, p_k$, and $k+1$ torus-invariant
divisors $D_0, D_1, \ldots, D_k$ which are smooth projective curves of
genus zero. For $i=1, \ldots, k$ the divisors $D_{i-1}$ and $D_i$
intersect at the point $p_i$. The curves $D_1, \ldots, D_{k-1}$ are
the irreducible components of the exceptional divisor
$\varphi_k^{-1}(0)$, where $\varphi_k \colon X_k\to \C^2/\Z_k$ is the resolution of singularities morphism. By the McKay correspondence, there is a one-to-one correspondence between the irreducible representations of $\Z_k$ and the divisors $D_1, \ldots, D_{k-1}$ \cite[Corollary~10.3.11]{book:coxlittleschenck2011}. By \cite[Equation~(10.4.3)]{book:coxlittleschenck2011}, the intersection matrix $(D_i\cdot D_j)_{1\leq i,j\leq k-1}$ is given by minus the Cartan matrix $C$ of type $A_{k-1}$, i.e. one has
\begin{equation}
\left( D_i \cdot D_j \right)_{1\leq i,j\leq k-1} = -C= 
\begin{pmatrix}
-2 & 1 & \cdots & 0 \\
1 & -2 & \cdots & 0 \\
\vdots & \vdots & \ddots & \vdots\\
0 & 0 & \cdots & -2
\end{pmatrix} \ .
\end{equation}

The coordinate ring of $\C^2/\Z_k$ is $\C\big[\C^2/\Z_k\big]:=\C\big[T_1, T_1^{k-1} \, T_2^k\big]$. On the other hand $\C\big[\C^2/\Z_k\big]=\C[t_1,t_2]^{\Z_k}=\C\big[t_1^k,t_2^k,t_1\,t_2\big]$. These two rings are isomorphic under the change of variables
\begin{equation}\label{eq:variablerelations}
T_1=t_1^k\qquad\mbox{and}\qquad T_2=t_2 \, t_1^{1-k}\ .
\end{equation}
Let $U_i$ be the torus-invariant affine open subset of $X_k$ which is
a neighbourhood of the torus-fixed point $p_i$ for $i=1, \ldots, k$; its coordinate ring is given by $\C[U_i]:=\C\big[T_1^{2-i}\, T_2^{1-i},T_1^{i-1}\, T_2^i\big]$. By the relations \eqref{eq:variablerelations} we have
\begin{equation}\label{eq:coordinatering-i}
\C[U_i]=\C\big[t_1^{k-i+1}\,
t_2^{1-i}\,,\,t_1^{i-k}\, t_2^{i}\big]\ .
\end{equation}
Thus in this case
\begin{equation}\label{eq:weights}
\varepsilon_1^{(i)}(\varepsilon_1,\varepsilon_2)=(k-i+1)\,
\varepsilon_1-(i-1)\,
\varepsilon_2\qquad\mbox{and}\qquad\varepsilon_2^{(i)}(\varepsilon_1,\varepsilon_2)=
-(k-i)\, \varepsilon_1+i\, \varepsilon_2\ .
\end{equation}

The master formula \eqref{eq:nekrasov} for pure $\Ncal=2$ $\mathrm{U}(r)$ gauge theory on $X_k$ reads (with some slight modification)
\begin{equation}\label{eq:masterALE}
\Zcal_{X_k}^{\mathrm{full}}\big(\varepsilon_1, \varepsilon_2,
\mbf{a} ; \qsf, \vec{\zeta}\ \big)=\sum_{(\vec{h}_1,\ldots,
  \vec{h}_r)\in (\frac1k \,\Z^{k-1})^r}\ \prod_{i=1}^k\,
\Zcal_{\C^2}^{\mathrm{full}}\big(\varepsilon_1^{(i)},
\varepsilon_2^{(i)}, \mbf{a}^{(i)}; \qsf\big)\ \prod_{l=1}^{k-1}\, \zeta_l^{\crm_1^{(l)}}\ ,
\end{equation}
where $\mbf{a}^{(i)}:=\big(a_1^{(i)}, \ldots, a_r^{(i)}\big)$ is
defined by
\begin{equation}\label{eq:shift}
a_\alpha^{(i)}:=
a_\alpha+(\vec{h}_\alpha)_{i}\,\varepsilon_1^{(i)}+(\vec{h}_\alpha)_{i-1}\,\varepsilon_2^{(i)}\qquad\mbox{for}
\quad \alpha=1, \ldots, r\ ,
\end{equation}
and we set $(\vec{h}_\alpha)_{0}=(\vec{h}_\alpha)_{k}=0$. Equation
\eqref{eq:shift} is superficially different from \cite[Equation
(2.2)]{art:bonellimaruyoshitanziniyagi2012} because there the
$T_t$-fixed points of $X_k$ are labelled $0,1,\dots,k-1$, while in
this paper we label them as $1,\dots,k$; this implies that our weights
in \eqref{eq:shift} are equivalent to those of
\cite{art:bonellimaruyoshitanziniyagi2012} after a suitable reparameterization.

By localization, a $T$-invariant $\mathrm{U}(r)$ instanton on $X_k$
decomposes into $T_t$-invariant point-like ``regular'' instantons over
each chart $U_i$ for $i=1, \ldots, k$, together with ``fractional''
instantons which carry the magnetic fluxes of the gauge fields on the
exceptional curves $D_1, \ldots, D_{k-1}$ and keep track of the first
Chern class of the instanton (see e.g. \cite{art:fucitomoralespoghossian2006,art:griguoloseminaraszabotanzini2007}). The vectors $(\vec{h}_1,\ldots,
\vec{h}_r)$, which appear in \eqref{eq:masterALE}, fix the first Chern
class. For this, let us denote by $R_l$ the $l$-th tautological line
bundle on $X_k$ for $l=1, \ldots, k-1$ introduced in \cite{art:gochonakajima1992}. Then the first Chern classes
$\crm_1(R_l)$ for $l=1, \ldots, k-1$ form a basis of $H^2(X_k;\Z)$
\cite[Proposition 2.2]{art:kronheimernakajima1990}. Therefore, any
first Chern class of a $T$-invariant $\mathrm{U}(r)$ instanton is of the form
\begin{equation}
\crm_1=\sum_{\alpha=1}^r \ \sum_{l=1}^{k-1}\, (\vec{u}_\alpha)_l\, \crm_1(R_l)
\end{equation}
for $\vec{u}_\alpha\in \Z^{k-1}$ and $\alpha=1, \ldots, r$. We set
$C^{-1}\,\vec{u}_\alpha=\vec{h}_\alpha$ for $\alpha=1,\ldots, r$, where the inverse of the Cartan
matrix $C$ is given by
\begin{equation}
\big(C^{-1}\big)^{ij}=\frac{i\, (k-j)}k \qquad \mbox{for} \quad i\leq
j \ .
\end{equation}
One
further introduces the chemical potentials $\zeta_l$ for $l=1, \ldots,
k-1$ for the fractional instantons to keep track of the first Chern classes
\begin{equation}
\crm_1^{(l)}:=\sum_{\alpha=1}^r\, (\vec{u}_\alpha)_l=\sum_{\alpha=1}^r\
\sum_{m=1}^{k-1}\, C_{l m}\,(\vec{h}_\alpha)_m\ .
\end{equation}
In order to determine the shifts $a_\alpha^{(i)}$ explicitly, one computes the weights
$\phi^{(i)}_l$ from the local Chern character of $R_l$ over $U_i$ for  $i=1, \ldots, k$ and $l=1, \ldots, k-1$;
using Equation \eqref{eq:charactertautological} below one immediately
arrives at Equation \eqref{eq:shift}. In \cite[Section
2]{art:bonellimaruyoshitanziniyagi2012}, the shifts \eqref{eq:shift}
are interpreted heuristically by studying the patching together of non-trivial magnetic fluxes of gauge fields through the exceptional curves $D_1, \ldots, D_{k-1}$ of $X_k$.

By subdividing the full partition function
$\Zcal_{X_k}^{\mathrm{full}}\big(\varepsilon_1, \varepsilon_2,
\boldsymbol{a} ; \qsf, \vec{\zeta}\ \big)$ with respect to the possible
holonomies at infinity of the fractional $\mathrm{U}(r)$ instantons one finds
\begin{equation}
\Zcal_{X_k}^{\mathrm{full}}\big(\varepsilon_1, \varepsilon_2,
\mbf{a} ; \qsf, \vec{\zeta}\
\big)=\Zcal_{X_k}^{\mathrm{cl}}\big(\varepsilon_1, \varepsilon_2,
\mbf{a} ; \tau_{\mathrm{cl}}\big)\ \sum_{{\mbf{I}}=({I}_1,
  \ldots, {I}_r)}\,
\Zcal_{X_k}^{\mathrm{pert}}\big(\varepsilon_1,\varepsilon_2,\mbf{a}
\big)_{{\mbf{I}}} \
\Zcal_{X_k}^{\mathrm{inst}}\big(\varepsilon_1, \varepsilon_2,
\mbf{a}; \qsf, \vec{\zeta}\ \big)_{{\mbf{I}}}\ ,
\end{equation}
where ${I}_\alpha\in \{0, 1, \ldots, k-1\}$ for $\alpha=1, \ldots, r$ and the vector ${\mbf{I}}$
parameterizes the holonomy class of a $T$-invariant $\mathrm{U}(r)$
instanton on $X_k$. Explicit expressions for the classical,
perturbative and instanton contributions $\Zcal_{X_k}^{\mathrm{cl}}$,
$\Zcal_{X_k}^{\mathrm{pert}}$ and $\Zcal_{X_k}^{\mathrm{inst}}$ are
derived in \cite{art:bonellimaruyoshitanziniyagi2012} in terms of the
corresponding partition functions $\Zcal_{\C^2}^{\mathrm{cl}}$,
$\Zcal_{\C^2}^{\mathrm{pert}}$ and
$\Zcal_{\C^2}^{\mathrm{inst}}$. The perturbative partition function is described below, while the classical contribution is given by
\begin{equation}
\Zcal_{X_k}^{\mathrm{cl}}(\varepsilon_1,
\varepsilon_2,\mbf{a} ;\tau_{\mathrm{cl}})=\Zcal_{\C^2}^{\mathrm{cl}}(\varepsilon_1,
\varepsilon_2,\mbf{a} ;\tau_{\mathrm{cl}})^{\frac{1}{k}} = \exp\Big(-\frac{\tau_{\mathrm{cl}}}{2\, k\, \varepsilon_1\, \varepsilon_2} \
\sum_{\alpha=1}^r\, a_\alpha^2\Big) \ .
\end{equation}
The explicit
form of the instanton contributions is given by
\begin{multline}
\Zcal_{X_k}^{\mathrm{inst}}\big(\varepsilon_1, \varepsilon_2,
\mbf{a} ; \qsf, \vec{\zeta}\
\big)_{{\mbf{I}}}:=\sum_{\stackrel{\scriptstyle
    (\vec{h}_1,\dots,\vec h_r)\in (\frac1k \,\Z^{k-1})^r}{\scriptstyle
    k\,(\vec{h}_\alpha)_1={I}_\alpha\bmod{k}}} \, \qsf^{\scriptstyle
  \frac{1}{2}\,\sum\limits_{\alpha=1}^r\,\sum\limits_{l,m=1}^{k-1}\,
  (\vec{h}_\alpha)_l \, C_{l m} \, (\vec{h}_\alpha)_m} \ \prod_{l=1}^{k-1}\, \zeta_l^{\scriptstyle \sum\limits_{\alpha=1}^r\, (\vec{u}_\alpha)_l}\\
\times \ \prod_{n=1}^{k-1} \ \prod_{\alpha\neq\beta} \,
g^{(n)}\big(a_{\alpha\beta}^{(n)},\varepsilon_1^{(n)},
\varepsilon_2^{(n)} , (\vec{h}_{\alpha\beta})_n,
(\vec{h}_{\alpha\beta})_{n+1}\big)^{-1}\
\prod_{i=1}^k\,\Zcal_{\C^2}^{\mathrm{inst}}\big(\varepsilon_1^{(i)},
\varepsilon_2^{(i)}, \mbf{a}^{(i)}; \qsf \big)\ ,
\end{multline}
where $a_{\alpha\beta}^{(n)}=a_\alpha^{(n)}-a_\beta^{(n)}$,
$\vec{h}_{\alpha\beta}=\vec{h}_\alpha-\vec{h}_\beta$ for
$\alpha,\beta=1, \ldots, r$, and
\begin{equation}
g^{(n)}(a, e_1, e_2, \mu, \nu):=\left\{
\begin{array}{cl}
\displaystyle\prod_{\stackrel{\scriptstyle m_1\geq 0\,,\, m_2\leq
    -1}{\scriptstyle n\, (\nu+m_1)\leq (n+1)\, (\mu+m_2)}}\, (a+m_1 \,
e_1+m_2\, e_2)  \ , & \qquad n\, \nu < (n+1)\, \mu \ , \\[12pt]
1 \ , & \qquad n \, \nu = (n+1)\, \mu \ ,\\[12pt]
\displaystyle\prod_{\stackrel{\scriptstyle m_1\leq -1\,,\, m_2 \geq
    0}{\scriptstyle n\, (\nu+m_1)> (n+1)\, (\mu+m_2)}} \, (a+m_1 \,
e_1+m_2 \,
e_2) \ , & \qquad n\, \nu > (n+1)\, \mu \ .
\end{array}
\right.
\end{equation}
By fixing the holonomy at infinity ${\mbf{I}}$, we have in addition the constraint
\begin{equation}\label{eq:firstchernclass-tanzini}
k\,(\vec{h}_\alpha)_1={I}_\alpha \ \bmod{k}
\end{equation}
on the allowed first Chern classes of the $T$-invariant
$\mathrm{U}(r)$ instantons on $X_k$. 

Following \cite[Appendix~A]{art:nekrasovokounkov2006}, we denote by $\Gamma_2(x|-\varepsilon_1,-\varepsilon_2)$ the Barnes double gamma-function \cite{art:barnes1901}
which is the double zeta-function
regularization of the infinite product
\begin{equation}
\prod_{i,j=0}^\infty \, \big(x-i\, \varepsilon_1-j\, \varepsilon_2
\big) \ .
\end{equation}
Then the perturbative partition function for $\C^2$ is
\begin{equation}\label{eq:pertpartC2}
\Zcal_{\C^2}^{\mathrm{pert}}(\varepsilon_1,\varepsilon_2,\mbf a) := \prod_{\alpha\neq\beta}\ \frac1{\Gamma_2(a_{\alpha\beta}|\varepsilon_1,\varepsilon_2)}\ .
\end{equation}
The ``edge factor'' $g^{(n)}$ depends only on the combinatorial data
of the fan $\Sigma_{X_k}$ of the toric variety $X_k$ and is induced by the difference between $k$ copies of the
perturbative partition functions
for
$\C^2$
\begin{equation}
\prod_{i=1}^k\, \Zcal_{\C^2}^{\mathrm{pert}}\big(\varepsilon_1^{(i)}, \varepsilon_2^{(i)}, \mbf{a}^{(i)}\, \big) 
\end{equation}
and the perturbative partition function
$\Zcal_{X_k}^{\mathrm{pert}}$ for $X_k$ \cite[Equation
(3.13)]{art:bonellimaruyoshitanziniyagi2012}. The partition function
$\Zcal_{X_k}^{\mathrm{pert}}$ is defined via a slight modification of
\eqref{eq:pertpartC2}, obtained by replacing the Barnes double gamma
function $\Gamma_2(a_{\alpha\beta}|\varepsilon_1,\varepsilon_2)$ with a
suitable $\Z_k$-invariant version and imposing the dependence on the
holonomy ${\mbf{I}}$ via \eqref{eq:firstchernclass-tanzini}
\cite[Equation (3.12)]{art:bonellimaruyoshitanziniyagi2012}; explicitly one finds
\begin{multline}
\Zcal_{X_k}^{\mathrm{pert}}\big(\varepsilon_1, \varepsilon_2,
\mbf{a}
\big)_{{\mbf{I}}} = \sum_{\stackrel{\scriptstyle
    (\vec{h}_1,\dots,\vec h_r)\in (\frac1k \,\Z^{k-1})^r}{\scriptstyle
    k\,(\vec{h}_\alpha)_1={I}_\alpha\bmod{k}}} \ \prod_{n=1}^{k-1} \ \prod_{\alpha\neq\beta} \,
g^{(n)}\big(a_{\alpha\beta}^{(n)}, \varepsilon_1^{(n)},
\varepsilon_2^{(n)} , (\vec{h}_{\alpha\beta})_n,
(\vec{h}_{\alpha\beta})_{n+1}\big) \\ \times \ 
\prod_{i=1}^k\,\Zcal_{\C^2}^{\mathrm{pert}}\big(\varepsilon_1^{(i)},
\varepsilon_2^{(i)}, \mbf{a}^{(i)}\, \big) \ .
\end{multline}

Comparison of the instanton partition functions $\Zcal_{X_k}^{\mathrm{inst}}\big(\varepsilon_1, \varepsilon_2,
\mbf{a} ; \qsf, \vec{\zeta}\
\big)_{{\mbf{I}}}$ with those of Section \ref{sec:Zkproj} is not
straightforward. Since the chambers $\Csf_0$ and $\Csf_\infty(\vec v)$
are distinct, the ALE spaces $X_{\xi_0^\circ}$ and $X_k$ are   diffeomorphic, but not isomorphic as algebraic varieties; hence one expects a non-trivial equivalence between the corresponding partition functions. Although some explicit low order comparisons of the combinatorial expansions confirm these expectations (see e.g.   \cite{art:belavinbershteinfeiginlitvinovtarnopolsky2011,art:bonellimaruyoshitanziniyagi2012,art:itomaruyoshiokuda2013,art:alfimovbelavintarnopolsky2013}), at present there is no general relationship established between the two partition functions.

\bigskip
\section{Sheaves on root stacks and fractional instantons on $X_k$\label{sec:rootstacks}}

In this section we summarise the main results of
\cite{art:bruzzopedrinisalaszabo2013}, which provided the first
attempt at putting the calculations of Section \ref{sec:levelinfinity}
on a rigorous footing and elucidating the geometry behind these constructions. The key development is a moduli theory of framed sheaves on a suitable \emph{orbifold compactification} of the ALE space $X_k$. A related but somewhat different construction in the rank one case for unframed sheaves is provided by \cite{art:ciraficikashanipoorszabo2011}.

\subsection{Stacks}

The geometric construction of $\Ncal=2$ gauge theories on the ALE space $X_k$ relies on the formalism of Deligne-Mumford stacks;
see e.g. \cite[Section 7]{art:vistoli1989} and \cite{art:edidin1997}
for an overview. The main class of Deligne-Mumford stacks that we will
encounter in the following are global quotient stacks $[X/G]$ where
$X$ is a variety, and $G$ is a reductive linear algebraic group acting
on $X$ properly and with finite stabilizers. The stack $[X/G]$ is
associated with the action groupoid $G\times X\rightrightarrows X$,
where one arrow is given by the $G$-action and the other arrow by
projection to the second factor. If in addition $X$ is smooth and the
generic stabilizer of the $G$-action is trivial (equivalently $[X/G]$
contains a variety as a dense open subset), we say that $[X/G]$ is an
(effective) orbifold. If $f \colon X\to Y$ is the geometric quotient of $X$ by $G$, then $Y$ is called the coarse moduli space of $[X/G]$; for
example, when the action of $G$ on $X$ is free then $Y$ is the orbit
space $X/G$,
i.e. $Y$ is the set of isomorphism classes of objects of the associated action
groupoid. If the variety $X$ has trivial Picard group, then line bundles on the global quotient stack $[X/G]$ are associated with characters of the group $G$.

A nice family of global quotient stacks are toric stacks. The theory of Deligne-Mumford tori and toric Deligne-Mumford stacks is developed in \cite{art:fantechimannnironi2010} (see also \cite{art:borisovchensmith2004} for an equivalent approach based on stacky fans). Here we just recall the relevant definitions. A Deligne-Mumford torus $\Tscr$ is the product of an ordinary torus $T$ and a global quotient stack of the form $\Bscr H:= [\mathrm{pt}/H]$, where $H$ is a finite group. One can define a group-like structure on $\Tscr$ and a notion of $\Tscr$-action by using the theory of Picard stacks, but we do not enter into such details here. For us a toric Deligne-Mumford stack is a global quotient stack $[X/G]$ as before, with a Deligne-Mumford torus $\Tscr$ embedded in it as a dense open substack such that the action of $\Tscr$ on itself extends to an action on the whole stack $[X/G]$. A toric Deligne-Mumford stack with a projective coarse moduli space is said to be projective.

\subsection{Orbifold compactification of $X_k$\label{sec:orbifold}}

We define a normal projective compactification $\bar X_k= X_k\cup D_\infty$ of the minimal resolution $X_k$ by adding a smooth rational curve $D_\infty$ in such a way that $\bar X_2$ is the second Hirzebruch surface $\F_2$. For $k\geq 3$, $\bar X_k$ is only normal: it has two singular points, which are the two $T_t$-fixed points $0,\infty$ of $D_\infty\simeq \PP^1$; the affine toric open neighbourhoods of these points in $\bar X_k$ are isomorphic to  $\C^2/\Z_{\tilde k}$, where $\tilde k=\frac k2$ if $k$ is even while $\tilde k=k$ if $k$ is odd. Since $\bar X_k$ is not smooth for $k\geq 3$, we replace it with its \emph{canonical} orbifold; as $\bar X_k$ is toric, such an orbifold also naturally has a toric structure. Thus the canonical orbifold of $\bar X_k$ is a two-dimensional projective toric orbifold $\Xscr_k^{\rm can}$ with Deligne-Mumford torus $T_t$ and coarse moduli space $\pi_k^{\mathrm{can}} \colon \Xscr_k^{\mathrm{can}}\to\bar X_k$. The morphism $\pi_k^{\mathrm{can}}$ is an isomorphism over the smooth locus of $\bar X_k$; hence, for $k=2$, we find that $\Xscr_2^{\mathrm{can}}$ is isomorphic to the Hirzebruch surface $\F_2$. The orbifold $\Xscr_k^{\mathrm{can}}$ is realised as the global quotient stack of an open subset $Z$ of $\C^{k+2}$ by an action of the torus $(\C^\ast)^k$. The ALE space $X_k$ is a dense open subset $\imath_k^{\mathrm{can}}\colon X_k \xrightarrow{\sim} \Xscr_k^{\mathrm{can}}\setminus \tilde\Dscr_\infty$, where $\tilde\Dscr_\infty$ is the (reduced) preimage of the divisor $D_\infty$ in $\Xscr_k^{\rm can}$ through $\pi_k^{\mathrm{can}}$. The stack $\tilde\Dscr_\infty$ is an orbifold curve over the projective line $\PP^1$ with two orbifold points (a ``football'').

By performing a $k$-th root construction on this orbifold along $\tilde\Dscr_\infty$, one extends the automorphism group of a generic point on $\tilde\Dscr_\infty$ by the cyclic group $\Z_k$ and obtains a two-dimensional projective toric orbifold $\Xscr_k$ with coarse moduli space $\pi_k\colon \Xscr_k\to\bar X_k$. The orbifold $\Xscr_k$ is obtained as the fibre product
\begin{equation}
  \begin{tikzpicture}[xscale=1.5,yscale=-1.2]
    \node (A0_0) at (0, 0) {$\Xscr_k$};
    \node (A0_2) at (2, 0) {$\big[\A^1/\C^\ast\big]$};
    \node (A1_1) at (1, 1) {$\square$};
    \node (A2_0) at (0, 2) {$\Xscr_k^{\rm can}$};
    \node (A2_2) at (2, 2) {$\big[\A^1/\C^\ast\big]$};
    \node (Comma) at (2.7, 1) {$,$};
    \path (A0_0) edge [->]node [auto] {$\scriptstyle{}$} (A2_0);
    \path (A0_0) edge [->]node [auto] {$\scriptstyle{}$} (A0_2);
    \path (A0_2) edge [->]node [auto] {$\scriptstyle{}$} (A2_2);
    \path (A2_0) edge [->]node [auto] {$\scriptstyle{}$} (A2_2);
  \end{tikzpicture} 
\end{equation}
where the bottom horizontal arrow is determined by the line bundle $\Ocal_{\Xscr_k^{\rm can}}(\tilde\Dscr_\infty)$ together with its tautological global section and the right vertical arrow is induced by the morphism $z\mapsto z^k$. The orbifold $\Xscr_k$ can also be realised as a global quotient stack of the same variety $Z$ as before by an action of the torus $(\C^\ast)^k$; the difference between the stacks $\Xscr_k$ and $\Xscr_k^{\mathrm{can}}$ lies in the different $(\C^\ast)^k$-actions on $Z$.
The ALE space $X_k$ is a dense open subset $\imath_k\colon X_k \xrightarrow{\sim} \Xscr_k\setminus \Dscr_\infty$, where the
(reduced) preimage $\Dscr_\infty$ of $D_\infty$ in $\Xscr_k$ though $\pi_k$ is a smooth Cartier divisor which has the structure of a $\Z_k$-gerbe over $\tilde\Dscr_\infty$.

The orbifold structure of $\Xscr_k$ is concentrated on the compactification gerbe $\Dscr_\infty$, which will enable the rigorous incorporation of contributions from fractional instantons. The stack $\Dscr_\infty$ is isomorphic to the
global quotient stack
\begin{equation}
\Dscr_\infty \ \simeq \ \left[\,
      \frac{\C^2\setminus\{0\}}{\C^\ast\times \Z_k}\,\right] \ ,
\end{equation}
where the group action in given in \cite[Equation (3.28)]{art:bruzzopedrinisalaszabo2013}; this characterises $\Dscr_\infty$ as a toric Deligne-Mumford stack with Deligne-Mumford torus $\Tscr= \C^\ast\times \Bscr\Z_k$ and coarse moduli space $D_\infty$. Hence the Picard group $\Pic(\Dscr_\infty)$ is isomorphic to $\Z\oplus \Z_k$, and it is generated by the line bundles $\Lcal_{\mathrm{free}}, \Lcal_{\mathrm{tor}}$ corresponding to the characters $\chi_{\mathrm{free}}, \chi_{\mathrm{tor}}\colon\C^\ast\times\Z_k \to \C^\ast$ given respectively by the projections $(t,\omega)\mapsto t$ and $(t,\omega)\mapsto\omega$, where $t\in\C^\ast$ and $\omega$ is a primitive $k$-th root of unity. In particular, $\Lcal_{\mathrm{tor}}^{\otimes k}$ is trivial. Define the degree zero line bundles $\Ocal_{\Dscr_\infty}(i):=\Lcal_{\mathrm{tor}}^{\otimes i}$ for $k$ even and $\Ocal_{\Dscr_\infty} (i):=\Lcal_{\mathrm{tor}}^{\otimes i\,(k+1)/2}$ for $k$ odd. As pointed out in \cite{art:eyssidieuxsala2013}, the fundamental group of the underlying topological stack of $\Dscr_\infty$ is isomorphic to $\Z_k$, and for any $i=0, 1, \ldots, k-1$ the line bundle $\Ocal_{\Dscr_\infty}(i)$ inherits a unitary flat connection associated with the $i$-th irreducible unitary representation $\rho_i$ of $\Z_k$. Hence by \cite[Theorem 6.9]{art:eyssidieuxsala2013} line bundles on $\Xscr_k$ which are isomorphic along $\Dscr_\infty$ to $\Ocal_{\Dscr_\infty}(i)$ correspond to $\mathrm{U}(1)$ instantons on $X_k$ with holonomy at infinity given by $\rho_i$. One can also consider more general line bundles by tensoring $\Ocal_{\Dscr_\infty}(i)$ with a power $\Lcal_{\mathrm{free}}^{\otimes s}$ to get line bundles of degree a rational multiple of $s$ \cite[Lemma 3.39]{art:bruzzopedrinisalaszabo2013}; however, here we shall set $s=0$ as we wish to concentrate on the line bundles corresponding to fractional instantons.

For $i=1,\dots,k-1$ let $\Dscr_i$ be the preimages of the
exceptional divisors $D_i$ equipped with the reduced scheme structure; their intersection
form is given by $-C$. Let $\Rcal_i=\Ocal_{\Xscr_k}(\omega_i)$ be the
line bundle associated to the dual class
\begin{equation}
\omega_i= -\sum_{j=1}^{k-1}\,
  \big(C^{-1} \big)^{ij}\, \Dscr_j \ ,
\end{equation}
which is an integral class in the Picard group $\Pic(\Xscr_k)$ for $i=1,\dots,k-1$ (cf.\ \cite[Lemma 3.21]{art:bruzzopedrinisalaszabo2013}). The
restrictions $\Rcal_i|_{X_k}$ coincide with the Kronheimer-Nakajima tautological line
bundles $R_i$, while $\Rcal_i|_{\Dscr_\infty}\simeq \Ocal_{\Dscr_\infty}(i)$. The line bundles $\Ocal_{\Xscr_k}(\Dscr_\infty)$ and $\Rcal_i$ for $i=1,\dots,k-1$ freely generate $\Pic(\Xscr_k)$ over $\Z$.

\subsection{Framed sheaves}

In order to construct moduli spaces of framed sheaves on the orbifold $\Xscr_k$
which are needed for the formulation of supersymmetric gauge theories
on $X_k$, we first have to choose a suitable framing sheaf, which
will encode the   holonomy at infinity of the fractional instantons. In light of the discussion of Section \ref{sec:orbifold}, we choose as framing sheaf the locally free sheaf
\begin{equation}
\Fcal_{\infty}^{\vec w}:=\bigoplus_{i=0}^{k-1} \,
\Ocal_{\Dscr_\infty}(i)^{\oplus w_i}
\end{equation}
for a fixed vector $\vec{w}:=(w_0,w_1, \ldots, w_{k-1})\in (\Z_{\geq0})^{k}$. By applying the general theory of framed sheaves on projective Deligne-Mumford stacks developed in \cite{art:bruzzosala2013}, one can construct a fine moduli space
$\Mcal_{\vec{u},\Delta,\vec w}$ parameterizing isomorphism classes of $(\Dscr_\infty,\Fcal_\infty^{
  \vec{w}}\, )$-framed sheaves
$(\Ecal,\phi_\Ecal\colon \Ecal\vert_{ \Dscr_\infty}\xrightarrow{\sim}
\Fcal_\infty^{\vec{w}}\, )$ on $\Xscr_k$ with fixed rank
$r:=\sum_{i=0}^{k-1}\, w_i$, first Chern class
$\crm_1(\Ecal)=\sum_{l=1}^{k-1} \, u_l \, \omega_l$ and discriminant 
\begin{equation}
\Delta(\Ecal):= \int_{\Xscr_k}\, \Big(\crm_2(\Ecal)-\frac{r-1}r\,
\crm_1(\Ecal)^2\Big) = \Delta\ .
\end{equation}

Due to the framing the vector $\vec{u}=(u_1,\dots,u_{k-1})\in \Z^{k-1}$ satisfies the constraint
\begin{equation}\label{eq:firstchernclass}
\sum_{i=1}^{k-1}\, i \, u_i = \sum_{i=1}^{k-1}\, i \, w_i \ \bmod{k}\ ,
\end{equation}
or equivalently
\begin{equation}
k \, h_{k-1} = \sum_{i=1}^{k-1}\, i \, w_i \ \bmod{k}
\end{equation}
where $\vec{h}:=C^{-1}\vec{u}$. As we now explain, this constraint has a representation theory meaning. For this, let us denote by $\vec{e}_i$ the $i$-th coordinate vector in $\Z^{k}$. Define the elements $\vec{\gamma}_i:=\vec{e}_i-\vec{e}_{i+1}$ for $i=1, \ldots, k-1$. Then $\vec{\gamma}_i\cdot \vec{\gamma}_j=C_{ij}$, where $C$ is the Cartan matrix of the Dynkin diagram of type $A_{k-1}$ which has root lattice $\Qfrak:=\bigoplus_{i=1}^{k-1}\, \Z\vec{\gamma}_i$ with a nondegenerate symmetric bilinear form $\langle-,- \rangle_\Qfrak$ induced by $C$. The elements of $\Qfrak$ are called {roots}, and $\vec{\gamma}_i$ is called the {$i$-th simple root} for $i=1, \ldots, k-1$. The {fundamental weights $\vec{\omega}_i$ of type $A_{k-1}$} are the vectors in $\Z^k$ given by
\begin{equation}
\vec{\omega}_i:=\sum_{l=1}^{i} \, \vec{e}_l-\frac{i}{k}\,
\sum_{l=1}^{k}\, \vec{e}_l
\end{equation}
for $i=1,\dots,k-1$. Let $\Pfrak:=\bigoplus_{i=1}^{k-1}\, \Z \vec{\omega}_i$ be the {weight lattice}; then $\Qfrak\subset \Pfrak$, as $\vec{\gamma}_i=\sum_{j=1}^{k-1} \, C_{ij}\, \vec{\omega}_j$. We subdivide the vectors $\vec{u}\in\Z^{k-1}$ according to Equation \eqref{eq:firstchernclass} as
\begin{equation}
\Ufrak_{\vec{w}}:=\Big\{\vec{u}\in\Z^{k-1}\ \Big\vert\ \mbox{$\sum\limits_{i=1}^{k-1}\, i \, u_i = \sum\limits_{i=1}^{k-1}$} \, i \, w_i \ \bmod{k}\Big\} \ .
\end{equation}
Define now a bijective map by
\begin{equation}
\psi\, \colon\, \Z^{k-1} \ \longmapsto \ \Pfrak\ ,\qquad
\vec{u}\ \longmapsto \  \sum_{i=1}^{k-1} \, u_i \, \vec{\omega}_i\ .
\end{equation}
Then $\psi^{-1}\big(\Qfrak + \sum_{i=1}^{k-1}\,w_i\,\vec{\omega}_i\big) = \Ufrak_{\vec{w}}$, which implies that $\psi(\vec{u}\, )$ for $\vec{u}\in \Ufrak_{\vec{w}}$ is naturally written as a sum of the weight $\sum_{i=1}^{k-1}\,w_i\,\vec{\omega}_i$ and the root $\vec{\gamma}_{\vec{u}}$ given by
\begin{equation}
\vec{\gamma}_{\vec{u}}:=\sum_{i,j=1}^{k-1}\, \big(C^{-1}\big)^{
ij}\, (u_j-w_j) \ \vec{\gamma}_i\ .
\end{equation}

Since the moduli space
$\Mcal_{\vec{u},\Delta,\vec w}$
is fine, there exists a {universal framed
  sheaf}
\begin{equation}
\big(\boldsymbol{\Ecal}_{\vec{u},\Delta,\vec w}\,,\, \boldsymbol{\phi}_{\boldsymbol{\Ecal}_{\vec{u},\Delta,\vec w}}\colon \boldsymbol{\Ecal}_{\vec{u},\Delta,\vec w}\big\vert_{
  \Mcal_{\vec{u},\Delta,\vec w}\times\Dscr_\infty}\xrightarrow{\sim}
p_{2}^\ast(\Fcal_\infty^{\vec{w}})\big) \ ,
\end{equation}
where
$\boldsymbol{\Ecal}_{\vec{u},\Delta,\vec w}$ is a coherent sheaf on $
\Mcal_{\vec{u},\Delta,\vec w}\times \Xscr_k$ which is flat over
$\Mcal_{\vec{u},\Delta,\vec w}$; here and in the following we use the notation $p_{i}$ for the projection of a product of varieties to the $i$-th factor. The fibre over $[(\Ecal,\phi_\Ecal)]\in\Mcal_{\vec{u},\Delta,\vec w}$ is itself the $(\Dscr_\infty,\Fcal_\infty^{\vec{w}}\, )$-framed
sheaf $(\Ecal,\phi_\Ecal)$ on $\Xscr_k$. 

The moduli space $\Mcal_{\vec{u},\Delta,\vec w}$ is a smooth quasi-projective variety of dimension 
\begin{equation}
\dim\Mcal_{\vec{u},\Delta,\vec w} =2\, r\, \Delta - \frac12\, \sum_{j=1}^{k-1}\, \big(C^{-1}\big)^{jj}\, \vec{w}\cdot\vec{w}(j)\ ,
\end{equation}
where for $j=1, \ldots, k-1$ we defined the vector $\vec{w}(j):=(w_j, \ldots, w_{k-1}, w_0, w_1, \ldots, w_{j-1})$; the Zariski tangent
space of
$\Mcal_{\vec{u},\Delta,\vec{w}}$ at a point $[(\Ecal,\phi_{\Ecal})]$ is $\mathrm{Ext}^1(\Ecal,\Ecal\otimes \Ocal_{\Xscr_k}(-\Dscr_\infty))$.
By \cite[Theorem 6.9]{art:eyssidieuxsala2013}, it contains as an open
subset the moduli space of $\mathrm{U}(r)$ instantons on $X_k$ with
first Chern class $\crm_1= \sum_{l=1}^{k-1} \, u_l \, \crm_1(R_l)$,
discriminant $\Delta$ and holonomy at infinity associated with the
unitary representation $\rho=\bigoplus_{i=0}^{k-1} \, \rho_i^{\oplus
  w_i}$ of the cyclic group $\Z_k$.

In the rank one case $r=1$, there is a non-canonical isomorphism of fine moduli spaces
\begin{equation}
\Mcal_{\vec{u},n,\vec w_j}\simeq \hilb{n}{X_k} \ ,
\end{equation}
where $j\in\{0,1,\dots,k-1\}$ is defined by $(\vec w_j)_i= \delta_{ij}$, and $\hilb{n}{X_k}$ is the Hilbert scheme of $n$ points on $X_k$, which is a smooth quasi-projective variety of dimension $2n$. Kuznetsov shows in \cite{art:kuznetsov2007} that this Hilbert scheme is isomorphic to the Nakajima quiver variety $\Mcal_{\xi_\infty(n\,\delta)}(n\,\delta,\vec w_0\,)$ for $\vec w_0=(1,0,\dots,0)$, $\xi_\infty(n\,\delta)=(0, \xi_\R)$ and $\xi_\R\in \Csf_\infty(n\,\delta\,)$, and hence our rank one moduli spaces are also non-canonically isomorphic to quiver varieties. It is natural to anticipate that this property generalises to the higher rank case, and hence we conjecture that the moduli spaces $\Mcal_{ \vec{u}, \Delta,\vec w}$ are isomorphic to 
Nakajima quiver varieties $\Mcal_{\xi_\infty(\vec{v}\,)}(\vec{v}, \vec{w}\, )$ for a suitable choice of dimension vector $\vec{v}\in(\Z_{\geq0})^k$.

\subsection{$\Ncal=2$ gauge theory on $X_k$}

We can now define the instanton partition functions for the $\Omega$-deformed pure $\Ncal=2$ gauge theory on $X_k$ as
generating functions for $T$-equivariant integrals over the moduli spaces
$\Mcal_{\vec{u},\Delta,\vec w}$. There is a natural $T$-action on
$\Mcal_{\vec{u},\Delta,\vec w}$ defined on a framed sheaf by pullback of the underlying torsion-free sheaf via the $T_t$-action on $\Xscr_k$ and by rotation of the framing by a constant gauge transformation from $T_{\mbf a}$. Then a torus-fixed point $[(\Ecal,\phi_\Ecal)]\in
(\Mcal_{\vec{u},\Delta,\vec w})^T$ decomposes as a direct sum of rank one framed sheaves
\begin{equation}
(\Ecal, \phi_{\Ecal})=\bigoplus_{\alpha=1}^r \, (\Ecal_\alpha, \phi_{\alpha})
\end{equation}
parameterized by combinatorial data $({\boldsymbol{\vec Y}},
{\boldsymbol{\vec u}}\, )$ consisting of collections of vectors of Young tableaux ${\boldsymbol{\vec Y}}=(\vec{Y}_1, \ldots, \vec{Y}_r)$, $\vec{Y}_\alpha=\{Y_\alpha^i\}_{i=1, \ldots, k}$, and vectors of integers ${\boldsymbol{\vec u}}=(\vec{u}_1, \ldots, \vec{u}_r)$ such that for each $i=0, 1, \ldots, k-1$ and $\sum_{j=0}^{i-1}\, w_j<\alpha\leq\sum_{j=0}^i\, w_j$:
\begin{itemize}
\item $\Ecal_\alpha={\imath_k}_\ast(I_\alpha)\otimes
  \Rcal^{\vec{u}_\alpha}$, where $I_\alpha$ is the
  ideal sheaf of a zero-dimensional subscheme $Z_\alpha$ of $X_k$ with
  length $n_\alpha= \sum_{i=1}^k\, \big|Y_\alpha^i\big|$
  supported at the $T_t$-fixed points $p_1,\ldots,p_k$, while
  $\vec{u}_\alpha\in\Z^{k-1}$ obeys $\sum_{\alpha=1}^r\, \vec
  u_\alpha=\vec u$ and $\vec{h}_\alpha:=C^{-1}\vec{u}_\alpha$ satisfies
\begin{equation}\label{eq:firstchernclass-fixedpoints}
k\, (\vec{h}_\alpha)_{k-1}=  i\, \bmod{k} \ ;
\end{equation}
\item ${\phi_\alpha}\colon \Ecal_\alpha\big\vert_{ \Dscr_\infty}
  \xrightarrow{\sim}\Ocal_{\Dscr_\infty}(i)$ is induced by the
  canonical isomorphism $\Rcal^{\vec{u}_\alpha}\big\vert_{ \Dscr_\infty}\simeq
  \Ocal_{\Dscr_\infty}(i)$;

\medskip

\item $\displaystyle{ \Delta =\sum_{\alpha=1}^r \, n_\alpha+\frac{1}{2}\,
\sum_{\alpha=1}^r\, \vec{h}_\alpha\cdot C\vec{h}_\alpha-\frac{1}{2r}\,
\sum_{\alpha,\beta=1}^r \, \vec{h}_\alpha\cdot C\vec{h}_\beta \ \in \
\mbox{$\frac{1}{2\, r\, k}$} \, \mathbb{Z}} \ . $
\end{itemize}
Here we introduced the shorthand $\Rcal^{\vec{u}_\alpha}:= \bigotimes_{l=1}^{k-1}\,
  \Rcal_l^{\otimes (\vec{u}_\alpha)_l}$.

Introduce topological couplings $\qsf\in\C^\ast$ with $|\qsf|<1$ and $\vec\xi=(\xi_1,\dots,\xi_{k-1}) \in (\C^\ast)^{k-1}$ with $|\xi_i|<1$. The instanton partition function in the topological sector labelled by $\vec h=C^{-1}\vec u$ is defined by
\begin{equation}
\Zcal_{\vec{h}}^{\rm inst}\big(\varepsilon_1, \varepsilon_2, \mbf{a}; \qsf \big)_{\vec w} = \sum_{\Delta\in \frac{1}{2\,r\, k}\, \mathbb{Z}}\,
  \qsf^{\Delta+\frac{1}{2r}\, \vec{h}\cdot C\vec{h}} \ 
  \int_{\Mcal_{C\vec{h},\Delta,\vec w}} \, \big[\Mcal_{C\vec{h},\Delta,\vec w} \big]_T \ ,
\end{equation}
while the full instanton partition function is given by a weighted sum over fractional instantons with chemical potentials $\vec\xi$ as
\begin{equation}
\Zcal_{X_k}^{\rm inst}\big(\varepsilon_1, \varepsilon_2,
\mbf{a} ; \qsf , \vec{\xi} \ \big)_{\vec w}=
\sum_{C\vec{h}\in\Ufrak_{\vec w}} \,
\vec{\xi}^{\ \vec{h}}\ \Zcal_{\vec{h}}^{\rm inst}(\varepsilon_1, \varepsilon_2,
\mbf{a} ; \qsf )_{\vec w}
\end{equation}
where $\vec\xi\ ^{\vec
  h}:=\prod_{l=1}^{k-1}\, \xi_l^{h_l}$. The $T$-equivariant
Euler class of the tangent bundle
$T\Mcal_{\vec{u},\Delta,\vec w}$ at a $T$-fixed point with combinatorial data $(\vec{\boldsymbol{Y}},\vec{\boldsymbol{u}})$ is given by
\begin{equation}
\eu_{T}\big(T_{(\vec{\boldsymbol{Y}},\vec{\boldsymbol{u}})}
\Mcal_{\vec{u},\Delta,\vec w}\big) =\prod_{\alpha,\beta=1}^r \ \prod_{i=1}^k \,
m_{Y_{\alpha}^{i}, {Y_{\beta}^{i}}}
\big(\varepsilon_1^{(i)},\varepsilon_2^{(i)}, a_{\beta\alpha}^{(i)} \,
\big) \ \prod_{n=1}^{k-1} \,
\ell^{(n)}_{\vec{h}_{\beta\alpha}}\big(\varepsilon_1^{(n)},\varepsilon_2^{(n)},
a_{\beta\alpha} \big)
\end{equation}
where $a_{\beta\alpha}=a_\beta-a_\alpha$ and
$\vec{h}_{\beta\alpha}=\vec{h}_\beta-\vec{h}_\alpha$. The contributions $m_{Y_{\alpha}^{i}, {Y_{\beta}^{i}}}$ from the open affine neighbourhoods $U_i$ are given by \eqref{eq:mYfactor}. 
The explicit
expressions for the ``edge factors" $\ell_{\vec
  h_{\beta\alpha}}^{(n)}$ are rather complicated and can be found in
Appendix \ref{app:edgecontributions}; they depend
explicitly on the Cartan matrix $C$. By using the localization theorem
in equivariant cohomology (see Appendix \ref{app:eqcoh}) we thus obtain the factorization formula
\begin{multline}
\Zcal^{\mathrm{inst}}_{X_k}\big(\varepsilon_1, \varepsilon_2, \mbf{a} ; \qsf, \vec{\xi}\ \big)_{\vec w}= \sum_{C\vec{h}\in\Ufrak_{\vec w}} \, \vec{\xi}^{\ \vec{h}}\
  \sum_{\stackrel{\scriptstyle {\boldsymbol{\vec h}}=(\vec h_1,\dots,\vec h_r)}{\scriptstyle \vec h= \sum_{\alpha=1}^r\, \vec h_\alpha}} \,\qsf^{\frac{1}{2}\,
    \sum\limits_{\alpha=1}^r \, \vec{h}_\alpha\cdot C\vec{h}_\alpha} \\ \times \ 
  \prod_{\alpha,\beta=1}^r\ \prod_{n=1}^{k-1} \,
  \ell^{(n)}_{\vec{h}_{\beta\alpha}}\big(\varepsilon_1^{(n)},\varepsilon_2^{(n)}, a_{\beta\alpha}\big)^{-1} \ 
\prod_{i=1}^k\, \Zcal_{\C^2}^{\mathrm{inst}}\big(\varepsilon_1^{(i)}, \varepsilon_2^{(i)}, \mbf{a}^{(i)}; \mathsf{q} \big)\ .
\end{multline}

Let us compare this result explicitly with the result of Section
\ref{sec:MasterALE}. First note that our $\vec{u}\in \Z^{k-1}$ are the
same as in \cite{art:bonellimaruyoshitanziniyagi2012} while
$\vec{I}=(k-1, \ldots, k-1, k-2, \ldots, k-2, \ldots, 0,\ldots, 0)$,
where $k-i-1$ appears with multiplicity $w_i$ for $i=0, 1,\ldots, k-1$;
hence the constraint \eqref{eq:firstchernclass-tanzini} is equivalent
to \eqref{eq:firstchernclass} because $k\, (\vec{h}_\alpha)_{k-1}=
k\, (\vec{h}_\alpha)_{1}\, \bmod{k}$. The ``fugacities'' are related by
$\xi_l=\zeta_{l-1}\, \zeta_{l+1}\,/\,\zeta_{l}$ for $l=1, \ldots, k-1$
(where we set $\zeta_0=\zeta_k=1$).

For $k=2$ our instanton partition function assumes the form
\begin{multline}
\Zcal^{\mathrm{inst}}_{X_2}(\varepsilon_1, \varepsilon_2,
\mbf{a}\, ; \qsf, \xi)_{\vec w} =\sum_{\stackrel{\scriptstyle h\in\frac{1}{2}\, \Z }{\scriptstyle 2
    h= \, w_1 \bmod{2}}} \, \xi^{h} \ \sum_{\stackrel{{\scriptstyle {\boldsymbol h}}=( h_1,\dots, h_r)}{\scriptstyle h=\sum_{\alpha=1}^r\, h_\alpha}} \,\qsf^{\sum\limits_{\alpha=1}^r \, h_\alpha^2} \ \prod_{\alpha,\beta=1}^r\,\ell_{h_{\beta\alpha}}\big(2\varepsilon_1, \varepsilon_2 - \varepsilon_1, a_{\beta\alpha} \big)^{-1} \\  
  \times \ \Zcal_{\C^2}^{\mathrm{inst}}\big(2\varepsilon_1,
  \varepsilon_2 -\varepsilon_1, \mbf{a}+2\varepsilon_1\,
  \mbf{h} ; \qsf\big) \
  \Zcal_{\C^2}^{\mathrm{inst}}\big(\varepsilon_1 - \varepsilon_2,
  2\varepsilon_2,\mbf{a}+2\varepsilon_2\, \mbf{h} ; \qsf \big)\ ,
\end{multline}
where from Appendix \ref{app:edgecontributions} the edge factors are given by
\begin{equation*}
\ell_{h}(e_1,e_2,
a)=\left\{
\begin{array}{cl}
\displaystyle\prod_{m_1=0}^{\lfloor h\rfloor-1}\limits\hspace{0.2cm}
\prod_{m_2=0}^{2m_1+2\{h\}}\limits\big(a+m_1\,
e_1+m_2\, e_2\big) \ , & \qquad \lfloor h\rfloor> 0\ ,\\[8pt]
1 \ , & \qquad \lfloor h\rfloor= 0\ ,\\[8pt]
\displaystyle\prod_{m_1=1}^{-\lfloor h\rfloor}\limits\hspace{0.2cm} \prod_{m_2=1}^{2m_1-2\{h\}-1}\limits
\big(a+(2\{h\}-m_1)\,
e_1-m_2\, e_2\big) \ , & \qquad \lfloor h\rfloor< 0 \ .
\end{array}
\right.
\end{equation*}
For $\{h\}=0$ these formulas coincide with the edge factors obtained in \cite{art:gasparimliu2010} up to a redefinition of the equivariant parameters (see also \cite{art:ciraficiszabo2012}), as they should, since the computation of the edge factors in this case is
equivalent to that carried out for the Hirzebruch surface $\F_2$ in
\cite[Section~4.2]{art:bruzzopoghossiantanzini2011}. Moreover, for $\lfloor h\rfloor>0$ they can be easily written in the form
\begin{equation}
\ell_{h}(e_1,e_2,
a ) = \prod_{\stackrel{\scriptstyle m_1,m_2 \geq1\,,\,m_1+m_2 \leq 2\lfloor h\rfloor}{\scriptstyle m_1+m_2 \ {\rm even}}}\, \big(a+(m_1-1)\, \tilde e_1+(m_2-1)\, \tilde e_2\big)
\end{equation}
with $\tilde e_1=\frac{e_1}2$ and $\tilde e_2=\frac{e_1}2+e_2$, which
coincide with the edge factors of
\cite{art:bonellimaruyoshitanzini2011,
  art:bonellimaruyoshitanzini2012,
  art:belavinbershteinfeiginlitvinovtarnopolsky2011} (similarly for
$\lfloor h\rfloor<0$ and/or $\{h\}=\frac12$).

However, for $k\geq3$ the edge factors $\ell_{\vec h}^{(n)}$ appear to be drastically different from $g^{(n)}$, and although some explicit checks for $r=2$, $k=3$ in \cite[Appendix D]{art:bruzzopedrinisalaszabo2013} demonstrate that our
partition functions agree with those in
\cite[Appendix C]{art:bonellimaruyoshitanziniyagi2012} at leading orders in the
$\qsf$-expansion, a complete proof of the equivalence between the two partition functions is currently lacking.

\bigskip
\section{$\Ncal=2$ superconformal quiver gauge theories on $X_k$\label{sec:quivergaugetheory}}

Most of the material in this section is new. We consider $\Omega$-deformed $\Ncal=2$ superconformal quiver gauge theories on the ALE space $X_k$, generalising the analysis of \cite{art:pedrinisalaszabo2014} which considered only the abelian cases. We shall also present a new computation of the perturbative partition functions for $\Ncal=2$ gauge theories on $X_k$ using our formalism, and hence provide a rigorous proof of the Nekrasov master formula for $X_k$. We also describe some aspects of the Seiberg-Witten geometry of these gauge theories.

\subsection{Hypermultiplet bundles}

Let $\quiv=(\quiv_0,\quiv_1,\source,\tail)$ be a quiver. The vertices $\quiv_0$ label vector multiplets in the corresponding $\Ncal=2$ quiver gauge theory on $X_k$, while the matter hypermultiplets are associated with representations of $\quiv$. We fix integer vectors $\boldsymbol{r},\boldsymbol m,\boldsymbol{\bar m}\in (\Z_{\geq0})^{\quiv_0}$. Let $E:=\bigoplus_{\upsilon\in\quiv_0}
\, E_\upsilon$,  $V:=\bigoplus_{\upsilon\in\quiv_0} \, V_\upsilon$ and $\bar V:=\bigoplus_{\upsilon\in\quiv_0} \, \bar V_\upsilon$ be $\quiv_0$-graded complex
vector spaces such that $\dim E_\upsilon=r_\upsilon$, $\dim V_\upsilon=m_\upsilon$ and $\dim \bar V_\upsilon=\bar m_\upsilon$ for each vertex $\upsilon\in\quiv_0$. Set
\begin{equation}
M_\quiv(\boldsymbol{r},\boldsymbol m,\boldsymbol{\bar m}):= \Big(\, \bigoplus_{\upsilon\in{\quiv}_0}\,\Hom\big(V_\upsilon,
E_\upsilon\big)\oplus\Hom\big(E_\upsilon,\bar V_\upsilon\big)\, \Big) \
\oplus \ \Big(\, \bigoplus_{e\in
  {\quiv}_1}\,\Hom\big(E_{\source(e)}, E_{\tail(e)}\big)\, \Big) \ .
\end{equation}
The gauge group
\begin{equation}
G_{\boldsymbol{r}}:=\prod_{\upsilon\in\quiv_0}\,\mathrm{GL}(r_\upsilon,\C)
\end{equation}
acts on $M_\quiv(\boldsymbol{r},\boldsymbol m,\boldsymbol{\bar m})$ in a way analogous  to the action defined in \eqref{eq:quivergroup}. The hypermultiplet space also carries a natural action of the total flavour symmetry group
\begin{equation}
G_{\boldsymbol m,\boldsymbol{\bar m}}:= 
\prod_{\upsilon\in\quiv_0}\, \mathrm{GL}(m_\upsilon,\C)\times \mathrm{GL}(\bar m_\upsilon,\C)  \
\times \ \prod_{e\in\quiv_1}\, \C^* \ .
\end{equation}
The matter bundle associated to the quiver representation $M_\quiv(\boldsymbol{r},\boldsymbol m,\boldsymbol{\bar m})$ is constructed as follows. Let $T_\mu=\C^\ast$ and $H^\ast_{T_\mu}(\mathrm{pt})=\C[\mu]$. Denote by $\Ocal_{\Xscr_k}(\mu)$ the trivial line bundle on the orbifold $\Xscr_k$ on which $T_\mu$ acts by scaling the fibres with weight $\mu$. 

As described in \cite[Section 4.5]{art:bruzzopedrinisalaszabo2013},
one can define Carlsson-Okounkov type Ext-bundles as the elements $\Ebf_{\mu}^{\vec u_1,
  \Delta_1, \vec w_1; \vec u_2, \Delta_2, \vec w_2}$ in
the K-theory group $K\big(\Mcal_{\vec{u}_1,\Delta_1,{\vec
  w}_1}\times\Mcal_{\vec{u}_2,\Delta_2,\vec w_2}\big)$
by
\begin{equation}
\Ebf_{\mu}^{\vec u_1,
  \Delta_1, \vec w_1; \vec u_2, \Delta_2, \vec w_2}:={p_{12}}_\ast\big(-\boldsymbol{\Ecal
}_1^\vee\otimes \boldsymbol{\Ecal}_2\otimes p_3^\ast(\Ocal_{\Xscr_k}(\mu)\otimes\Ocal_{\Xscr_k}(-\Dscr_\infty))
\big) \ ,
\end{equation}
where $
\boldsymbol{\Ecal}_i:=p_{i 3}^\ast\big(\boldsymbol{\Ecal}_{\vec
  u_i,\Delta_i,\vec w_i} \big)$ in $K\big(\Mcal_{\vec{u}_1,\Delta_1,\vec w_1}
\times\Mcal_{\vec{u}_2,\Delta_2,\vec w_2}\times \Xscr_k\big)$ with $\boldsymbol{\Ecal}_{\vec{u},\Delta,\vec w}$ the {universal sheaf} on $
\Mcal_{\vec{u},\Delta,\vec w}\times\Xscr_k$; here $p_{ij}$ is the projection onto the product of the $i$-th and $
j$-th factors. Its fibre over $\big([(\Ecal_1,\phi_{\Ecal_1})]\,,\,[(\Ecal_2,\phi_{\Ecal_2})]\big)$ is 
\begin{equation}
\Ebf_{\mu}^{\vec u_1, \Delta_1, \vec w_1; \vec u_2, \Delta_2, \vec w_2}\big|_{\left([(\Ecal_1,\phi_{\Ecal_1})]\,,\,[(\Ecal_2,\phi_{\Ecal_2})]\right)}=\Ext^1\big(\Ecal_1\,,\,\Ecal_2 \otimes
\Ocal_{\Xscr_k}(\mu)\otimes\Ocal_{\Xscr_k}(-\Dscr_\infty) \big)\ .
\end{equation}
One can compute the dimension of this vector space by a straightforward generalization of the   computations of \cite[Appendix A]{art:bruzzopedrinisalaszabo2013} to get the rank
\begin{align}
\rk\big(\Ebf_{\mu}^{\vec u_1, \Delta_1, \vec w_1; \vec u_2,
  \Delta_2, \vec w_2} \big) & = r_1\,\Delta_2+r_2 \, \Delta_1+\frac{r_2}{2r_1} \, \vec{h}_1\cdot C\vec{h}_1+\frac{r_1}{2r_2}\, \vec{h}_2\cdot C\vec{h}_2 -\vec{h}_1\cdot C\vec{h}_2 \\ & \qquad -\, \frac{1}{2}\,\sum_{j=1}^{k-1}\,\big(C^{-1}\big)^{jj}\,\vec{w}_1\cdot\vec{w}_2(j) \ ,
\end{align}
where $r_l=\sum_{j=0}^{k-1}\, (\vec w_l)_j$ and $\vec{h}_l:=C^{-1}\vec{u}_l$ for $l=1,2$.

Let $\Vbf_{\mu}^{\vec u, \Delta, \vec w}$ be the natural bundle on $\Mcal_{\vec{u},\Delta,\vec w}$ defined by the derived pushforward (cf.\ \cite[Section 4.4]{art:bruzzopedrinisalaszabo2013})
\begin{equation}
\Vbf_{\mu}^{\vec u, \Delta, \vec w} := R^1 p_{1\ast}\big({\boldsymbol\Ecal}_{\vec{u},\Delta,\vec w} \otimes
p_{2}^\ast(\Ocal_{\Xscr_k}(\mu)\otimes\Ocal_{\Xscr_k}(-\Dscr_\infty) )\big) \ .
\end{equation}
It is a vector bundle on $\Mcal_{\vec{u},\Delta,\vec w}$ whose fibre over $[(\Ecal,\phi_\Ecal)]$ is given by
\begin{equation}
\Vbf_{\mu}^{\vec u, \Delta, \vec w}\big|_{[(\Ecal,\phi_\Ecal)]}=H^1\big(\Xscr_k\,;\,\Ecal\otimes
\Ocal_{\Xscr_k}(\mu)\otimes\Ocal_{\Xscr_k}(-\Dscr_\infty) \big)\ .
\end{equation}
Its rank was computed in \cite[Proposition 4.18]{art:bruzzopedrinisalaszabo2013} to be
\begin{equation}
\rk\big( \Vbf_{\mu}^{\vec u, \Delta, \vec w}\, \big) = \Delta+\frac{1}{2r} \, \vec{h}\cdot C\vec{h} -\frac{1}{2}\,\sum_{j=1}^{k-1}\,\big(C^{-1}\big)^{jj}\, w_j \ .
\end{equation}
We denote by $\bar\Vbf{}_{\mu}^{\vec u, \Delta, \vec w}$ the analogous vector bundle on $\Mcal_{\vec{u},\Delta,\vec w}$ defined using the dual ${\boldsymbol\Ecal}_{\vec{u},\Delta,\vec w}^\vee$ of the universal sheaf.

Let us now fix topological data $(\vec u_\upsilon, \Delta_\upsilon,
\vec w_\upsilon)_{\upsilon\in\quiv_0}$ associated with the moduli
spaces $\Mcal_{\vec{u}_\upsilon,\Delta_\upsilon,\vec w_\upsilon }$ at
the vertices $\quiv_0$ with $\vec u_\upsilon\in\Z^{k-1}$ satisfying
the constraint \eqref{eq:firstchernclass}, $\Delta_\upsilon\in
\frac1{2\, r_\upsilon\,k}\, \Z$, and $\vec w_\upsilon\in(\Z_{\geq 0})^k$. The fundamental (resp. antifundamental) hypermultiplets of masses $\mu^s_{\upsilon}$, $s=1,\dots,m_\upsilon$ (resp. $\bar\mu^{\bar s}_{\upsilon}$, $\bar s=1,\dots,\bar m_\upsilon$) at the nodes $\upsilon\in\quiv_0$ correspond to the vector bundles $\Vbf_{\mu^s_{\upsilon}}^{\vec u_\upsilon, \Delta_\upsilon, \vec w_\upsilon}$ (resp.\ $\bar\Vbf{}_{\bar\mu^{\bar s}_{\upsilon}}^{\vec u_\upsilon, \Delta_\upsilon, \vec w_\upsilon}$) on $\Mcal_{\vec{u}_\upsilon,\Delta_\upsilon,\vec w_\upsilon }$. The bifundamental hypermultiplets of masses $\mu_e$ at the edges $e\in \quiv_1$ correspond to the vector bundles $\Ebf_{\mu_e}^{\vec u_{\source(e)}, \Delta_{\source(e)}, \vec w_{\source(e)}; \vec u_{\tail(e)}, \Delta_{\tail(e)}, \vec w_{\tail(e)}}$ on $\Mcal_{\vec{u}_{\source(e)},\Delta_{\source(e)},\vec w_{\source(e)} } \times \Mcal_{\vec{u}_{\tail(e)},\Delta_{\tail(e)},\vec w_{\tail(e)}\, }$; in particular, for a vertex loop edge $e$, i.e. $\source(e)=\tail(e)$, the restriction of the bundle $\Ebf_{\mu_e}^{\vec u_{\source(e)}, \Delta_{\source(e)}, \vec w_{\source(e)}; \vec u_{\source(e)}, \Delta_{\source(e)}, \vec w_{\source(e)}}$ to the diagonal of $\Mcal_{\vec{u}_{\source(e)},\Delta_{\source(e)},\vec w_{\source(e)}} \times \Mcal_{\vec{u}_{\source(e)},\Delta_{\source(e)},\vec w_{\source(e)}}$ describes an adjoint hypermultiplet of mass $\mu_e$. The total matter field content of the $\Ncal=2$ quiver gauge theory on $X_k$ associated to $\quiv$ in the sector labelled by $(\vec u_\upsilon, \Delta_\upsilon, \vec w_\upsilon)_{\upsilon\in\quiv_0}$ is then described by the bundle on $\prod_{\upsilon\in\quiv_0}\,\Mcal_{\vec{u}_\upsilon,\Delta_\upsilon,\vec w_\upsilon}$ given by
\begin{align}
\boldsymbol{M}^{\boldsymbol{\vec u}, \boldsymbol\Delta, \boldsymbol{\vec w}}_{\boldsymbol\mu}:= & \ \bigoplus_{\upsilon\in\quiv_0}\, p_\upsilon^* \Big(\, \bigoplus_{s=1}^{m_\upsilon}\,
\Vbf_{\mu^s_{\upsilon}}^{\vec u_\upsilon, \Delta_\upsilon, \vec w_\upsilon} \
\oplus \ \bigoplus_{\bar s=1}^{\bar m_\upsilon}\,
\bar\Vbf{}_{\bar\mu^{\bar s}_{\upsilon}}^{\vec u_\upsilon, \Delta_\upsilon, \vec w_\upsilon}\, \Big) \\ & \ \qquad \ \oplus \ \bigoplus_{e\in\quiv_1}\,
p_e^*\Ebf_{\mu_e}^{\vec u_{\source(e)}, \Delta_{\source(e)}, \vec w_{\source(e)}; \vec u_{\tail(e)}, \Delta_{\tail(e)}, \vec w_{\tail(e)}} \ ,
\end{align}
where $p_\upsilon$ is the projection of $\prod_{\upsilon\in\quiv_0}\,\Mcal_{\vec{u}_\upsilon,\Delta_\upsilon,\vec w_\upsilon }$ to the $\upsilon$-th factor while $p_e$ is the projection to the product $\Mcal_{\vec{u}_{\source(e)},\Delta_{\source(e)},\vec w_{\source(e)} } \times \Mcal_{\vec{u}_{\tail(e)},\Delta_{\tail(e)},\vec w_{\tail(e)} }$; for vertex loop edges $p_e$ is understood as projection to the diagonal of $\Mcal_{\vec{u}_{\source(e)},\Delta_{\source(e)},\vec w_{\source(e)} } \times \Mcal_{\vec{u}_{\source(e)},\Delta_{\source(e)},\vec w_{\source(e)}}$.

In this paper we consider only $\Ncal=2$ superconformal quiver gauge theories. On $\R^4$, this implies that $M_\quiv(\boldsymbol{r},\boldsymbol{m}, \boldsymbol{\bar m})$ is in the conformal class of quiver representations whose dimension vectors satisfy the constraint
\begin{equation} \label{eq:confconstr}
m_\upsilon+\bar m_\upsilon= 2 r_\upsilon-\sum_{\stackrel{\scriptstyle e\in\quiv_1}{\scriptstyle \source(e)=\upsilon}}\, r_{\tail(e)}-\sum_{\stackrel{\scriptstyle e\in\quiv_1}{\scriptstyle \tail(e)=\upsilon}}\, r_{\source(e)}
\end{equation}
at any vertex $\upsilon\in\quiv_0$ (see e.g. \cite[Chapter 3]{art:nekrasovpestun2012} and \cite[Section 2]{art:nekrasovpestunshatashvili2013}). In an analogous way to the pure $\Ncal=2$ gauge
theory on $X_k$, we shall wish to relate our quiver gauge theory partition
functions to the corresponding partition functions on $\R^4$
via suitable factorization formulas. Hence we shall impose the same general
conformal constraints \eqref{eq:confconstr}. When $\bar{\mbf m}=\mbf 0$, this restricts the
admissible quivers into three classes according to the ADE classification of \cite[Chapter 3]{art:nekrasovpestun2012} in the following way:
\begin{itemize}
\item When $\mbf m\neq \mbf0$, the underlying graph of $\quiv$ coincides with the Dynkin diagram of a finite-dimensional simply-laced Lie algebra of ADE type; these theories are called `Class I' in \cite{art:nekrasovpestun2012}.

\smallskip

\item When $\mbf m=\mbf 0$ and $\mu_e=0$ for all $e\in\quiv_1$, the underlying graph of $\quiv$ is a simply-laced affine Dynkin diagram of type $\widehat{A}_n$, $\widehat{D}_n$ or $\widehat{E}_n$; these theories are called `Class II' in \cite{art:nekrasovpestun2012}. In this case the dimension vectors $\mbf r$ are uniquely determined by a single integer $N$ as
\begin{equation}
r_\upsilon= N\, d_\upsilon \ ,
\end{equation}
where $d_\upsilon$ are the Dynkin indices of the affine roots at the vertices $\quiv_0$.

\smallskip

\item When $\mbf m=\mbf 0$ and $\mu_e\neq0$ for some $e\in\quiv_1$, the underlying graph of $\quiv$ is an affine Dynkin diagram of type $\widehat{A}_n$; these theories are called `Class II$^*$' in \cite{art:nekrasovpestun2012}.
\end{itemize}

To formulate the analogous conformal constraint on $X_k$, we compute the degree of the Euler class of the hypermultiplet bundle $\boldsymbol{M}^{\boldsymbol{\vec u}, \boldsymbol\Delta, \boldsymbol{\vec w}}_{\boldsymbol\mu}$ which is given by
\begin{multline}
{\rm deg}\, \eu\big(\boldsymbol{M}^{\boldsymbol{\vec u}, \boldsymbol\Delta, \boldsymbol{\vec w}}_{\boldsymbol\mu}\, \big):= \sum_{\upsilon\in \quiv_0}\, \dim \Mcal_{\vec{u}_\upsilon,\Delta_\upsilon,\vec w_\upsilon}- \rk\big(\boldsymbol{M}^{\boldsymbol{\vec u}, \boldsymbol\Delta, \boldsymbol{\vec w}}_{\boldsymbol\mu}\,\big)\\[4pt] \shoveleft{=\sum_{\upsilon\in \quiv_0}\,\Big( 2r_\upsilon\, \Delta_\upsilon-\frac{1}{2}\,\sum_{j=1}^{k-1}\,\big(C^{-1}\big)^{jj}\,\vec w_\upsilon\cdot \vec w_\upsilon(j) \Big) } \\
-\, \sum_{\upsilon\in \quiv_0}\, \big(m_\upsilon+\bar m_\upsilon\big)\, \Big(\Delta_\upsilon+\frac{1}{2r_\upsilon}\, \vec{h}_\upsilon\cdot C \vec{h}_\upsilon-\frac{1}{2}\,\sum_{j=1}^{k-1}\, \big(C^{-1}\big)^{jj}\, w_{\upsilon, j} \Big)-\sum_{e\in \quiv_1}\, \Big(\, r_{\source(e)}\, \Delta_{\tail(e)}+r_{\tail(e)}\, \Delta_{\source(e)}\\
+\, \frac{r_{\tail(e)}}{2r_{\source(e)}} \, \vec{h}_{\source(e)}\cdot C\vec{h}_{\source(e)}+\frac{r_{\source(e)}}{2r_{\tail(e)}}\, \vec{h}_{\tail(e)}\cdot C\vec{h}_{\tail(e)} -\vec{h}_{\source(e)}\cdot C\vec{h}_{\tail(e)}-\frac{1}{2}\,\sum_{j=1}^{k-1}\,\big(C^{-1}\big)^{jj}\,\vec{w}_{\source(e)}\cdot\vec{w}_{\tail(e)}(j)\, \Big)\ ,
\end{multline}
where $r_\upsilon=\sum_{j=0}^{k-1}\, (\vec w_{\upsilon})_{ j}$ and $\vec{h}_\upsilon:=C^{-1}\vec{u}_\upsilon$. Using \eqref{eq:confconstr} the degree becomes
\begin{equation}
{\rm deg}\, \eu\big(\boldsymbol{M}^{\boldsymbol{\vec u}, \boldsymbol\Delta, \boldsymbol{\vec w}}_{\boldsymbol\mu}\,\big) = 
\sum_{\upsilon\in \quiv_0}\,d^{X_k}_\upsilon(\vec{\boldsymbol h}_\upsilon,\vec{\boldsymbol w}_\upsilon) \ ,
\end{equation}
where we defined
\begin{multline}
d^{X_k}_\upsilon(\vec{\boldsymbol h}_\upsilon,\vec{\boldsymbol w}_\upsilon) :=\frac{1}{2}\,\sum_{j=1}^{k-1}\, \big(C^{-1}\big)^{jj}\, w_{\upsilon, j} \,\Big(2 r_\upsilon-\sum_{\stackrel{\scriptstyle e\in\quiv_1}{\scriptstyle \source(e)=\upsilon}}\, r_{\tail(e)}-\sum_{\stackrel{\scriptstyle e\in\quiv_1}{\scriptstyle \tail(e)=\upsilon}}\, r_{\source(e)}\Big)\\ \shoveright{
-\, \frac{1}{4}\,\sum_{j=1}^{k-1}\,\big(C^{-1}\big)^{jj}\, \Big( 2 \vec w_\upsilon\cdot \vec w_\upsilon(j)-\sum_{\stackrel{\scriptstyle e\in\quiv_1}{\scriptstyle \source(e)=\upsilon}} \,\vec{w}_{\upsilon}\cdot\vec{w}_{\tail(e)}(j)
- \sum_{\stackrel{\scriptstyle e\in\quiv_1}{\scriptstyle \tail(e)=\upsilon}} \,\vec{w}_{\source(e)}\cdot\vec{w}_{\upsilon}(j) \Big) } \\
- \vec{h}_\upsilon\cdot C \vec{h}_\upsilon+\frac{1}{2}\, \sum_{\stackrel{\scriptstyle e\in\quiv_1}{\scriptstyle \source(e)=\upsilon}}\, \vec{h}_\upsilon\cdot C \vec{h}_{\tail(e)}+\frac{1}{2}\, \sum_{\stackrel{\scriptstyle e\in\quiv_1}{\scriptstyle\tail(e)=\upsilon}}\, \vec{h}_\upsilon\cdot C \vec{h}_{\source(e)}
\end{multline}
for each vertex $\upsilon\in\quiv_0$; here $\vec{\boldsymbol h}_\upsilon:=\big(\vec h_\upsilon,(\vec h_{\tail(e)})_{e\in\quiv_1\,:\, \source(e)=\upsilon},(\vec h_{\source(e)})_{e\in\quiv_1\,:\, \tail(e)=\upsilon}\big)$ and similarly for $\vec{\boldsymbol w}_\upsilon$. By analogy with the case of gauge theories on $\R^4$, we say that the $\Ncal=2$ quiver gauge theory on $X_k$ is conformal if $d^{X_k}_\upsilon(\vec{\boldsymbol h}_\upsilon,\vec{\boldsymbol w}_\upsilon)=0$ for all $\upsilon\in\quiv_0$; this is formally the requirement of vanishing beta-function for the running of the $\upsilon$-th gauge coupling constant. For any vertex $\upsilon\in \quiv_0$, define the set of conformal fractional instanton charges by
\begin{equation}
\Ufrak_{\vec{\boldsymbol w}_\upsilon}^{\mathrm{conf}}:=\big\{\vec{\boldsymbol h}_\upsilon \ \big\vert\ d^{X_k}_\upsilon(\vec{\boldsymbol h}_\upsilon,\vec{\boldsymbol w}_\upsilon)=0 \big\}\ .
\end{equation}
These charge sets generalise the conformal restrictions derived in \cite[Section 5.4]{art:bruzzopedrinisalaszabo2013} for the $A_0$-theory with fundamental matter, whose quiver $\quiv=\circ$ consists of a single node with no arrows. 

\subsection{Instanton partition functions}

Introduce topological couplings $\qsf_\upsilon\in\C^\ast$ with $|\qsf_\upsilon|<1$ and $\vec\xi_v=\big((\xi_v)_1,\dots, (\xi_v)_{k-1}\big) \in (\C^\ast)^{k-1}$ with $|(\xi_v)_i|<1$ at each vertex $\upsilon\in\quiv_0$. Let $T:=T_t\times T_{\vec{\boldsymbol a}}$, where $T_{\vec{\boldsymbol a}}$ is the maximal torus of the gauge group $G_{\boldsymbol r}$ with $H^*_{T_{\vec{\boldsymbol a}}}({\rm
  pt}) =\C\big[(a^\upsilon_1,\dots,a^\upsilon_{r_\upsilon})_{\upsilon\in\quiv_0}\big]$. For a vertex $\upsilon\in\quiv_0$ we write $T_\upsilon:=T_t\times T_{\boldsymbol a^\upsilon}$, where $T_{\mbf a^\upsilon}$ is the maximal torus of the gauge group ${\rm GL}(r_\upsilon,\C)$ at the node with $H^*_{T_{\mbf{a}^\upsilon}}({\rm
  pt}) =\C[a^\upsilon_1,\dots,a^\upsilon_{r_\upsilon}]$, and for an arrow $e\in\quiv_1$ we write $T_e:=T_t\times T_{\mbf a^{\source(e)}}\times T_{\mbf a^{\tail(e)}}$; for a vertex loop edge we set $T_e:=T_t\times T_{\mbf a^{\source(e)}}$.
  Let $T_{\mubf}$ be the maximal torus of the flavour symmetry group $
G_{\boldsymbol m,\boldsymbol{\bar m}}$
with $H_{T_{\mubf}}^*({\rm
  pt})=\C\big[(\mu_e)_{e\in\quiv_1}\,,\,(\mu_\upsilon^1,\dots,\mu_\upsilon^{m_\upsilon})_{\upsilon\in\quiv_0}
\,,\, (\bar\mu_\upsilon^1,\dots, \bar\mu_\upsilon^{\bar m_\upsilon})_{\upsilon\in\quiv_0} \big]$.
  
The instanton partition function for the $\Ncal=2$ superconformal quiver gauge theory on $X_k$ is then defined by the generating function
\begin{multline}
\Zcal_{X_k}^{\quiv,{\rm inst}}\big(\varepsilon_1,\varepsilon_2,\vec{\boldsymbol a},\mubf;\qbf, \vec\xibf \ \big)_{\vec{\boldsymbol{w}}} :=\sum_{\stackrel{\scriptstyle \vec{\boldsymbol{h}}\,,\, \boldsymbol{\Delta}}{\scriptstyle C\vec h_\upsilon\in\Ufrak_{\vec w_\upsilon}\,,\, \vec{\boldsymbol h}_\upsilon\in\Ufrak_{\vec{\boldsymbol w}_\upsilon}^{\mathrm{conf}}}} \,\vec\xibf{}^{\ \vec{\boldsymbol{h}}}\  \qbf^{\boldsymbol{\Delta} + \frac{1}{2\boldsymbol{r}}\, \vec{\boldsymbol{h}}\cdot C\vec{\boldsymbol{h}}} \\ \shoveright{\times \  
  \int_{\prod\limits_{\upsilon\in\quiv_0}\,\Mcal_{C\vec{h}_\upsilon,\Delta_\upsilon,\vec w_\upsilon}} \, \eu_{T\times
  T_{\mubf}}\big(\boldsymbol{M}^{C\boldsymbol{\vec h},
  \boldsymbol{\Delta}, \boldsymbol{\vec w}}_{\boldsymbol\mu} \big) } \\[4pt] 
\shoveleft{ = \sum_{\stackrel{\scriptstyle \vec{\boldsymbol{h}}\,,\,
      \boldsymbol{\Delta}}{\scriptstyle C\vec
      h_\upsilon\in\Ufrak_{\vec w_\upsilon}\,,\, \vec{\boldsymbol
        h}_\upsilon\in\Ufrak_{\vec{\boldsymbol
          w}_\upsilon}^{\mathrm{conf}}}} \, \vec\xibf{}^{\
    \vec{\boldsymbol{h}}}\ \qbf^{\boldsymbol{\Delta} +
    \frac{1}{2\boldsymbol{r}}\, \vec{\boldsymbol{h}}\cdot
    C\vec{\boldsymbol{h}}} } 
\\ 
\times \ \int_{\prod\limits_{\upsilon\in\quiv_0}\,\Mcal_{C\vec{h}_\upsilon,\Delta_\upsilon,\vec w_\upsilon}} \ 
  \prod\limits_{\upsilon\in\quiv_0}\, p_\upsilon^\ast \Big(\, \prod\limits_{s=1}^{m_\upsilon}\, \eu_{T_\upsilon \times T_{\mu^s_{\upsilon}}}\big(\Vbf_{\mu^s_{\upsilon}}^{C\vec h_\upsilon, \Delta_\upsilon, \vec w_\upsilon}\big) \ \prod\limits_{\bar s=1}^{\bar m_\upsilon}\, \eu_{T_\upsilon \times T_{\bar\mu^{\bar s}_{\upsilon}}}\big(\bar\Vbf{}_{\bar\mu^{\bar s}_{\upsilon}}^{C\vec h_\upsilon, \Delta_\upsilon, \vec w_\upsilon}\big)\, \Big) \\ 
\times \ \prod\limits_{e\in\quiv_1}\, p_e^\ast\eu_{T_e \times T_{\mu_e}}\big(\Ebf_{\mu_e}^{C\vec h_{\source(e)}, \Delta_{\source(e)}, \vec w_{\source(e)}; C\vec h_{\tail(e)}, \Delta_{\tail(e)}, \vec w_{\tail(e)}}\big)\ ,
\end{multline}
where $\qbf^{\boldsymbol{\Delta} + \frac{1}{2\boldsymbol{r}}\,
  \vec{\boldsymbol{h}}\cdot
  C\vec{\boldsymbol{h}}}:=\prod_{\upsilon\in\quiv_0}\,
\qsf_\upsilon^{\Delta_\upsilon+\frac{1}{2r_\upsilon}\, \vec
  h_\upsilon\cdot C\vec h_\upsilon}$ and $\vec\xibf{}^{\
  \vec{\boldsymbol{h}}} := \prod_{\upsilon\in\quiv_0}\,
\vec\xi_\upsilon^{\ \vec h_\upsilon}$. The equivariant Euler class of
the Ext-bundle at a fixed point with combinatorial data
$\big((\boldsymbol{\vec Y},\boldsymbol{\vec u}\, )\,,\,
(\boldsymbol{\vec Y}{}',\boldsymbol{\vec u}\,'\, )\big)$ is computed in \cite[Section 4.7]{art:bruzzopedrinisalaszabo2013} with the result
\begin{multline}
{\rm eu}_{{T_t\times T_{\vec a}\times T_{\vec a\,'}\times T_\mu}}\big(\boldsymbol E_\mu^{\vec u, \Delta, \vec w; \vec u^{\,\prime}, \Delta', \vec w^{\, \prime}} \big|_{((\boldsymbol{\vec Y},\boldsymbol{\vec u}\, )\,,\, (\boldsymbol{\vec Y}{}',\boldsymbol{\vec u}\,'\, ))}\big) \\ =
\prod\limits_{\alpha=1}^r \ \prod\limits_{\alpha'=1}^{r'} \ \prod\limits_{i=1}^k \, m_{Y_{\alpha}^{i},
  {Y_{\alpha'}^{\prime\, i}}} \big(\varepsilon_1^{(i)},\varepsilon_2^{(i)},
a^{(i)}_{\alpha'\alpha} +\mu \big) \ \prod\limits_{n=1}^{k-1}\,
\ell^{(n)}_{\vec{h}_{\alpha'\alpha}}
\big(\varepsilon_1^{(n)},\varepsilon_2^{(n)}, a^{(n)}_{\alpha'\alpha} +\mu\big) \ ,
\end{multline}
where  
\begin{equation}
a_{\alpha'\alpha}^{(i)}:=a_{\alpha'\alpha}+(\vec{h}_{\alpha'\alpha})_{i}\,
\varepsilon_1^{(i)}+(\vec{h}_{\alpha'\alpha})_{i-1}\, \varepsilon_2^{(i)}
\end{equation}
for $i=1, \ldots, k$ and for $\alpha=1, \ldots, r$, $\alpha'=1, \ldots, r'$, and $a_{\alpha'\alpha}:=a_{\alpha'}'-a_\alpha$, $\vec{h}_{{\alpha'}\alpha}:=\vec{h}_{\alpha'}^{\, \prime}-\vec{h}_\alpha$ (we 
set $(\vec{h}_{{\alpha'}\alpha})_0=(\vec{h}_{{\alpha'}\alpha})_k=0$). The
equivariant Euler class of the bundle $\Vbf_{\mu}^{\vec u, \Delta,
  \vec w}$ (resp. $\bar\Vbf{}_{\mu}^{\vec u, \Delta, \vec w}$) is
obtained from this formula by setting $(\boldsymbol{\vec
  Y},\boldsymbol{\vec u}\, )$ (resp. $(\boldsymbol{\vec Y}{}',\boldsymbol{\vec u}\,'\,)$) to $(\emptyset,\boldsymbol{\vec 0}\,)$. By applying 
the localization theorem we   obtain
\begin{multline}
\Zcal_{X_k}^{\quiv, {\rm inst}}\big(\varepsilon_1,\varepsilon_2,\vec{\boldsymbol a},\mubf;\qbf, \vec\xibf \ \big)_{\vec{\boldsymbol{w}}}= \sum_{\stackrel{\scriptstyle \vec{{\boldsymbol{h}}}\,,\, \boldsymbol{\vec Y}}{\scriptstyle C\vec h_\upsilon\in\Ufrak_{\vec w_\upsilon}\,,\, \vec{\boldsymbol h}_\upsilon\in\Ufrak_{\vec{\boldsymbol w}_\upsilon}^{\mathrm{conf}}}} \, \vec\xibf{}^{\ \vec{\boldsymbol{h}}}\  \qbf^{\boldsymbol{\Delta} + \frac{1}{2\boldsymbol{r}}\, \vec{\boldsymbol{h}}\cdot C\vec{\boldsymbol{h}}}\\ \times \ \sum_{\stackrel{\scriptstyle (\vec{h}_{\upsilon,1}, \ldots, \vec{h}_{\upsilon, r_\upsilon})_{\upsilon\in\quiv_0}}{\scriptstyle \vec{h}_\upsilon=\sum_{\alpha=1}^{r_\upsilon}\,\vec{h}_{\upsilon,\alpha}}}\,  
m_{\boldsymbol{\vec Y}, (\vec{h}_{\upsilon,\alpha})}(\varepsilon_1, \varepsilon_2, \vec{\boldsymbol a}, \mubf)\, \ell_{(\vec{h}_{\upsilon,\alpha})}(\varepsilon_1, \varepsilon_2, \vec{\boldsymbol a}, \mubf) \ ,
\end{multline}
where
\begin{multline}
m_{\boldsymbol{\vec Y}, (\vec{h}_{\upsilon,\alpha})}(\varepsilon_1, \varepsilon_2, \vec{\boldsymbol a}, \mubf) := \prod\limits_{i=1}^k \, \frac{\prod\limits_{e\in\quiv_1}\ \prod\limits_{\alpha=1}^{r_{\source(e)}} \ \prod\limits_{\beta=1}^{r_{\tail(e)}} \  m_{Y_{\source(e), \alpha}^{i},
  {Y_{\tail(e), \beta}^{i}}} \big(\varepsilon_1^{(i)},\varepsilon_2^{(i)},
a^{(i)}_{e;\beta\alpha}+\mu_e \big)}{\prod\limits_{\upsilon\in\quiv_0}\ \prod\limits_{\alpha,\beta=1}^{r_\upsilon} \, m_{Y_{\upsilon,\alpha}^{i},
  {Y_{\upsilon, \beta}^{i}}} \big(\varepsilon_1^{(i)},\varepsilon_2^{(i)},
a^{(i)}_{\upsilon;\beta\alpha} \big)} \\
\times\ \prod\limits_{\upsilon\in\quiv_0}\ \prod_{\alpha=1}^{r_\upsilon} \ \prod_{i=1}^k \ \prod\limits_{s=1}^{m_\upsilon} \,
m_{Y_{\upsilon, \alpha}^{i}}\big(\varepsilon_1^{(i)},\varepsilon_2^{(i)},
a_{\upsilon;\alpha}^{(i)}+\mu^s_{\upsilon} \big) \ \prod\limits_{\bar s=1}^{\bar m_\upsilon} \,
m_{Y_{\upsilon,\alpha}^{i}}\big(\varepsilon_1^{(i)},\varepsilon_2^{(i)},
a_{\upsilon;\alpha}^{(i)}+\bar\mu^{\bar s}_{\upsilon} \big) 
\end{multline}
and
\begin{multline}
\ell_{(\vec{h}_{\upsilon,\alpha})}(\varepsilon_1, \varepsilon_2, \vec{\boldsymbol a}, \mubf)
:= \prod\limits_{n=1}^{k-1}\, \frac{\prod\limits_{e\in\quiv_1}\ \prod\limits_{\alpha=1}^{r_{\source(e)}} \ \prod\limits_{\beta=1}^{r_{\tail(e)}}\,
\ell^{(n)}_{\vec{h}_{e;\beta\alpha}}
\big(\varepsilon_1^{(n)},\varepsilon_2^{(n)}, a^e_{\beta\alpha}+\mu_e \big)}{\prod\limits_{\upsilon\in\quiv_0}\ \prod\limits_{\alpha,\beta=1}^{r_\upsilon}\,
\ell^{(n)}_{\vec{h}_{\upsilon;\beta\alpha}}
\big(\varepsilon_1^{(n)},\varepsilon_2^{(n)}, a^\upsilon_{\beta\alpha} \big)}\\
\times\ \prod\limits_{\upsilon\in\quiv_0}\ \prod_{\alpha=1}^{r_\upsilon} \ \prod_{n=1}^{k-1} \ \prod\limits_{s=1}^{m_\upsilon} \,
\ell^{(n)}_{\vec{h}_{\upsilon,\alpha}}\big(\varepsilon_1^{(n)},\varepsilon_2^{(n)},
a^\upsilon_\alpha+\mu^s_{\upsilon} \big)\ \prod\limits_{\bar s=1}^{\bar m_\upsilon} \,
\ell^{(n)}_{\vec{h}_{\upsilon,\alpha}}\big(\varepsilon_1^{(n)},\varepsilon_2^{(n)},
a^\upsilon_\alpha+\bar\mu^{\bar s}_{\upsilon} \big) \ .
\end{multline}
Here
\begin{equation}
m_{Y}(e_1,e_2,a):= \prod_{(s_1,s_2)\in Y}\, \big(a-(s_1-1)\, e_1-(s_2-1)\, e_2\big)
\end{equation}
for a Young tableau $Y\subset \Z_{>0}\times\Z_{>0}$, while
\begin{align}
a_{\upsilon;\alpha}^{(i)} =a^\upsilon_\alpha+(\vec{h}_{\upsilon,\alpha})_i\,\varepsilon_1^{(i)}+(\vec{h}_{\upsilon,\alpha})_{i-1}\,\varepsilon_1^{(i)} \qquad \mbox{and} \qquad 
a^{(i)}_{e;\beta\alpha} = a^e_{\beta\alpha}+(\vec{h}_{e;\beta\alpha})_i\,\varepsilon_1^{(i)}+(\vec{h}_{e;\beta\alpha})_{i-1}\,\varepsilon_1^{(i)}\ ,
\end{align}
with $\vec{h}_{\upsilon, \alpha}:=C^{-1}\vec{u}_{\upsilon, \alpha}$, $\vec{h}_{e; \beta\alpha}:=C^{-1}(\vec{u}_{\tail(e),\beta}-\vec{u}_{\source(e),\alpha})$ and $a_{\beta\alpha}^e=a_\beta^{\tail(e)}-a_\alpha^{\source(e)}$; for vertex loop edges with $\tail(e)=\source(e)=\upsilon$ we also set $a^{(i)}_{\upsilon;\beta\alpha}:=a^{(i)}_{e;\beta\alpha}$, $\vec{h}_{\upsilon; \beta\alpha}:= \vec{h}_{e; \beta\alpha}$ and $a^\upsilon_{\beta\alpha}:=a_{\beta\alpha}^e$.

It follows that the instanton partition function $\Zcal_{X_k}^{\quiv, {\rm inst}}$ factorises in terms of the corresponding partition functions $\Zcal_{\C^2}^{\quiv, {\rm inst}}$ for the $\Ncal=2$ superconformal quiver gauge theory on $\R^4$ (cf.\ \cite[Section 2]{art:nekrasovpestunshatashvili2013}) as
\begin{multline}
\Zcal_{X_k}^{\quiv, {\rm inst}}\big(\varepsilon_1,\varepsilon_2,\vec{\boldsymbol a},\mubf;\qbf, \vec\xibf \ \big)_{\vec{\boldsymbol{w}}}
= \sum_{\stackrel{\scriptstyle \vec{{\boldsymbol{h}}}}{\scriptstyle C\vec h_\upsilon\in\Ufrak_{\vec w_\upsilon}\,,\, \vec{\boldsymbol h}_\upsilon\in\Ufrak_{\vec{\boldsymbol w}_\upsilon}^{\mathrm{conf}}}} \, \vec\xibf{}^{\ \vec{\boldsymbol{h}}}\ \sum_{\stackrel{\scriptstyle (\vec{h}_{\upsilon,1}, \ldots, \vec{h}_{\upsilon, r_\upsilon})_{\upsilon\in\quiv_0}}{\scriptstyle \vec{h}_\upsilon=\sum_{\alpha=1}^{r_\upsilon}\,\vec{h}_{\upsilon,\alpha}}} \ \prod_{\upsilon\in\quiv_0}\, \qsf_\upsilon^{\frac{1}{2}\,\sum\limits_{\alpha=1}^{r_\upsilon}\, \vec{h}_{\upsilon,\alpha}\cdot C\vec{h}_{\upsilon,\alpha}}\\
\times\ \ell_{(\vec{h}_{\upsilon,\alpha})}(\varepsilon_1, \varepsilon_2, \vec{\boldsymbol a}, \mubf)\ \prod_{i=1}^k\, \Zcal_{\C^2}^{\quiv, {\rm inst}}\big(\varepsilon_1^{(i)},\varepsilon_2^{(i)},\vec{\boldsymbol{a}}^{(i)},\mubf;\qbf \big)\ .
\end{multline}

\subsection{Perturbative partition functions}

The perturbative partition function for the pure $\Ncal=2$ gauge theory on $X_k$, which is independent of the topological couplings $\qsf$ and $\vec\xi$, is also defined in \cite[Section 6]{art:bruzzopedrinisalaszabo2013}. Here we define a version of it for the $\Ncal=2$ superconformal quiver gauge theories on $X_k$.

The perturbative partition function is determined by defining the perturbative part of the equivariant Chern character of the vector bundle $\boldsymbol{M}^{\boldsymbol{\vec u}, \boldsymbol\Delta, \boldsymbol{\vec w}}_{\boldsymbol\mu}$ at a fixed point
\begin{equation}
\big([(\Ecal_\upsilon,\phi_{\Ecal_\upsilon})]\big)_{\upsilon\in\quiv_0}=\bigoplus_{\alpha=1}^r \,\big([( {\imath_k}_\ast(I_{\upsilon,\alpha})\otimes\Rcal^{\vec{u}_{\upsilon,\alpha}}, \phi_{\upsilon,\alpha})]\big)_{\upsilon\in\quiv_0}
\end{equation}
of $\prod_{\upsilon\in\quiv_0}\,\Mcal_{\vec{u}_\upsilon,\Delta_\upsilon,\vec w_\upsilon}$ as
\begin{multline}\label{eq:perturbative}
\ch_{T\times T_{\boldsymbol\mu}}^{\mathrm{pert}} \boldsymbol{M}^{\boldsymbol{\vec u}, \boldsymbol\Delta, \boldsymbol{\vec w}}_{\boldsymbol\mu} \big\vert_{([(\Ecal_\upsilon,\phi_{\Ecal_\upsilon})])_{\upsilon\in\quiv_0} }
:= \sum_{\upsilon\in \quiv_0}\, \Big(\, \sum_{s=1}^{m_\upsilon}\, \e^{\mu^s_\upsilon}+\sum_{\bar s=1}^{\bar m_\upsilon}\, \e^{\bar \mu^{\bar s}_\upsilon}\, \Big)  \\ \shoveright{ \times \ \sum_{\alpha=1}^{r_\upsilon}\, \e^{a^\upsilon_\alpha}\, \Big( \chi_{T_t}\big(\bar{X}_k\,,\, {\pi_k}_\ast(\Rcal^{C\vec h_{\upsilon,\alpha}}\otimes \Ocal_{\Xscr_k}(-\Dscr_\infty))\big)-\chi_{T_t}\big(X_k \,,\,
R^{C \vec{h}_{\upsilon,\alpha}} \big)\Big) } \\
+\, \sum_{e\in\quiv_1}\ \sum_{\alpha=1}^{r_{\source(e)}}\ \sum_{\beta=1}^{r_{\tail(e)}}\, \e^{a^e_{\beta\alpha}+\mu_e} \,\Big( \chi_{T_t}\big(\bar{X}_k\,,\, {\pi_k}_\ast(\Rcal^{C\vec h_{e;\beta\alpha}}\otimes \Ocal_{\Xscr_k}(-\Dscr_\infty))\big)-\chi_{T_t}\big(X_k \,,\,R^{C\vec h_{e;\beta\alpha}} \big)\Big)
\end{multline}
where $\pi_k:\Xscr_k\to\bar X_k$ is the coarse moduli space morphism and $R^{\vec u}:=\Rcal^{\vec u}\, \big|_{X_k}$.
One way to compute it is to cancel the common factors between the two equivariant Euler characteristics, as was done in \cite[Section 6]{art:bruzzopedrinisalaszabo2013}. In this way $\ch_{T\times T_{\boldsymbol\mu}}^{\mathrm{pert}} \boldsymbol{M}^{\boldsymbol{\vec u}, \boldsymbol\Delta, \boldsymbol{\vec w}}_{\boldsymbol\mu} \big\vert_{([(\Ecal_\upsilon,\phi_{\Ecal_\upsilon})])_{\upsilon\in\quiv_0} }$ reduces to a sum depending on the equivariant Euler characteristics, over the two open affine toric neighbourhoods of the two fixed points $0,\infty$ of the compactification divisor $D_\infty$, of a Weil divisor on $\bar{X}_k$ which is a linear combination of $D_\infty$, $D_0$ and $D_k$ with integer coefficients. The result
depends only on the holonomies at infinity, i.e. on the framing vectors $\vec{w}_\upsilon$ for $\upsilon\in \quiv_0$.
    
In this paper we explicitly compute $\chi_{T_t}\big(X_k \,,\, R^{\vec{u}} \, \big)$ for any $\vec{u}\in \Z^{k-1}$ in \eqref{eq:perturbative} and arrive at an equivalent yet more transparent formulation of the perturbative partition function. By using some of the arguments from the proof of \cite[Theorem~4.1]{art:brionvergne1997}, we have
\begin{equation}
\chi_{T_t}\big(X_k \,,\, R^{\vec{u}} \, \big)=\sum_{i=1}^k\, \ch_{T_t}^{i}\big(R^{\vec{u}}\, \big)\, \ \big(1-\e^{\varepsilon_1^{(i)}}\big)^{-1}\,\big(1-\e^{\varepsilon_2^{(i)}}\big)^{-1} \ ,
\end{equation}
where $\ch_{T_t}^{i}\big(R^{\vec{u}} \, \big)$ is the local Chern character of $R^{\vec{u}}$ over the open affine toric neighbourhood of the fixed point $p_i$ for $i=1, \ldots, k$ (cf.\ \cite[Section 2.1]{art:brionvergne1997}).
Since $X_k$ is smooth, by \cite[Section 2.3]{art:brionvergne1997} for any Weil divisor $D=\sum_{j=0}^{k}\,d_j\,D_j$ on $X_k$ one has
\begin{equation}
\ch_{T_t}^{i}\big(\Ocal_{X_k}(D) \big)=\e^{-d_i\,\varepsilon_1^{(i)}-d_{i-1}\,\varepsilon_2^{(i)}}\ .
\end{equation}
It follows that in our case for $i=1, \ldots, k$ we have
\begin{equation}\label{eq:charactertautological}
\ch_{T_t}^{i}\big(R^{\vec{u}} \, \big)=\e^{{h}_i\,\varepsilon_1^{(i)}+{h}_{i-1}\,\varepsilon_2^{(i)}}\ ,
\end{equation}
where $\vec{h}:=C^{-1} \vec{u}$.

To write down the perturbative partition function, we follow the same arguments as in \cite[Section 6]{art:gasparimliu2010}: the perturbative partition function is given by a sum, over the torus-fixed points, of the Euler classes obtained formally from the perturbative characters. Since the Euler class associated with a term $\chi_{T_t}\big(X_k \,,\, R^{\vec{u}}\, \big)$ involves infinite products, we regularise it by using  Barnes' double gamma function $\Gamma_2$. Then we define the {perturbative partition function} for the $\Ncal=2$ superconformal quiver gauge theory on $X_k$ as
\begin{equation}
\Zcal_{X_k}^{\quiv, \mathrm{pert}}(\varepsilon_1, \varepsilon_2, \vec{\boldsymbol a}, \mubf)_{\vec{\boldsymbol{w}}}\\
:= \sum_{\stackrel{\scriptstyle \vec{\boldsymbol{h}}}{\scriptstyle C\vec h_\upsilon\in\Ufrak_{\vec w_\upsilon}\,,\, \vec{\boldsymbol h}_\upsilon\in\Ufrak_{\vec{\boldsymbol w}_\upsilon}^{\mathrm{conf}}}} \ \sum_{\stackrel{\scriptstyle (\vec{h}_{\upsilon,1}, \ldots, \vec{h}_{\upsilon, r_\upsilon})_{\upsilon\in\quiv_0}}{\scriptstyle \vec{h}_\upsilon=\sum_{\alpha=1}^{r_\upsilon}\,\vec{h}_{\upsilon,\alpha}}} \, \Zcal_{X_k}^{\quiv, \mathrm{pert}}(\varepsilon_1, \varepsilon_2, \vec{\boldsymbol{a}}, \mubf)_{\vec{\boldsymbol{w}}, (\vec{h}_{\upsilon,\alpha})}\ ,
\end{equation}
where
\begin{multline}
\Zcal_{X_k}^{\quiv, \mathrm{pert}}(\varepsilon_1, \varepsilon_2,
\vec{\boldsymbol{a}},\mubf)_{\vec{\boldsymbol{w}},
  (\vec{h}_{\upsilon,\alpha})}:=\ell_{(\vec{h}_{\upsilon,\alpha})}(\varepsilon_1,
\varepsilon_2, \vec{\boldsymbol a}, \mubf)^{-1} \\ \shoveright{ \times
  \ 
\prod_{i=1}^k\, \frac{\prod\limits_{e\in\quiv_1} \ \prod\limits_{\alpha=1}^{r_{\source(e)}}\ \prod\limits_{\beta=1}^{r_{\tail(e)}}\, \Gamma_2\big(a_{e;\beta\alpha}^{(i)}+\mu_e\,\big|\, \varepsilon_1^{(i)}, \varepsilon_2^{(i)}\big)}{\prod\limits_{\upsilon\in\quiv_0}\ \prod\limits_{\alpha\neq \beta}\, \Gamma_2\big(a_{\upsilon;\beta\alpha}^{(i)}\,\big|\,\varepsilon_1^{(i)}, \varepsilon_2^{(i)}\big)} } \\ \times \ \prod_{\upsilon\in\quiv_0}\ \prod_{\alpha=1}^{r_\upsilon}\ \prod_{i=1}^k\ \prod_{s=1}^{m_\upsilon} \, \Gamma_2\big(a_{\upsilon,\alpha}^{(i)}+\mu_\upsilon^s\,\big|\, \varepsilon_1^{(i)}, \varepsilon_2^{(i)} \big) \ \prod_{\bar s=1}^{\bar m_\upsilon}\,  \Gamma_2\big(a_{\upsilon,\alpha}^{(i)}+\bar \mu_\upsilon^{\bar s}\,\big|\, {\varepsilon_1^{(i)}, \varepsilon_2^{(i)}} \big) \ ,
\end{multline}
and we observe once again the ensuing factorization formula
\begin{equation}
\Zcal_{X_k}^{\quiv, \mathrm{pert}}(\varepsilon_1, \varepsilon_2, \vec{\boldsymbol{a}},\mubf)_{\vec{\boldsymbol{w}}, (\vec{h}_{\upsilon,\alpha})}= \ell_{(\vec{h}_{\upsilon,\alpha})}(\varepsilon_1, \varepsilon_2, \vec{\boldsymbol a}, \mubf)^{-1}\ \prod_{i=1}^k\, \Zcal_{\C^2}^{\quiv, \mathrm{pert}}\big(\varepsilon_1^{(i)}, \varepsilon_2^{(i)}, \vec{\boldsymbol{a}}^{(i)}, \mubf \big)\ .
\end{equation}

\subsection{Proof of the master formula}

Let us define the generating function for correlators of $p$-observables ($p=0,2$) in the $\Ncal=2$ superconformal quiver gauge theory on $X_k$ in terms of integrals of equivariant cohomology classes over the moduli spaces as
\begin{multline}
\Zcal_{X_k}^\quiv\big(\varepsilon_1, \varepsilon_2, \vec{\boldsymbol a},\mubf; \qbf,
\vec{\boldsymbol\xi}, \vec{\boldsymbol\tau}, \vec{\boldsymbol t}{}^{\;(1)}, \ldots, \vec{\boldsymbol t}{}^{\:(k-1)}
\big)_{\vec{\boldsymbol w}} \\ \shoveleft{
:= \sum_{\stackrel{\scriptstyle \vec{\boldsymbol{h}}\,,\, \boldsymbol{\Delta}}{\scriptstyle C\vec h_\upsilon\in\Ufrak_{\vec w_\upsilon}\,,\, \vec{\boldsymbol h}_\upsilon\in\Ufrak_{\vec{\boldsymbol w}_\upsilon}^{\mathrm{conf}}}} \,\vec\xibf{}^{\ \vec{\boldsymbol{h}}}\  \qbf^{\boldsymbol{\Delta} + \frac{1}{2\boldsymbol{r}}\, \vec{\boldsymbol{h}}\cdot C\vec{\boldsymbol{h}}} \  
  \int_{\prod\limits_{\upsilon\in\quiv_0}\,\Mcal_{C\vec{h}_\upsilon,\Delta_\upsilon,\vec w_\upsilon}} \, \eu_{T\times
  T_{\mubf}}\big(\boldsymbol{M}^{C\boldsymbol{\vec h}, \boldsymbol\Delta, \boldsymbol{\vec w}}_{\boldsymbol\mu} \big) } \\ \times \ \prod_{\upsilon\in\quiv_0}\, p_\upsilon^* \exp \bigg(\, \sum_{s=0}^\infty\, \Big(\,
  \sum_{i=1}^{k-1}\, t_s^{\upsilon,(i)} \,
  \big[\ch_{T}(\boldsymbol{\Ecal}_{C\vec{h}_\upsilon,\Delta_\upsilon,\vec w_\upsilon})/[\Dscr_i]\big]_s+\tau^\upsilon_s \,
  \big[\ch_T(\boldsymbol{\Ecal}_{C\vec{h}_\upsilon,\Delta_\upsilon,\vec w_\upsilon})/[X_k]\big]_{s-1}\, \Big)\, \bigg) \ ,
\end{multline}
where $\tau_0^\upsilon=\frac1{2\pi\ii}\, \log \qsf_\upsilon$ is the $\upsilon$-th bare gauge coupling. Here $\ch_{T}(\boldsymbol{\Ecal}_{C\vec{h},\Delta,\vec w})/[\Dscr_i]$ is the {slant product} between the equivariant Chern character of the universal sheaf $\boldsymbol{\Ecal}_{C\vec{h},\Delta,\vec w}$ of $\Mcal_{C\vec{h},\Delta,\vec w}$ and the divisor class $[\Dscr_i]$, and the class $\ch_T(\boldsymbol{\Ecal}_{C\vec{h},\Delta,\vec w})/[X_k]$ is defined by localization as
\begin{equation}
\ch_T(\boldsymbol{\Ecal}_{C\vec{h},\Delta,\vec w})/[X_k]:=\sum_{i=1}^k \,
\frac{1}{\mathrm{eu}_{T_t}(T_{p_i} X_k)}\, \imath_{
  \Mcal_{C\vec{h},\Delta,\vec w}\times\{p_i\}}^\ast \ch_T(\boldsymbol{\Ecal}_{C\vec{h},\Delta,\vec w})
\end{equation}
with $\imath_{
 \Mcal_{C\vec{h},\Delta,\vec
      w}\times\{p_i\}}$ the inclusion map of $
\Mcal_{C\vec{h},\Delta,\vec
      w}\times\{p_i\}$ in $
\Mcal_{C\vec{h},\Delta,\vec
      w}\times X_k$. The brackets $[-]_s$ indicate the degree $s$ part. For further discussion about these kinds of partition functions and their physical origins, see \cite[Section 1.2]{art:bruzzopedrinisalaszabo2013}. By applying the localization theorem, it is straightforward to generalise the computations leading to \cite[Proposition 5.9]{art:bruzzopedrinisalaszabo2013} for the pure $\Ncal=2$ gauge theory on $X_k$ to the present case and obtain
\begin{multline}
\Zcal_{X_k}^\quiv\big(\varepsilon_1, \varepsilon_2, \vec{\boldsymbol a},\mubf; \qbf,
\vec{\boldsymbol\xi}, \vec{\boldsymbol\tau}, \vec{\boldsymbol t}{}^{\;(1)}, \ldots, \vec{\boldsymbol t}{}^{\:(k-1)}
\big)_{\vec{\boldsymbol w}} = \sum_{\stackrel{\scriptstyle \vec{{\boldsymbol{h}}} \,,\, \boldsymbol{\vec Y}}{\scriptstyle C\vec h_\upsilon\in\Ufrak_{\vec w_\upsilon}\,,\, \vec{\boldsymbol h}_\upsilon\in\Ufrak_{\vec{\boldsymbol w}_\upsilon}^{\mathrm{conf}}}} \, \vec\xibf{}^{\ \vec{\boldsymbol{h}}}\  \qbf^{\boldsymbol{\Delta} + \frac{1}{2\boldsymbol{r}}\, \vec{\boldsymbol{h}}\cdot C\vec{\boldsymbol{h}}}\\ \shoveright{ \times \ \sum_{\stackrel{\scriptstyle (\vec{h}_{\upsilon,1}, \ldots, \vec{h}_{\upsilon, r_\upsilon})_{\upsilon\in\quiv_0}}{\scriptstyle \vec{h}_\upsilon=\sum_{\alpha=1}^{r_\upsilon}\,\vec{h}_{\upsilon,\alpha}}}\,  
m_{\boldsymbol{\vec Y}, (\vec{h}_{\upsilon,\alpha})}(\varepsilon_1, \varepsilon_2, \vec{\boldsymbol a}, \mubf)\, \ell_{(\vec{h}_{\upsilon,\alpha})}(\varepsilon_1, \varepsilon_2, \vec{\boldsymbol a}, \mubf) } \\ 
\shoveright{ \times \ \prod_{\upsilon\in\quiv_0} \ \prod_{i=1}^k\, \exp\bigg(\,
\sum_{s=0}^\infty\, \Big(\big(t_s^{\upsilon,(i)}\,
\varepsilon_1^{(i)}+t_s^{\upsilon, (i-1)}\, \varepsilon_2^{(i)}+\tau^\upsilon_s\big)\,
\Big[\ch_{\boldsymbol{Y}^i_\upsilon}\big(\varepsilon_1^{(i)},\varepsilon_2^{(i)},
\mbf{a}_\upsilon^{(i)} \big)\Big]_{s-1}\, \Big)\bigg) } \\
\shoveleft{ \times\ \prod_{\upsilon\in\quiv_0}\, \exp\Bigg(\, \sum_{s=0}^\infty\, \bigg(\,
  \sum_{l=1}^k\, \Big(\, \sum_{i=1}^{l-2}\,
  t_s^{\upsilon,(i)}+\sum_{i=l+1}^{k-1}\, t_s^{\upsilon,(i)}\Big)\, \Big[\ch_{\mbf{Y}_\upsilon^l}\big(\varepsilon_1^{(l)},\varepsilon_2^{(l)}, \mbf{a}_\upsilon^{(l)} \big)\Big]_{s}}\\
+\, \sum_{i=1}^{k-1}\, t_s^{\upsilon,(i)} \ \sum_{l=2}^{k-1}\,
\Big(\, \Big[\big(\varepsilon_1^{(l)}\big)^{\delta_{l,i}}\,
\ch_{\mbf{Y}_\upsilon^l}\big(\varepsilon_1^{(l)},\varepsilon_2^{(l)},
\mbf{a}_\upsilon^{(l)}-\big(\vec{h}_\upsilon \big)_{l-1}\, \varepsilon_2^{(l)}\big)\Big]_s \\
+\, \Big[\big(\varepsilon_2^{(l)}\big)^{\delta_{l-1,i}}\,
\ch_{\mbf{Y}_\upsilon^l}\big(\varepsilon_1^{(l)},\varepsilon_2^{(l)},
\mbf{a}_\upsilon^{(l)}-\big(\vec{h}_\upsilon \big)_{l}\, \varepsilon_1^{(l)}\big)
\Big]_{s}\, \Big)\bigg) \Bigg)
\end{multline}
where we set $t_s^{\upsilon,(0)}=t_s^{\upsilon, (k)}=0$ for any $s\geq0$ and $\upsilon\in\quiv_0$, and we introduced the notation
\begin{multline}
\ch_{\mbf{Y}_\upsilon^i}\big(\varepsilon_1^{(i)}, \varepsilon_2^{(i)},
\mbf{a}_\upsilon^{(i)}\big):= \sum_{\alpha=1}^{r_\upsilon} \, \frac{{\, \rm
    e}\,^{a^{(i)}_{\upsilon;\alpha}}}{\varepsilon_1^{(i)}\, \varepsilon_2^{(i)}}
\\ \times \ \Big(1-\big(1-\e^{-\varepsilon_1^{(i)}}\big)\,
\big(1-\e^{-\varepsilon_2^{(i)}}\big)\, \sum_{(s_1,s_2) \in Y_{\upsilon,\alpha}^i}\,
\e^{-(s_1-1)\, \varepsilon_1^{(i)}-(s_2-1)\, \varepsilon_2^{(i)}}\Big)
\end{multline}
for $i=1,\dots,k$.
Apart from the case $k=2$, generically there seems to be no nice factorizations of this expression into corresponding generating functions for 0-observables on $\R^4$; for $k\geq
3$ the apparent lack of a factorization property for $\Zcal^{\quiv}_{X_k}$ is
due to the fact that it involves terms which depend on pairs of
exceptional divisors intersecting at the fixed points of $X_k$, and
such terms do not split into terms each depending on a single affine
toric subset of $X_k$. In the following we consider various specializations of this generating function.
      
Here we specialise to the generating function for correlators of quadratic 0-observables on $X_k$, and hence provide a rigorous derivation of the Nekrasov master formula for all $\Ncal=2$ quiver gauge theories on $X_k$; a similar computation is done in \cite[Section 4.4]{book:nakajimayoshioka2004} in the context of pure $\Ncal=2$ gauge theory on the blowup of $\C^2$ at a point. For this, we set $\vec{\boldsymbol\tau}=(0,-\tau_{\rm cl}^\upsilon,0,\dots)_{\upsilon\in\quiv_0}$ and $\vec{\boldsymbol t}{}^{\;(1)}= \cdots= \vec{\boldsymbol t}{}^{\:(k-1)}=\vec{\boldsymbol 0}$; we denote the resulting partition function by $\Zcal_{X_k}^{\quiv,\circ}$. Then one finds 
\begin{multline}\label{eq:deformedinstanton}
\Zcal^{\quiv, \circ}_{X_k}\big(\varepsilon_1, \varepsilon_2, \vec{\boldsymbol a},\mubf;\qbf,\boldsymbol\tau_{\mathrm{cl}}, \vec\xibf \ \big)_{\vec{\boldsymbol{w}}}
\\ =\sum_{\stackrel{\scriptstyle \vec{{\boldsymbol{h}}}}{\scriptstyle C\vec h_\upsilon\in\Ufrak_{\vec w_\upsilon}\,,\, \vec{\boldsymbol h}_\upsilon\in\Ufrak_{\vec{\boldsymbol w}_\upsilon}^{\mathrm{conf}}}} \ \sum_{\stackrel{\scriptstyle (\vec{h}_{\upsilon,1}, \ldots, \vec{h}_{\upsilon, r_\upsilon})_{\upsilon\in\quiv_0}}{\scriptstyle \vec{h}_\upsilon=\sum_{\alpha=1}^{r_\upsilon}\,\vec{h}_{\upsilon,\alpha}}}\, \Zcal^{\quiv, \circ}_{X_k}\big(\varepsilon_1, \varepsilon_2, \vec{\boldsymbol{a}},\mubf;\qbf,{\boldsymbol\tau}_{\mathrm{cl}}, \vec\xibf \ \big)_{\vec{\boldsymbol{w}}, (\vec{h}_{\upsilon,\alpha})} \ ,
\end{multline}
where
\begin{multline}
\Zcal^{\quiv, \circ}_{X_k}\big(\varepsilon_1, \varepsilon_2, \vec{\boldsymbol{a}},\mubf;\qbf,{\boldsymbol\tau}_{\mathrm{cl}}, \vec\xibf \ \big)_{\vec{\boldsymbol{w}}, (\vec{h}_{\upsilon,\alpha})}
:=\prod_{\upsilon\in\quiv_0}\, \vec\xi{}_\upsilon^{\ \  \sum\limits_{\alpha=1}^{r_\upsilon}\vec{h}_{\upsilon,\alpha}}\,\qsf_\upsilon^{\frac{1}{2}\,\sum\limits_{\alpha=1}^{r_\upsilon}\, \vec{h}_{\upsilon,\alpha}\cdot C\vec{h}_{\upsilon,\alpha}}\ \ell_{(\vec{h}_{\upsilon,\alpha})}(\varepsilon_1, \varepsilon_2, \vec{\boldsymbol a}, \mubf)\\ 
\times \
\prod_{i=1}^k\, \Zcal_{\C^2}^{\quiv,{\rm inst}}\big(\varepsilon_1^{(i)},\varepsilon_2^{(i)},\vec{\boldsymbol{a}}^{(i)},\mubf;\qbf_{\mathrm{eff}}\big)\ \Zcal_{\C^2}^{\quiv,\mathrm{cl}}\big(\varepsilon_1^{(i)},
\varepsilon_2^{(i)},\vec{\boldsymbol{a}}^{(i)};{\boldsymbol\tau}_{\mathrm{cl}} \big)\ ,
\end{multline}
with $\qbf_{\mathrm{eff}}:=(\qsf_\upsilon \, \e^{\tau^\upsilon_{\mathrm{cl}}})_{\upsilon\in\quiv_0}$ and
\begin{equation}
\Zcal_{\C^2}^{\quiv,\mathrm{cl}}\big(\varepsilon_1^{(i)},
\varepsilon_2^{(i)},\vec{\boldsymbol{a}}^{(i)};{\boldsymbol\tau}_{\mathrm{cl}} \big):= \prod_{\upsilon\in\quiv_0}\, \Zcal_{\C^2}^{\mathrm{cl}}\big(\varepsilon_1^{(i)},
\varepsilon_2^{(i)},\mbf{a}_\upsilon^{(i)};{\tau}^\upsilon_{\mathrm{cl}} \big) \ .
\end{equation}

Define now the full generating function for correlators of quadratic 0-observables for the $\Ncal=2$ superconformal quiver gauge theory on $X_k$ as
\begin{multline}
\Zcal^{\quiv, \mathrm{full}}_{X_k}\big(\varepsilon_1, \varepsilon_2, \vec{\boldsymbol a},\mubf;\qbf,{\boldsymbol\tau}_{\mathrm{cl}}, \vec\xibf \ \big)_{\vec{\boldsymbol{w}}} \\ := 
\sum_{\stackrel{\scriptstyle \vec{{\boldsymbol{h}}}}{\scriptstyle C\vec h_\upsilon\in\Ufrak_{\vec w_\upsilon}\,,\, \vec{\boldsymbol h}_\upsilon\in\Ufrak_{\vec{\boldsymbol w}_\upsilon}^{\mathrm{conf}}}} \ \sum_{\stackrel{\scriptstyle (\vec{h}_{\upsilon,1}, \ldots, \vec{h}_{\upsilon, r_\upsilon})_{\upsilon\in\quiv_0}}{\scriptstyle \vec{h}_\upsilon=\sum_{\alpha=1}^{r_\upsilon}\,\vec{h}_{\upsilon,\alpha}}} \, \Zcal_{X_k}^{\quiv,\mathrm{pert}}(\varepsilon_1, \varepsilon_2, \vec{\boldsymbol{a}},\mubf\,)_{\vec{\boldsymbol{w}}, (\vec{h}_{\upsilon,\alpha})}\\ \times \ \Zcal^{\quiv,\circ}_{X_k}\big(\varepsilon_1, \varepsilon_2, \vec{\boldsymbol a},\mubf;\qbf,{\boldsymbol\tau}_{\mathrm{cl}}, \vec\xibf \ \big)_{\vec{\boldsymbol{w}}, (\vec{h}_{\upsilon,\alpha} )} \ .
\end{multline}
Then
\begin{multline}
\Zcal^{\quiv, \mathrm{full}}_{X_k}\big(\varepsilon_1, \varepsilon_2, \vec{\boldsymbol a},\mubf;\qbf,{\boldsymbol\tau}_{\mathrm{cl}}, \vec\xibf \ \big)_{\vec{\boldsymbol{w}}}\\
=  \sum_{\stackrel{\scriptstyle \vec{{\boldsymbol{h}}}}{\scriptstyle C\vec h_\upsilon\in\Ufrak_{\vec w_\upsilon}\,,\, \vec{\boldsymbol h}_\upsilon\in\Ufrak_{\vec{\boldsymbol w}_\upsilon}^{\mathrm{conf}}}} \ \sum_{\stackrel{\scriptstyle (\vec{h}_{\upsilon,1}, \ldots, \vec{h}_{\upsilon, r_\upsilon})_{\upsilon\in\quiv_0}}{\scriptstyle \vec{h}_\upsilon=\sum_{\alpha=1}^{r_\upsilon}\,\vec{h}_{\upsilon,\alpha}}} \,\prod_{\upsilon\in\quiv_0}\, \vec\xi{}_\upsilon^{\ \  \sum\limits_{\alpha=1}^{r_\upsilon}\vec{h}_{\upsilon,\alpha}}\,\qsf_\upsilon^{\frac{1}{2}\,\sum\limits_{\alpha=1}^{r_\upsilon}\, \vec{h}_{\upsilon,\alpha}\cdot C\vec{h}_{\upsilon,\alpha}} \\ \times 
\ \prod_{i=1}^k\, \Zcal_{\C^2}^{\quiv, \mathrm{full}}\big(\varepsilon_1^{(i)}, \varepsilon_2^{(i)}, \vec{\boldsymbol{a}}^{(i)}, \mubf; \qbf_{\mathrm{eff}} \big)\ .
\end{multline}
Therefore we obtain a factorization of the full generating function $\Zcal^{\quiv, \mathrm{full}}_{X_k}$ in terms of $k$ copies of $\Zcal_{\C^2}^{\quiv, \mathrm{full}}$. This proves the Nekrasov master formula for $\Ncal=2$ quiver gauge theories on $X_k$.

The master formula suggests in particular that the $\Ncal=2$ quiver gauge theories on the ALE space $X_k$, like their counterparts on $\R^4$, are some sort of quantizations of Hitchin systems; we confirm this expectation in an explicit example below. Recall that the low energy limit of $\Ncal=2$ gauge theories on $\R^4$
is completely characterised by the (punctured) Seiberg-Witten curve $\Sigma$ of genus $r:=\sum_{\upsilon\in\quiv_0}\, r_\upsilon$ \cite{art:seibergwitten1994-I}. The
curve $\Sigma$ is equiped with a meromorphic differential
$\lambda_{\rm SW}$, called the {Seiberg-Witten differential}, and
its periods determine the {Seiberg-Witten prepotential} $\Fcal^{\quiv}_{\C^2}(\vec{\mbf
a},\mbf\mu;\qbf)$ which is a holomorphic function of all parameters. Setting $\bar{\mbf m}=\mbf 0$ for brevity, in a
symplectic basis $\{A^\upsilon_{1},B^\upsilon_{1},\dots, A^\upsilon_{r_\upsilon},B^\upsilon_{r_\upsilon},S^\upsilon_1,\dots, S^\upsilon_{m_\upsilon} \}_{\upsilon\in\quiv_0}\cup \{T_e\}_{e\in\quiv_1}$ for the
homology group $H_1(\Sigma;\Z)$, the periods of the Seiberg-Witten
differential determine the quantities
\begin{equation}
a_\alpha^\upsilon= \oint_{A^\upsilon_\alpha}\, \lambda_{\rm SW} \qquad \mbox{and} \qquad \frac{\partial\Fcal^{\quiv}_{\C^2}}{\partial a^\upsilon_\alpha}(\vec{\mbf a},\mbf\mu;\qbf)
= -2\pi\ii \,\oint_{B^\upsilon_\alpha}\, \lambda_{\rm SW} \ ,
\end{equation}
together with the mass parameters
\begin{equation}
\mu^s_\upsilon=\oint_{S^\upsilon_s} \,\lambda_{\rm SW} \qquad \mbox{and}
\qquad  \mu_e=\oint_{T_e}\,\lambda_{\rm SW} \ .
\end{equation}
It follows that the period matrix $\mbf\tau=(\tau^{\upsilon\upsilon'}_{\alpha\beta})$ of the
Seiberg-Witten curve
$\Sigma$ is related to the prepotential by
\begin{equation}
\tau^{\upsilon\upsilon'}_{\alpha\beta} = -\frac1{2\pi\ii}\
\frac{\partial^2\Fcal^{\quiv}_{\C^2}}{\partial a^{\upsilon}_\alpha\, \partial a^{\upsilon'}_\beta}(\vec{\mbf
a},\mubf;\qbf)\ ,
\end{equation}
and it determines the infrared effective gauge couplings. The prepotential is conjecturally recovered from the partition function $\Zcal_{\C^2}^{\quiv, \mathrm{full}}(\varepsilon_1, \varepsilon_2, \vec{\boldsymbol{a}}, \mubf; \qbf \big)$ in the low energy limit in which the equivariant parameters $\varepsilon_1,\varepsilon_2$ vanish. Recall that $\Sigma$ is also the spectral curve of an algebraic integrable system, the Donagi-Witten integrable system \cite{art:donagiwitten1996}, which can be described as a particular Hitchin system on a punctured Riemann surface $C$ such that $\Sigma$ is a branched cover of $C$. For the ADE quivers $\quiv$ we label the vertices $\quiv_0$ of the Dynkin diagram by $\upsilon=0,1,\dots,n$, where $n$ is the rank of the corresponding simply-laced finite-dimensional Lie algebra. Then for the various theories in the ADE classification the spectral data can be described in the following way \cite{art:nekrasovpestun2012} (see also \cite[Section 8]{art:pedrinisalaszabo2014}):
\begin{itemize}
\item For Class I theories $\Sigma$ has $n+1$ punctures and is the spectral curve of a Hitchin system on the Riemann sphere with $n+4$ punctures at $\infty,1,z_1,\dots,z_{n+1},0$, where $z_\upsilon:= \qsf_0\,
\qsf_1\cdots \qsf_{\upsilon-1}$ for $\upsilon=1,\dots,n+1$.

\smallskip

\item For Class II theories $\Sigma$ has no punctures and is the spectral curve of a ${\rm U}(N)$-Hitchin system on an elliptic curve with nome $\qsf:=\qsf_0\,
\qsf_1\cdots\qsf_n$.

\smallskip

\item For Class II$^*$ theories $\Sigma$ has $n+1$ punctures and is the spectral curve of a ${\rm U}(N)$-Hitchin system on an elliptic curve with nome $\qsf$ as above and $n+1$ punctures at $0,z_1,\dots,z_n$, where $z_\upsilon:=  \qsf_1\cdots \qsf_\upsilon$
for $\upsilon=1,\dots,n$.
\end{itemize}

\subsection{$\widehat{A}_{0}$-theory}

Let us consider the quiver $\quiv$ consisting of a single node with a vertex loop edge
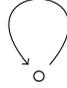
\begin{equation}
  \begin{tikzpicture}[auto]
\node (A0_0) at (0,0) {$\circ$};
    \path[->] (A0_0) edge[loop above, in=130,out=50,looseness=10, shorten >=-2pt, shorten <=-2pt] node {$\scriptstyle{}$} (A0_0);
  \end{tikzpicture}
\end{equation}
The corresponding quiver gauge theory is the $\Ncal=2$ gauge theory with an adjoint hypermultiplet of mass $\mu$, also known as the $\Ncal=2^\ast$ gauge theory. In this case the superconformal constraint is trivially satisfied by any charge vector $\vec{h}$ for a fixed choice of framing vector $\vec{w}$.

In the following we characterise the low energy limit of the $\Ncal=2^\ast$ ${\rm U}(r)$ gauge theory on $X_k$. The spectral curve $\Sigma$ in this case has
genus $r$ and one puncture, and its period matrix $\tau=(\tau_{\alpha\beta})$ is related to the {Seiberg-Witten prepotential} $\Fcal^{\widehat{A}_0}_{\C^2}(\mbf
a,\mu;\qsf)$ by
\begin{equation}
\tau_{\alpha\beta} = -\frac1{2\pi\ii}\
\frac{\partial^2\Fcal^{\widehat{A}_0}_{\C^2}}{\partial a_\alpha\, \partial a_\beta}(\mbf
a,\mu;\qsf)\ .
\end{equation}
The Donagi-Witten integrable system is given by the elliptic Calogero-Moser model of type $A_r$, which can also be described as a ${\rm U}(r)$-Hitchin system on
an elliptic curve with nome $\qsf$ and one puncture at $z=0$. The prepotential can be recovered from the partition function for the $\Omega$-deformed $\Ncal=2^{*}$ gauge theory on $\R^4$ in the low energy limit in which the equivariant parameters $\varepsilon_1,\varepsilon_2$ vanish; this result was originally conjectured by Nekrasov \cite{art:nekrasov2003} and subsequently proven in
\cite{art:nakajimayoshioka2005-I,art:nekrasovokounkov2006}. In \cite[Section 7]{art:bruzzopedrinisalaszabo2013} we prove analogous results for gauge theory on $X_k$. For this, let us define 
\begin{equation}
F_{X_k}^{{\widehat{A}_0}, \mathrm{inst}}(\varepsilon_1,\varepsilon_2,\mbf{a}, \mu;
\qsf, \vec{\xi} \ )_{\vec w} :=- \tilde{k}\, \varepsilon_1\, \varepsilon_2\, \log
\Zcal_{X_k}^{{\widehat{A}_0}, \mathrm{inst}}(\varepsilon_1,\varepsilon_2,\mbf{a},
\mu; \qsf, \vec{\xi} \ )_{\vec w} \ .
\end{equation}
Then $F_{X_k}^{{\widehat{A}_0}, \mathrm{inst}}$ is analytic in $\varepsilon_1,\varepsilon_2$ near $\varepsilon_1=\varepsilon_2=0$ and
\begin{equation}
\lim_{\varepsilon_1,\varepsilon_2\to 0}\, F_{X_k}^{{\widehat{A}_0},
  \mathrm{inst}}\big(\varepsilon_1,\varepsilon_2,\mbf{a}, \mu; \qsf,
\vec{\xi} \ \big)_{\vec w} = \frac{1}{k} \, \Fcal_{\C^2}^{{\widehat{A}_0}, \mathrm{inst}}(\mbf{a}, \mu; \qsf)\ ,
\end{equation}
where $\Fcal_{\C^2}^{{\widehat{A}_0}, \mathrm{inst}}(\mbf{a}, \mu; \qsf)$ is the instanton part of the Seiberg-Witten prepotential of $\Ncal=2^\ast$ gauge theory on $\R^4$; the same result is true of the perturbative contributions. We can also define a prepotential associated with the generating function for correlators of quadratic 0-observables by letting
\begin{equation}
F_{X_k}^{{\widehat{A}_0}, \circ}(\varepsilon_1,\varepsilon_2,\mbf{a}, \mu; \qsf,
\vec{\xi}, \tau_{\mathrm{cl}})_{\vec w} :=- \tilde{k}\, \varepsilon_1\, \varepsilon_2\, \log
\Zcal_{X_k}^{{\widehat{A}_0}, \circ}(\varepsilon_1,\varepsilon_2,\mbf{a}, \mu ; \qsf, \vec{\xi}, \tau_{\mathrm{cl}})_{\vec w} \ .
\end{equation}
Then $F_{X_k}^{{\widehat{A}_0}, \circ}$ is also analytic in $\varepsilon_1,\varepsilon_2$ near $\varepsilon_1=\varepsilon_2=0$ and
\begin{equation}
\lim_{\varepsilon_1,\varepsilon_2\to 0}\, F_{X_k}^{{\widehat{A}_0},
  \circ}\big(\varepsilon_1,\varepsilon_2,\mbf{a}; \qsf,
\vec{\xi},\tau_{\mathrm{cl}})_{\vec w} = \frac1k\, \Big(\,
\frac{\tilde{k} \, \tau_{\mathrm{cl}}}{2}\, \sum_{\alpha=1}^r \, a_\alpha^2 +
\Fcal_{\C^2}^{{\widehat{A}_0}, \mathrm{inst}}(\mbf{a}, \mu ; \qsf_{\rm eff})
\Big) \ .
\end{equation}
These results confirm that the $\Ncal=2^*$ gauge theory on the ALE space $X_k$ is a quantization of a Hitchin system with spectral curve $\Sigma$.

Let us focus now on the case $k=2$, and denote by $\Zcal_{X_2}^{\widehat{A}_0,\bullet}(\varepsilon_1, \varepsilon_2, \mbf{a},
\mu ; \qsf, \vec{\xi}, t)_{\vec w}$ the generating function for correlators of quadratic 2-observables of $\Ncal=2^*$ gauge theory on $X_2$ which is obtained from the generating function
$\Zcal_{X_2}^{\widehat{A}_0}(\varepsilon_1,\varepsilon_2, \mbf{a}, \mu ; \qsf,
\vec{\xi}, \vec{\tau}, \vec{t} \ )_{\vec w}$ specialised at $\vec{\tau}=\vec 0$
and $\vec{t}:=(0, -t, 0, \ldots)$. Then we have the factorization formula (cf.\ \cite[Example 5.11 and Section 7.3]{art:bruzzopedrinisalaszabo2013})
\begin{multline}
\Zcal_{X_2}^{\widehat{A}_0,\bullet}(\varepsilon_1, \varepsilon_2, \mbf{a}, \mu; \qsf, \xi, t)_{\vec w} =
\sum_{\genfrac{}{}{0pt}{}{\scriptstyle h\in\frac{1}{2}\, \Z}{\scriptstyle 2h=  w_1\bmod{2} }}
 \, \xi^{h} \\
\times \ 
  \sum_{\stackrel{\scriptstyle {\boldsymbol{h}}=(h_1, \ldots, h_r)}{h=\sum_{\alpha=1}^r\, h_\alpha}} \, \big(\qsf \,
  \e^{2t\, (\varepsilon_1+\varepsilon_2)} \big)^{\sum_{\alpha=1}^r\limits
    \, h_\alpha^2} \, \e^{-2t\, \sum_{\alpha=1}^r\limits\, h_\alpha\,
    a_\alpha} \ \prod\limits_{\alpha,\beta=1}^r\,
\frac{\ell_{h_{\beta\alpha}}(2\varepsilon_1, \varepsilon_2-\varepsilon_1,
  a_{\beta\alpha}+\mu)}{\ell_{h_{\beta\alpha}}(2\varepsilon_1, \varepsilon_2-\varepsilon_1 , a_{\beta\alpha})} \\
\times \ \Zcal_{\C^2}^{\widehat{A}_0 ,
  \mathrm{inst}}\big(2\varepsilon_1,\varepsilon_2-\varepsilon_1,
\mbf{a}-2\varepsilon_1\, {\boldsymbol{h}}, \mu ; \qsf \,
\e^{2\varepsilon_1\, t}\big)\, \Zcal_{\C^2}^{\widehat{A}_0 ,
  \mathrm{inst}}\big(\varepsilon_1-\varepsilon_2 ,2\varepsilon_2 ,
\mbf{a}-2\varepsilon_2\, {\boldsymbol{h}}, \mu; \qsf \,
\e^{2\varepsilon_2\, t}\big) \ .
\end{multline}
We shall describe blowup equations which relate $\Zcal_{X_2}^{{\widehat{A}_0},\bullet}$ to the instanton
partition function $\Zcal_{X_2}^{{\widehat{A}_0},{\rm inst}}$ in the low energy limit. Let
$\Theta\big[\genfrac{}{}{0pt}{}{\mbf{\mu}}{\mbf{\nu}}\big](\mbf{\zeta}\,\vert\,
\tau)$ be the Riemann theta-function with characteristic $\big[\genfrac{}{}{0pt}{}{\mbf{\mu}}{\mbf{\nu}}\big]$ on the Seiberg-Witten curve
$\Sigma$ for $\Ncal=2^\ast$ gauge theory on $\R^4$. Then the ratio
\begin{equation}
\Zcal_{X_2}^{{\widehat{A}_0},\bullet}(\varepsilon_1, \varepsilon_2, \mbf{a},
\mu; \qsf, \xi, t)_{\vec w} \, \big/ \, \Zcal^{{\widehat{A}_0}, \mathrm{inst}}_{X_2}(\varepsilon_1,
\varepsilon_2, \mbf{a},\mu ; \qsf, \xi)_{\vec w}
\end{equation}
is analytic in
$\varepsilon_1, \varepsilon_2$ near $\varepsilon_1=\varepsilon_2=0$, and
\begin{multline}
\lim_{\varepsilon_1,\varepsilon_2\to0}\ \frac{\Zcal_{X_2}^{{\widehat{A}_0},\bullet}(\varepsilon_1, \varepsilon_2, \mbf{a},
\mu; \qsf, \xi, t)_{\vec w}}{\Zcal^{{\widehat{A}_0}, \mathrm{inst}}_{X_2}(\varepsilon_1,
\varepsilon_2, \mbf{a},\mu ; \qsf, \xi)_{\vec w}} \\
=\exp\bigg(\Big(\qsf \, \frac{\partial}{\partial \qsf}\Big)^2
\Fcal_{\C^2}^{{\widehat{A}_0}, \mathrm{inst}}(\mbf{a}, \mu ; \qsf)\, t^2+2\pi
\operatorname{i}\, \sum_{\alpha=w_0+1}^r \, \zeta_\alpha \bigg)\,
\frac{\Theta\big[\genfrac{}{}{0pt}{}{0}{C\, \mbf{\nu}}\big](C\,
  (\mbf{\zeta}+\mbf{\kappa})\,\vert\, C\,
  \tau)}{\Theta\big[\genfrac{}{}{0pt}{}{0}{C\, \mbf{\nu}}\big](C\,
  \mbf{\kappa}\,\vert\, C\, \tau)}\ ,
\end{multline}
where
\begin{equation}
\zeta_\alpha:=-\frac{t}{2\pi \operatorname{i}} \, \Big(a_\alpha+\qsf
\, \frac{\partial^2\Fcal_{\C^2}^{{\widehat{A}_0},
    \mathrm{inst}}}{\partial\qsf\, \partial a_\alpha}(\mbf{a}, \mu ;
\qsf) \Big) \ ,
\end{equation}
while $\kappa_\alpha:=\frac{1}{4\pi\operatorname{i}}\, \log(\xi)$ for
$\alpha=1,\ldots,r$ and
\begin{equation*}
\nu_\alpha:=\left\{
\begin{array}{ll}
\displaystyle
\sum_{\beta=w_0+1}^r\, \log
\left((a_{\beta}-a_{\alpha})^2-\mu^2\right)-\frac{2\pi\operatorname{i}w_1}{r}\,
\tau_0 & \\[5pt]
\displaystyle
\hskip20mm + \, \sum_{\beta=w_0+1}^r \, \frac{\partial^2 \Fcal_{\C^2}^{{\widehat{A}_0},
    \mathrm{inst}}}{\partial a_\alpha \, \partial a_\beta}(\mbf{a},
\mu ; \qsf) & \mbox{ for } \alpha=1, \ldots, w_0\ , \\[10pt]
\displaystyle
-\sum_{\beta=1}^{w_0}\, \log
\left((a_{\beta}-a_{\alpha})^2-\mu^2\right)+\frac{4\pi\operatorname{i}w_0}{r}\,
\tau_0 & \mbox{ for } \alpha=w_0+1, \ldots, r\ .
\end{array}\right.
\end{equation*}
If the fixed holonomy at infinity is trivial, i.e.
$\vec{w}=(w_0,w_1)=(r,0)$, the characteristic vector $\mbf{\nu}\in
\C^r$ vanishes and this result resembles
\cite[Theorem~8.1]{art:nakajimayoshioka2005-I} and
\cite[Equation~(2.25)]{art:bonellimaruyoshitanziniyagi2012}. In
general, the nontrivial holonomy at infinity is encoded in
$\mbf{\nu}$. It would be interesting to extend this analysis to the low energy limit of all $\Ncal=2$ superconformal quiver gauge theories on ALE spaces in the spirit of \cite{art:nekrasovpestun2012}.

This blowup equation underlies the ${\rm Sp}(2r,\Z)$ modularity properties of the
partition function and of the correlators of
quadratic $2$-observables on the Seiberg-Witten curve for $\Ncal=2^{\ast}$
gauge theory on $X_2$ with period matrix $\tau$ twisted by the $A_1$
Cartan matrix $C$. It
generalises the representation of the Vafa-Witten partition function for the $\Ncal=4$ superconformal gauge theory on $X_k$ \cite{art:vafawitten1994}
at $\mu=0$ in terms of modular forms as (cf.\ \cite[Section 5.3]{art:bruzzopedrinisalaszabo2013})
\begin{equation}
\Zcal^{\mathrm{VW}}_{X_k}\big(\qsf, \vec{\xi}\
\big)_{\vec w} :=\lim_{\mu \to 0}\,\Zcal^{\widehat{A}_0,\mathrm{inst}}_{X_k}\big(\varepsilon_1, \varepsilon_2,
\mbf{a}, \mu ; \qsf, \vec{\xi}\ \big)_{\vec w} = \qsf^{\frac{r\,k}{24}}\,
\prod_{j=0}^{k-1}\, \Big(\, \frac{\chi^{\widehat{\omega}_{j}}(\zeta, \tau_0 )}{\eta(\tau_0)}\, \Big)^{w_j}\ ,
\end{equation}
Here $\chi^{\widehat{\omega}_{j}}(\zeta, \tau_0)$ is the character of the integrable highest weight representation of the affine Kac-Moody algebra 
$\widehat{\slfrak}(k)$ at level one with highest weight the $i$-th fundamental weight $\widehat{\omega}_i$ of type $\widehat{A}_{k-1}$ for
$i=0,1,\dots,k-1$, which can be expressed as a combination of string functions and theta functions as \cite[Section~14.4]{book:difrancescomathieusenechal1997}
\begin{equation}
\chi^{\widehat{\omega}_{j}}(\zeta, \tau_0) =\eta(\tau_0)^{-k+1}\, \sum_{\vec\gamma\,^\vee \in
  \Qfrak^\vee}\, \qsf^{\frac{1}{2}\,
  \vert\vec\gamma\,^\vee+\vec\omega_i\vert^2}\, \e^{2\pi \ii
  (\vec\gamma\,^\vee+\vec\omega_i) \cdot \zeta}\ ,
\end{equation}
where $\xi_j=\exp(2\pi
\operatorname{i}\, (2\zeta_{j}-\zeta_{j-1}-\zeta_{j+1}))$ (we set
$\zeta_0=\zeta_k=0$) and $\Qfrak^\vee:= \bigoplus_{i=1}^{k-1}\, \Z\vec\gamma_i{}^\vee$ is the coroot lattice of the Dynkin diagram of type $A_{k-1}$ with
$\vec\gamma_i{}^\vee$ the $i$-th simple coroot defined by
$\vec\gamma_i\cdot \vec\gamma_j{}^\vee =\delta_{ij}$ for $i,j=1, \ldots,
k-1$. The inverse of the Dedekind eta function
\begin{equation}
\eta(\tau_0)=\qsf^{\frac{1}{24}}\, \prod_{n=1}^\infty \, \big(1-\qsf^n \big)
\end{equation}
is the character of the
Heisenberg algebra $\hfrak$. Thus in this limit we correctly reproduce the
character of the representation of the affine Lie algebra
$\widehat{\mathfrak{gl}}(k)_r$, and hence confirm the ${\rm SL}(2,\Z)$ modularity
(S-duality) of the partition function. Note that in this case the prepotential is $\frac12\, \tau_0 \, \sum_{\alpha=1}^r a_\alpha^2$ and there are no quantum corrections to the Hitchin system.

\appendix

\bigskip
\section{Equivariant cohomology\label{app:eqcoh}}

In this appendix we summarise various aspects of equivariant
cohomology that have been used to compute instanton partition functions in
the main text.

\subsection{Definitions}

Let us fix a complex algebraic torus $T:=\big(\C^\ast\big)^n$. Denote by
$M:=\Hom_\Z(T, \C^\ast)$ the lattice of characters $\chi\colon T\to\C^\ast$ and by
$N=\Hom_\Z(M, \Z)$ the dual lattice of one-parameter subgroups of $T$. 

Let $E_T$ be a contractible space on which $T$ acts freely and set
$BT:=E_T/T$; in the following we shall choose
$E_T=\big(\C^\infty\setminus \{0\}\big)^n$ and
$BT=\big(\PP^\infty\big)^n$. Suppose that $T$ acts on a topological
space $X$. The Borel equivariant cohomology of $X$ is defined as
\begin{equation}
H_T^\ast(X):=H^\ast(X\times_T E_T; \C)\ .
\end{equation}
If $X$ is a point, $H_T^\ast(\mathrm{pt})=H^\ast(BT; \C)$. By
using the pullback via the collapsing map $X\to \mathrm{pt}$, one
makes  $H_T^\ast(X)$ into a module over the ring $H_T^\ast(\mathrm{pt})=H^\ast(BT; \C)$. 

Let us denote by $L_\chi$ the
  line bundle on $\big(\PP^\infty\big)^n$ associated to the principal $T$-bundle 
  $\big(\C^\infty\setminus \{0\}\big)^n\to\big(\PP^\infty\big)^n$ via a character $\chi$. The assignment
$\chi\mapsto -\crm_1(L_\chi)$ defines an isomorphism $\psi\colon
M\xrightarrow{\sim} H^2(BT; \C)$ which induces a ring isomorphism
$\mathrm{Sym}(M)\simeq H^\ast(BT; \C)$. For a character $\chi$, we
call $\psi(\chi)$ the weight of $\chi$. In particular, if $\chi_i$ is the character of $T$ defined by $(t_1,\ldots, t_n)\mapsto t_i$ for $i=1, \ldots, n$, then
\begin{equation}
H^\ast_T(\mathrm{pt})=H^\ast(BT;\C)=\C\big[\psi(\chi_1), \ldots,
\psi(\chi_n) \big] \ .
\end{equation}

Let $V$ be an equivariant vector bundle on $X$, i.e. a vector bundle
over $X$ such that the action of $T$ on $X$ lifts to an action on $V$
which is linear on its fibres. Then $V_T:=V\times_T E_T$ is a vector bundle over $X_T:= X\times_T E_T$
and the equivariant Chern classes $(\crm_l)_T(V)\in H_T^\ast(X)$ are
defined to be the ordinary Chern classes $\crm_l(V_T)\in H^\ast(X_T;\C)$. The top Chern
class of $V_T$ is called the equivariant Euler class of $V$ and is denoted by $\eu_T(V)$.

\subsection{Localization theorem}

Assume that the fixed point locus $X^T:=\{p_1, \ldots, p_k\}$ of $X$
consists of a finite number of points with $k\geq 1$. The equivariant inclusion
$\imath_l\colon \{p_l\}\hookrightarrow X$ induces a pullback
$\imath_l^\ast: H_T^\ast(X)\to H_T^\ast(p_l)$ and a Gysin map ${\imath_l}_!\colon H_T^\ast(p_l)\to H_T^\ast(X)$.

Let $\mathrm{Frac}\big(H_T^\ast(\mathrm{pt}) \big)$ be the field of fractions
of the ring $H_T^\ast(\mathrm{pt})$. For $l=1, \ldots, k$ the equivariant Euler
class $\eu_T(T_{p_l} X)$ of the tangent space to $X$ at the fixed
point $p_l$ is invertible in
$H_T^\ast(p_l)\otimes_{H_T^\ast(\mathrm{pt})}
\mathrm{Frac}\big(H_T^\ast(\mathrm{pt}) \big)$.
There is an isomorphism
\begin{equation}
H_T^\ast(X)\otimes_{H_T^\ast(\mathrm{pt})}
\mathrm{Frac}\big(H_T^\ast(\mathrm{pt})\big) \ \xrightarrow{ \ \sim \
} \ \bigoplus_{l=1}^k \, H_T^\ast(p_l)\otimes_{H_T^\ast(\mathrm{pt})}
\mathrm{Frac}\big(H_T^\ast(\mathrm{pt}) \big)
\end{equation}
induced by the map
\begin{equation}
\alpha \ \longmapsto \ \Big(\,
\frac{\imath_l^\ast(\alpha)}{\eu_T(T_{p_l} X)}\, \Big)_{l=1,\dots,k} \ .
\end{equation}
The inverse map is induced by $(\alpha_l)_{l=1,\dots,k} \mapsto
\sum_{l=1}^k \, {\imath_{l}}_!(\alpha_l)$.

\bigskip 
\section{Edge contributions\label{app:edgecontributions}}

In this appendix we list the edge contributions
$\ell^{(n)}_{\vec{h}}
\big(e_1,e_2,a \big)$
to the $T$-equivariant Euler class of the tangent bundle $T\Mcal_{\vec{u},\Delta,\vec
  w}$
which were derived in \cite[Section
4.7 and Appendix C]{art:bruzzopedrinisalaszabo2013}. For this, we
first introduce some notation. Let
$c\in\{0,1,\dots,k-1\}$ be the equivalence class of $k\,
h_{k-1}$ modulo $k$. Set $(C^{-1})^{n ,0}=0$ for
$n\in\{1, \ldots, k-1\}$ and $(C^{-1})^{k, c}=0$. We denote by $\lfloor x\rfloor\in\Z$
  the integer part and by $\{x\}:=x-\lfloor
  x\rfloor\in[0,1)$ the fractional part of a rational 
  number $x$.

If $h_n-(C^{-1})^{n, c}> 0$ for $n\in\{1, \ldots, k-1\}$, consider the equation
\small
\begin{multline}\label{eq:cond-index+}
i^2-i\, \Big(\vec{h}-\sum_{p=1}^{n-1}\,
\big(h_p-(C^{-1})^{p, c}\big)\,
\vec{e}_p\Big)\cdot C\vec{e}_n +\frac{1}{2}\,
\bigg(\Big(\vec{h}-\sum_{p=1}^{n-1}\,
\big(h_p-(C^{-1})^{p, c}\big)\,
\vec{e}_p\Big) \\
\cdot C\Big(\vec{h}-\sum_{p=1}^{n-1}\,
\big(h_p-(C^{-1})^{p ,c}\big)\,
\vec{e}_p\Big) -(C^{-1})^{c, c}\bigg)=0\ ,
\end{multline}
\normalsize
and define the set 
\begin{equation}
\splus_n:=\big\{i\in\Z_{>0}\ \big\vert\ i\leq
h_n-(C^{-1})^{n, c} \mbox{ is a
  solution of Equation \eqref{eq:cond-index+}} \big\}\ .
\end{equation}
Let $\dplus_n=\min(\splus_n)$ if $\splus_n\neq \emptyset$, otherwise $\dplus_n:=h_n-(C^{-1})^{n, c}$.

If $h_n-(C^{-1})^{n, c}< 0$, consider the equation
\small
\begin{multline}\label{eq:cond-index-}
i^2+i\, \Big(\vec{h}-\sum_{p=1}^{n-1}\,
\big(h_p-(C^{-1})^{p, c}\big)\,
\vec{e}_p\Big)\cdot C\vec{e}_n +\frac{1}{2}\,
\bigg(\Big(\vec{h}-\sum_{p=1}^{n-1}\,
\big(h_p-(C^{-1})^{p, c}\big)\,
\vec{e}_p\Big) \\
\cdot C\Big(\vec{h}-\sum_{p=1}^{n-1}\,
\big(h_p-(C^{-1})^{p, c}\big)\,
\vec{e}_p\Big)-(C^{-1})^{c, c}\bigg) =0\ ,
\end{multline}
\normalsize
and define the set 
\begin{equation}
\sminus_n:=\big\{i\in\Z_{>0} \ \big\vert\ i\leq
-h_n+(C^{-1})^{n ,c} \mbox{ is a
  solution of Equation \eqref{eq:cond-index-}} \big\}\ .
\end{equation}
Let $\dminus_n=\min(\sminus_n)$ if $\sminus_n\neq \emptyset$, otherwise $\dminus_n:=-h_n+(C^{-1})^{n, c}$.

Define $m$ to be the smallest index $n\in\{1, \ldots, k-1\}$ such that $\splus_n$ or $\sminus_n$ is nonempty, otherwise $m:=k-1$.

Then for fixed $n=1, \ldots, m$ we set:
\begin{itemize} 
\item[\scriptsize$\blacksquare$] If $h_n-(C^{-1})^{n, c}> 0$:
\smallskip
\begin{itemize} 
\item[$\bullet$] For $\delta_{n,c} -h_{n+1}+(C^{-1})^{n+1, c}+2(h_{n}-(C^{-1})^{n, c}-\dplus_n)\geq 0$:
\begin{multline}
\ell^{(n)}_{\vec{h}}\big(e_1,
e_2, a \big)=
\prod_{i=h_{n}-(C^{-1})^{n,
    c}-\dplus_n}^{h_n-(C^{-1})^{n
    , c}-1}\hspace{4mm} \prod_{j=0}^{2i+\delta_{n,c} -h_{n+1}+(C^{-1})^{n+1, c}}\\ 
\bigg(a+\Big(i+\left\lfloor\frac{\delta_{n,c}
    -h_{n+1}+(C^{-1})^{n+1, c}}{2}
\right\rfloor\Big)\, e_1+j\, e_2\bigg)\ .
\end{multline}
\smallskip
\item[$\bullet$]For $2\leq \delta_{n,c}-h_{n+1}+(C^{-1})^{n+1, c}+2(h_n-(C^{-1})^{n, c})<2 \dplus_n$:
\begin{multline}
\ell^{(n)}_{\vec{h}}\big(e_1,
e_2, a\big)=
\prod_{i=h_n-(C^{-1})^{n, c}-\dplus_n}^{-\big\lfloor \frac{\delta_{n,c} -h_{n+1}+(C^{-1})^{n+1, c}}{2}\big\rfloor - 1}\hspace{4mm} \prod_{j=1}^{2i- (\delta_{n,c} -h_{n+1}+(C^{-1})^{n+1, c})-1}\\
\shoveright{\bigg(a+\Big(i-\left\lfloor
    -\frac{\delta_{n,c}
      -h_{n+1}+(C^{-1})^{n+1,
        c}}{2}\right\rfloor\Big)\,
  e_1-j\, e_2\bigg)^{-1}\times}   \\[4pt]
\shoveleft{ \ \prod_{i=-\big\lfloor \frac{\delta_{n,c} -h_{n+1}+(C^{-1})^{n+1, c}}{2}\big\rfloor}^{2(h_n-(C^{-1})^{n, c})+\delta_{n,c} -h_{n+1}+(C^{-1})^{n+1, c} -2} \hspace{6mm}\prod_{j=0}^{2i+\delta_{n,c}-h_{n+1}+(C^{-1})^{n+1, c}}} \\[4pt]
  \bigg(a+\Big(i+\left\lfloor \frac{\delta_{n,c} -h_{n+1}+(C^{-1})^{n+1, c}}{2}\right\rfloor\Big)\, e_1-j\, e_2\bigg)\ .
\end{multline}
\smallskip
\item[$\bullet$]For $\delta_{n,c}
  -h_{n+1}+(C^{-1})^{n+1,
    c}<2-2(h_n-(C^{-1})^{n , c})$:
\begin{multline}
\ell^{(n)}_{\vec{h}}\big(e_1,
e_2, a \big)=\prod_{i=h_{n}-(C^{-1})^{n, c}-\dplus_n}^{h_n-(C^{-1})^{n, c}-1}\hspace{0.2cm} \prod_{j=1}^{-2i-\delta_{n,c} +h_{n+1}-(C^{-1})^{n+1, c}-1}\\[4pt]
\bigg(a+\Big(i-\left\lfloor -
  \frac{\delta_{n,c}
    -h_{n+1}+(C^{-1})^{n+1, c}}{2}
\right\rfloor\Big)\, e_1-j\, e_2\bigg)^{-1} \ .
\end{multline}
\smallskip
\end{itemize}
\item[\scriptsize$\blacksquare$] If $h_n-(C^{-1})^{n, c}= 0$:
\begin{equation}
\qquad\ell^{(n)}_{\vec{h}}\big(e_1,
e_2, a \big)=1\ .
\end{equation}
\item[\scriptsize$\blacksquare$] If $h_n-(C^{-1})^{n, c}< 0$:
\smallskip
\begin{itemize}
\item[$\bullet$]For $\delta_{n,c} -h_{n+1}+(C^{-1})^{n+1, c}+2h_n-2(C^{-1})^{n, c} < 2-2\dminus_n$:
\begin{multline}
\ell^{(n)}_{\vec{h}}\big(e_1,
e_2, a \big)=\prod_{i=1-h_{n}+(C^{-1})^{n, c}-\dminus_n}^{-h_{n}+(C^{-1})^{n, c}}\hspace{6mm} \prod_{j=1}^{2i-(\delta_{n,c} -h_{n+1}+(C^{-1})^{n+1, c})-1}
\\[4pt]
 \bigg(a-\Big(i+\left\lfloor
   -\frac{\delta_{n,c}
     -h_{n+1}+(C^{-1})^{n+1,
       c}}{2}\right\rfloor\Big)\,
 e_1-j\, e_2\bigg)\ .
\end{multline}
\item[$\bullet$]For $2-2 \dminus_n\leq \delta_{n,c} -h_{n+1}+(C^{-1})^{n+1, c}+2h_n-2(C^{-1})^{n, c}<0$:
\smallskip
\begin{multline}
\ell^{(n)}_{\vec{h}}\big(e_1,
e_2, a\big) =
\prod_{i=\big\lfloor \frac{\delta_{n,c} -h_{n+1}+(C^{-1})^{n+1, c}}{2} \big\rfloor+1}^{-h_{n}+(C^{-1})^{n, c}}\hspace{6mm} \prod_{j=1}^{2i-(\delta_{n,c} -h_{n+1}+(C^{-1})^{n+1, c})-1 }\\[4pt]
\shoveright{\bigg(a-\Big(i+\left\lfloor
    -\frac{\delta_{n,c}
      -h_{n+1}+(C^{-1})^{n+1,
        c}}{2}\right\rfloor\Big)\,
  e_1-j\, e_2\bigg)\times}\\[4pt]
\shoveleft{ \ \prod_{i=1-h_{n}+(C^{-1})^{n, c}-\dminus_n}^{\big\lfloor \frac{\delta_{n,c} -h_{n+1}+(C^{-1})^{n+1, c}}{2} \big\rfloor}\hspace{0.2cm} \prod_{j=0}^{-2i+\delta_{n,c} -h_{n+1}+(C^{-1})^{n+1, c}}}\\[4pt]
\bigg(a+\Big(-i+\left\lfloor
  \frac{\delta_{n,c}
    -h_{n+1}+(C^{-1})^{n+1, c}}{2}
\right\rfloor\Big)\, e_1+j\, e_2\bigg)^{-1}\ .
\end{multline}
\smallskip
\item[$\bullet$]For $\delta_{n,c} -h_{n+1}+(C^{-1})^{n+1, c}\geq -2h_n+2(C^{-1})^{n, c}$:
\begin{multline}
\ell^{(n)}_{\vec{h}}\big(e_1,
e_2, a \big) =\prod_{i=1-{n}+(C^{-1})^{n, c}-\dminus_n}^{-(\vec{h})_{n}+(C^{-1})^{n, c}}\hspace{6mm} \prod_{j=0}^{-2i+\delta_{n,c}-(\vec{h})_{n+1}+(C^{-1})^{n+1, c}}\\[4pt]
\bigg(a+\Big(-i+\left\lfloor
  \frac{\delta_{n,c}
    -(\vec{h})_{n+1}+(C^{-1})^{n+1,
      c}}{2}\right\rfloor\Big)\, e_1+j\,
e_2\bigg)^{-1}\ .
\end{multline}
\smallskip
\end{itemize}
\end{itemize}
For $n=m+1, \ldots, k-1$ we set
\begin{equation}
\ell^{(n)}_{\vec{h}}\big(e_1,
e_2, a \big)=1\ .
\end{equation}
Note that for any fixed $n\in\{1,\dots,k-1\}$,
$d_n^\pm=0$ implies $\ell^{(n)}_{\vec{h}}
\big(e_1,e_2, a
\big)=1$.

\end{document}